\documentclass[12pt]{article}

\usepackage{epsf}

\makeatletter
 
\def\section{\@startsection {section}{1}{\z@}{+3.0ex plus +1ex minus
  +.2ex}{2.3ex plus .2ex}{\normalsize\bf}}
\def\subsection{\@startsection{subsection}{2}{\z@}{+2.5ex plus +1ex
minus +.2ex}{1.5ex plus .2ex}{\normalsize\bf}}
\def\subsubsection{\@startsection{subsubsection}{3}{\z@}{+3.25ex plus
 +1ex minus +.2ex}{1.5ex plus .2ex}{\normalsize\bf}}
 
\expandafter\ifx\csname mathrm\endcsname\relax\def\mathrm#1{{\rm #1}}\fi
 
\makeatletter

\newcount\@tempcntc
\def\@citex[#1]#2{\if@filesw\immediate\write\@auxout{\string\citation{#2}}\fi
  \@tempcnta\z@\@tempcntb\m@ne\def\@citea{}\@cite{\@for\@citeb:=#2\do
    {\@ifundefined
       {b@\@citeb}{\@citeo\@tempcntb\m@ne\@citea
        \def\@citea{,\penalty\@m\ }{\bf ?}\@warning
       {Citation `\@citeb' on page \thepage \space undefined}}%
    {\setbox\z@\hbox{\global\@tempcntc0\csname
b@\@citeb\endcsname\relax}%
     \ifnum\@tempcntc=\z@ \@citeo\@tempcntb\m@ne
       \@citea\def\@citea{,\penalty\@m}
       \hbox{\csname b@\@citeb\endcsname}%
     \else
      \advance\@tempcntb\@ne
      \ifnum\@tempcntb=\@tempcntc
      \else\advance\@tempcntb\m@ne\@citeo
      \@tempcnta\@tempcntc\@tempcntb\@tempcntc\fi\fi}}\@citeo}{#1}}

\def\@citeo{\ifnum\@tempcnta>\@tempcntb\else\@citea
  \def\@citea{,\penalty\@m}%
  \ifnum\@tempcnta=\@tempcntb\the\@tempcnta\else
   {\advance\@tempcnta\@ne\ifnum\@tempcnta=\@tempcntb \else
\def\@citea{--}\fi
    \advance\@tempcnta\m@ne\the\@tempcnta\@citea\the\@tempcntb}\fi\fi}

\def\nl{\nonumber\\}

\def\asymp#1%
{\mathrel{\raisebox{-.4em}{$\widetilde{\scriptstyle #1}$}}}
\def\Nlim#1{\mathrel{\raisebox{-.4em}
{$\stackrel{\disp\longrightarrow}{\scriptstyle#1}$}}}
\def\Nequal#1%
{\mathrel{\raisebox{-.5em}{$\stackrel{=}{\scriptstyle\rm#1}$}}}

\def\beq{\begin{equation}}
\def\eeq{\end{equation}}
\def\beqar{\begin{eqnarray}}
\def\eeqar{\end{eqnarray}}
\def\barr#1{\begin{array}{#1}}
\def\earr{\end{array}}
\def\bfi{\begin{figure}}
\def\efi{\end{figure}}
\def\btab{\begin{table}}
\def\etab{\end{table}}
\def\bce{\begin{center}}
\def\ece{\end{center}}
\def\nn{\nonumber}
\def\disp{\displaystyle}
\def\text{\textstyle}
\def\fs{\footnotesize}

\def\al{\alpha}
\def\be{\beta}
\def\Ga{\Gamma}
\def\ga{\gamma}
\def\de{\delta}
\def\De{\Delta}

\def\si{\sigma}
\def\Si{\Sigma}

\def\refeq#1{\mbox{(\ref{#1})}}
\def\refeqs#1{\mbox{(\ref{#1})}}
\def\refeqf#1{\mbox{(\ref{#1})}}
\def\reffi#1{\mbox{Fig.~\ref{#1}}}
\def\reffis#1{\mbox{Figs.~\ref{#1}}}
\def\refta#1{\mbox{Table~\ref{#1}}}
\def\reftas#1{\mbox{Tables~\ref{#1}}}
\def\refse#1{\mbox{Section~\ref{#1}}}
\def\citere#1{\mbox{Ref.~\cite{#1}}}
\def\citeres#1{\mbox{Refs.~\cite{#1}}}
 
\newcommand{\TeV}{\unskip\,\mathrm{TeV}}
\newcommand{\GeV}{\unskip\,\mathrm{GeV}}
\newcommand{\MeV}{\unskip\,\mathrm{MeV}}
\newcommand{\pb}{\unskip\,\mathrm{pb}}
\newcommand{\fb}{\unskip\,\mathrm{fb}}

\newcommand{\rd}{{\mathrm{d}}}
\newcommand{\rU}{{\mathrm{U}}}
\newcommand{\rL}{{\mathrm{L}}}
\newcommand{\rT}{{\mathrm{T}}}

\newcommand{\Oa}{\mathswitch{{\cal{O}}(\alpha)}}

\newcommand{\M}{{\cal{M}}}

\def\mathswitchr#1{\relax\ifmmode{\mathrm{#1}}\else$\mathrm{#1}$\fi}

\newcommand{\PW}{\mathswitchr W}
\newcommand{\PZ}{\mathswitchr Z}

\newcommand{\PH}{\mathswitchr H}
\newcommand{\Pe}{\mathswitchr e}

\newcommand{\Pd}{\mathswitchr d}
\newcommand{\Pu}{\mathswitchr u}
\newcommand{\Ps}{\mathswitchr s}
\newcommand{\Pc}{\mathswitchr c}
\newcommand{\Pb}{\mathswitchr b}
\newcommand{\Pt}{\mathswitchr t}

\newcommand{\Pep}{\mathswitchr {e^+}}
\newcommand{\Pem}{\mathswitchr {e^-}}
\newcommand{\PWp}{\mathswitchr {W^+}}
\newcommand{\PWm}{\mathswitchr {W^-}}

\def\mathswitch#1{\relax\ifmmode#1\else$#1$\fi}

\newcommand{\MW}{\mathswitch {M_\PW}}

\newcommand{\MZ}{\mathswitch {M_\PZ}}
\newcommand{\MH}{\mathswitch {M_\PH}}
\newcommand{\Me}{\mathswitch {m_\Pe}}

\newcommand{\Md}{\mathswitch {m_\Pd}}
\newcommand{\Mu}{\mathswitch {m_\Pu}}
\newcommand{\Ms}{\mathswitch {m_\Ps}}
\newcommand{\Mc}{\mathswitch {m_\Pc}}
\newcommand{\Mb}{\mathswitch {m_\Pb}}
\newcommand{\Mt}{\mathswitch {m_\Pt}}
 
\newcommand{\sw}{\mathswitch {s_\PW}}
\newcommand{\cw}{\mathswitch {c_\PW}}

\newcommand{\GF}{\mathswitch {G_\mu}}
 
\def\Li{\mathop{\mathrm{Li}}\nolimits}
\def\Re{\mathop{\mathrm{Re}}\nolimits}
\def\ie{i.e.\ }
\def\eg{e.g.\ }

\newcommand{\se}{self-energy}
\newcommand{\ses}{self-energies}
\newcommand{\ct}{counterterm}
\newcommand{\cts}{counterterms}
\newcommand{\cs}{cross-section}
\newcommand{\css}{cross-sections}

\newcommand{\AAWW}{\gamma\gamma\to\PWp\PWm}
\newcommand{\eeff}{\Pep\Pem\to f\bar{f}}
\newcommand{\eeWW}{\Pep\Pem\to\PWp\PWm}
\newcommand{\eAnW}{\Pem\gamma\to\PWm\nu}
\newcommand{\Born}{\mathrm{Born}}

\newcommand{\ECMS}{E_{\mathrm{CMS}}}

\newcommand{\self}{\mathrm{self}}

\newcommand{\bos}{\mathrm{bos}}
\newcommand{\ferm}{\mathrm{ferm}}
\newcommand{\Coul}{\mathrm{Coul.}}
\newcommand{\counter}{\mathrm{counter}}
\newcommand{\cut}{\mathrm{cut}}
\newcommand{\SB}{\mathrm{SB}}
\newcommand{\unpol}{\mathrm{unpol}}
\newcommand{\gauge}{\mathrm{gauge}}
\newcommand{\CMS}{\mathrm{CMS}}
\newcommand{\NL}{\mathrm{NL}}
\newcommand{\tHF}{\mathrm{tHF}}
\renewcommand{\min}{\mathrm{min}}
\renewcommand{\max}{\mathrm{max}}

\begin{document}

\thispagestyle{empty}
\def\thefootnote{\fnsymbol{footnote}}
\setcounter{footnote}{1}
\null
\strut\hfill  BI-TP 95/04 \\
\strut\hfill WUE-ITP-95-002\\
\strut\hfill hep-ph/9503442
\vskip 0cm
\vfill
\begin{center}
{\Large \bf 
\boldmath{Radiative Corrections to $\gamma\gamma\to\PWp\PWm$ \\
    in the Electroweak Standard Model}
\par} \vskip 2.5em
{\large
{\sc A.~Denner%
\footnote{On leave of absence from 
Institut f\"ur Theoretische Physik, Universit\"at W\"urzburg,
Am Hubland,\\ \hspace*{.5cm} D-97074 W\"urzburg, Germany.}
}\\[1ex]
{\normalsize \it Institut f\"ur Theoretische Physik, Universit\"at Leipzig\\
Augustusplatz 10, D-04109 Leipzig, Germany}
\\[2ex]
{\sc S.~Dittmaier%
\footnote{Supported by the Bundesministerium f\"ur Forschung und
Technologie, Bonn, Germany.} }\\[1ex]
{\normalsize \it Theoretische Physik, Universit\"at Bielefeld\\ 
Postfach 100131, D-33501 Bielefeld, Germany}
\\[2ex]
{\sc R.~Schuster%
\footnote{Supported by the Deutsche Forschungsgemeinschaft.}
} \\[1ex]
{\normalsize \it Institut f\"ur Theoretische Physik, Universit\"at W\"urzburg\\
Am Hubland, D-97074 W\"urzburg, Germany}
}
\par \vskip 1em
\end{center} \par
\vskip 1cm 
\vfill
{\bf Abstract:} \par
The \cs\ for $\gamma\gamma\to\PW^+\PW^-$ with arbitrary
polarized photons and W bosons is calculated within the electroweak
Standard Model including the complete virtual and soft-photonic 
${\cal O}(\alpha)$ corrections. We present a detailed numerical discussion
of the complete radiative corrections and an analytical investigation of
the leading corrections. It turns out that in the on-shell
renormalization scheme for fixed $\MW$ no leading corrections 
associated with the 
running of $\alpha$ or heavy top-quark 
and Higgs-boson masses occur.
The corrections are typically of the order of 10\%. 
They reach, however, larger values where the lowest-order \css\ are suppressed.
\par
\vskip 1cm 
\noindent BI-TP 95/04 \\
WUE-ITP-95-002\par
\vskip .15mm
\noindent March 1995 \par
\null
\setcounter{page}{0}
\clearpage
\def\thefootnote{\arabic{footnote}}
\setcounter{footnote}{0}

\section{Introduction}

The $\mathrm{SU}(2)\times\mathrm{U}(1)$ standard electroweak theory 
has passed many
precision tests during the last years. In particular, measurements of
the muon decay constant $G_\mu$, the gauge-boson masses \MW\ and \MZ,
and the decay widths and asymmetries of the \PZ\ boson at LEP have 
provided stringent constraints which are successfully fulfilled by 
the Standard Model (SM) evaluated at one-loop level. 
The experimental data favor 
a value for the top-quark mass which is in accordance with the
direct measurements \cite{mt} of CDF, $\Mt=176\pm16\GeV$, and D\O,
$\Mt=199\pm30\GeV$. Nevertheless, further precision tests of the SM
are required. Up to now, only weak direct experimental information exists on
the non-Abelian self-interaction of the gauge bosons \cite{WWZ}.
Moreover, no experimental evidence on the mechanism of spontaneous 
symmetry breaking, which is responsible for mass generation and
postulates the existence of the scalar Higgs boson, has been found yet. 
For such
investigations, energies of several hundred GeV or even few TeV are
needed, since the sensitivity to deviations from the SM gauge-boson
self-interaction grows strongly with energy, and the existence of the
Higgs particle can be proven only by direct production. To this end, 
a ``Next Linear Collider'' (NLC) for $\Pe\Pe$,
$\Pe\ga$, and $\ga\ga$ collisions was proposed \cite{nlc} which offers
a unique environment for such precision experiments owing to the 
comparably small background.

A particularly interesting process in $\ga\ga$ collisions is $\AAWW$.
Its total \cs\ approaches a constant of about
$80\pb$ at high energies 
corresponding to $8\times10^6$ \PW~pairs for $10\fb^{-1}$.
This large \cs\ is due to the massive $t$-channel exchange and is
drastically reduced by angular cuts. But even for $|\!\cos\theta|< 0.8$
the \cs\ is still 15 and $4\pb$ at a center-of-mass energy of 500 and
$1000\GeV$, respectively, and thus much larger as \eg the one for $\eeWW$.
Hence $\AAWW$ is very well-suited for precision investigations of the SM.

Several features of $\AAWW$ have already been discussed in the literature.
Most of the existing works concentrated on tree-level predictions
\cite{Pe73}, in particular on the influence of anomalous
non-Abelian gauge couplings \cite{Ki73,Ye91,Be92}.
The process $\AAWW$  depends at tree level both on the triple
$\ga\PW\PW$ and the quartic $\ga\ga\PW\PW$ coupling, and
no other vertices are involved in the unitary gauge at lowest order.
The sensitivity to the $\ga\PW\PW$ coupling is comparable and 
complementary to the reactions $\eeWW$ and $\eAnW$: the first
involves a mixture of the $\ga\PW\PW$ and the $\PZ\PW\PW$ coupling, the
second involves the $\ga\PW\PW$ alone but is less sensitive \cite{Ye91}.
Because the sensitivity to the $\ga\ga\PW\PW$ coupling is much larger 
than the one in $\Pep\Pem$ processes, $\AAWW$ is the ideal process to study 
this coupling \cite{Be92}. 

The one-loop diagrams involving a resonant Higgs boson have been
calculated in order to 
study the possible investigation of the Higgs boson via
$\ga\ga\to\PH^*\to\PWp\PWm$ \cite{Va79,Bo92,Mo94,Ve94}.
Based on our complete one-loop calculation, we have
supplemented these investigations by a discussion of the heavy-Higgs
effects in \citere{HH}. As a matter of fact, only the (suppressed)
channels of longitudinal \PW-boson production are sensitive to the Higgs
mechanism, but the (dominant) channels of purely transverse \PW-boson 
production are extremely insensitive. 
This insensitivity to the Higgs sector renders $\AAWW$ even more
suitable for the investigation of the non-Abelian self couplings.

In this paper, we focus on the complete SM one-loop corrections to $\AAWW$.
One reason why these have not
been calculated so far is certainly their analytical
complexity. We have calculated the numerous Feynman graphs (roughly
300--550 depending on the gauge fixing) by using {\it Mathematica\/}
\cite{math}. More precisely, we have generated and drawn the Feynman
graphs by {\it FeynArts} \cite{fa} and performed three different
calculations, one in 't~Hooft--Feynman gauge using {\it FeynCalc}
\cite{fc} and two in a non-linear gauge with and without using
{\it FeynCalc}.
As the final result exhibits a very complicated and
untransparent analytical form, we refrain from writing
it down in full detail. Instead, we indicate its general structure and
present a detailed discussion of
the numerical results for the $\Oa$ (virtual and real
soft-photonic) corrections to the polarized as well as unpolarized
\css. We restrict the presentation of the analytical results
to the lowest-order \css\ and the most important $\Oa$ corrections.
In particular, we discuss the Higgs resonance, the heavy-Higgs effects,
the Coulomb singularity, and the leading
effects from light fermions and a heavy top quark.

The paper is organized as follows: After fixing our notation and
conventions in \refse{se:notcon}, we consider the lowest-order \css\ for
various polarizations in \refse{se:born}. In \refse{se:RC} we discuss 
the evaluation and general features of the radiative corrections and in
\refse{se:num} the numerical results. 

\section{Notation and conventions}
\label{se:notcon}

We consider the reaction
$$ \gamma(k_1,\lambda_1) + \gamma(k_2,\lambda_2) \rightarrow
   {\PWp}(k_3,\lambda_3) + {\PWm}(k_4,\lambda_4) \; ,
$$
where $\lambda_{1,2} = \pm 1$ and $\lambda_{3,4} = 0,\pm 1$
denote the helicities of the incoming photons and outgoing W bosons,
respectively.

In the center-of-mass system (CMS)
the momenta read 
in terms of the beam energy $E$ of the incoming photons and 
the scattering angle $\theta$
\begin{eqnarray}
k^\mu_1 &=& E(1,  0,  0,  -1) \; ,
\nonumber \\
k^\mu_2 &=& E(1,  0,  0, 1) \; ,
\nonumber \\
k^\mu_3 &=& E(1,  -\beta  \sin\theta,
                             0,  -\beta \cos\theta) \; ,
\nonumber \\
k^\mu_4 &=& E(1,  \beta  \sin\theta,
                             0,  \beta \cos\theta) \; ,
\end{eqnarray}
where $\beta = \sqrt{1 -   \MW^2 / E^2}$
is the velocity of the \PW\ bosons in the CMS.
We define the Mandelstam variables
\begin{eqnarray}
s &=& (k_1 + k_2)^2 = (k_3 + k_4)^2 = 4  E^2 \; ,
\nonumber \\
t &=& (k_1 - k_3)^2 = (k_2 - k_4)^2 = \MW^2 - \frac{s}{2}
                                      (1 - \beta \cos \theta) \; ,
\nonumber \\
u &=& (k_1 - k_4)^2 = (k_2 - k_3)^2 = \MW^2 - \frac{s}{2}
                                      (1 + \beta \cos \theta)  \; .
\end{eqnarray}

In order to calculate polarized \css, we introduce explicit polarization
vectors for the photons and W bosons as follows
\begin{eqnarray}
\varepsilon_1^\mu(k_1, \lambda_1 =\pm 1) &=& \frac{-1}{\sqrt{2}}  (0,   1, \mp i, 0) \; ,
\nonumber \\
\varepsilon_2^\mu(k_2, \lambda_2 =\pm 1) &=& \frac{1}{\sqrt{2}}  (0,   1, \pm i, 0) \; ,
\nonumber \\
{\varepsilon_3^*}^\mu(k_3, \lambda_3 = \pm 1) &=& \frac{-1}{\sqrt{2}} 
                   (0,  \cos\theta,  \pm i,  -\sin\theta) \; ,
\nonumber \\
{\varepsilon_4^*}^\mu(k_4, \lambda_4 =\pm 1) &=& \frac{1}{\sqrt{2}} 
                   (0,  \cos\theta,  \mp i, -\sin\theta) \; ,
\nonumber \\
{\varepsilon_3^*}^\mu(k_3, \lambda_3 =0)\;\;\; &=& \frac{E}{\MW} 
                   (\beta,   -\sin\theta,  0,  -\cos\theta) \; ,
\nonumber \\
{\varepsilon_4^*}^\mu(k_4, \lambda_4 =0)\;\;\; &=& \frac{E}{\MW} 
                   (\beta,   \sin\theta,  0,  \cos\theta) \; .
\end{eqnarray}

We decompose the amplitude ${\cal M}$ into invariant functions $F_{ijkl}$
and standard matrix elements (SME) ${\cal M}_{ijkl}$, which
contain the whole information about the boson polarizations.
Using the transversality condition for the polarization vectors
and Schouten's identity,
the amplitude ${\cal M}$ can be reduced to
\begin{eqnarray}
{\cal M}(\lambda_1,\lambda_2,\lambda_3,\lambda_4,s,t) &=& 
         \sum_{i,j = \{0,3,4\}}\sum_{k,l = \{0,1,2\}}
          F_{ijkl}(s,t)  
       {\cal M}_{ijkl}(\lambda_1,\lambda_2,\lambda_3,\lambda_4,s,t) 
\nonumber \\
       & &{}+ F^{(t)}_{0000}(s,t) 
       {\cal M}_{0000}^{(t)}(\lambda_1,\lambda_2,\lambda_3,\lambda_4,s,t) 
\nonumber \\
       & &{}+ F^{(u)}_{0000}(s,t) 
       {\cal M}_{0000}^{(u)}(\lambda_1,\lambda_2,\lambda_3,\lambda_4,s,t)
\label{decom}
\end{eqnarray}
with
($i,j = \{3,4\}$, $k,l = \{1,2\}$)
\begin{eqnarray}
{\cal M}_{ijkl} &=& (\varepsilon_1 \cdot  k_i)  (\varepsilon_2 \cdot  k_j) 
           (\varepsilon_3^* \cdot  k_k)  (\varepsilon_4^* \cdot  k_l) / s^2,
\label{5}
\\[.3em]
{\cal M}_{0jkl} &=& (\varepsilon_1 \cdot  A)  (\varepsilon_2 \cdot  k_j) 
           (\varepsilon_3^* \cdot  k_k)  (\varepsilon_4^* \cdot  k_l) / s,
\nonumber \\
{\cal M}_{i0kl} &=& (\varepsilon_1 \cdot k_i)  (\varepsilon_2 \cdot  A) 
           (\varepsilon_3^* \cdot  k_k)  (\varepsilon_4^* \cdot  k_l) / s,
\nonumber \\
{\cal M}_{ij0l} &=& (\varepsilon_1 \cdot  k_i)  (\varepsilon_2 \cdot  k_j) 
           (\varepsilon_3^* \cdot  A)  (\varepsilon_4^* \cdot  k_l) / s,
\nonumber \\
{\cal M}_{ijk0} &=& (\varepsilon_1 \cdot  k_i)  (\varepsilon_2 \cdot  k_j) 
           (\varepsilon_3^* \cdot  k_k)  (\varepsilon_4^* \cdot  A) / s,
\\[.3em]
{\cal M}_{00kl} &=& (\varepsilon_1 \cdot  \varepsilon_2) 
           (\varepsilon_3^* \cdot  k_k)  (\varepsilon_4^* \cdot  k_l) / s,
\nonumber \\
{\cal M}_{0j0l} &=& (\varepsilon_1 \cdot  \varepsilon_3^*) 
           (\varepsilon_2 \cdot  k_j)  (\varepsilon_4^* \cdot  k_l) / s,
\nonumber \\
{\cal M}_{0jk0} &=& (\varepsilon_1 \cdot  \varepsilon_4^*) 
           (\varepsilon_2 \cdot  k_j)  (\varepsilon_3^* \cdot  k_k) / s,
\nonumber \\
{\cal M}_{i00l} &=& (\varepsilon_2 \cdot  \varepsilon_3^*) 
           (\varepsilon_1 \cdot  k_i)  (\varepsilon_4^* \cdot  k_l) / s,
\nonumber \\
{\cal M}_{i0k0} &=& (\varepsilon_2 \cdot  \varepsilon_4^*) 
           (\varepsilon_1 \cdot  k_i)  (\varepsilon_3^* \cdot  k_k) / s,
\nonumber \\
{\cal M}_{ij00} &=& (\varepsilon_3^* \cdot  \varepsilon_4^*) 
           (\varepsilon_1 \cdot  k_i)  (\varepsilon_2 \cdot  k_j) / s,
\\[.3em]
{\cal M}_{000l} &=& (\varepsilon_1 \cdot  \varepsilon_2) 
           (\varepsilon_3^* \cdot  A)  (\varepsilon_4^* \cdot  k_l),
\nonumber \\
{\cal M}_{00k0} &=& (\varepsilon_1 \cdot  \varepsilon_2) 
           (\varepsilon_3^* \cdot  k_k) (\varepsilon_4^* \cdot  A),
\nonumber \\
{\cal M}_{0j00} &=& (\varepsilon_3^* \cdot  \varepsilon_4^*) 
           (\varepsilon_1 \cdot  A)  (\varepsilon_2 \cdot  k_j),
\nonumber \\
{\cal M}_{i000} &=& (\varepsilon_3^* \cdot  \varepsilon_4^*) 
           (\varepsilon_1 \cdot  k_i)  (\varepsilon_2 \cdot  A),
\\[.3em]
{\cal M}_{0000}  &=& (\varepsilon_1 \cdot  \varepsilon_2) 
           (\varepsilon_3^* \cdot  \varepsilon_4^*)  ,
\nonumber \\
{\cal M}^{(t)}_{0000}  &=& (\varepsilon_1 \cdot  \varepsilon_3^*) 
           (\varepsilon_2 \cdot  \varepsilon_4^*)  ,
\nonumber \\
{\cal M}^{(u)}_{0000}  &=& (\varepsilon_1 \cdot  \varepsilon_4^*) 
           (\varepsilon_2 \cdot  \varepsilon_3^*)  , \label{9}
\end{eqnarray}
and 
\beq
A_\mu =  \frac{i}{ut - \MW^4} 
       \varepsilon_{\mu\nu\rho\sigma} k_1^\nu k_2^\rho k_3^\sigma,
\qquad \varepsilon_{0123} = -1 \;.
\eeq
Our choice of polarization vectors for the photons implies
\begin{eqnarray}
\varepsilon_i k_j = 0, \qquad i,j = 1,2 \;,
\end{eqnarray}
and thus by virtue of momentum conservation
\begin{eqnarray}
\varepsilon_i k_3 = - \varepsilon_i k_4,
 \qquad i = 1,2 \;.
\end{eqnarray}
We use this relation to eliminate all SME
involving $\varepsilon_1 k_4$ and $\varepsilon_2 k_3$.
This reduces the 83 SME defined in (\ref{5}) -- 
(\ref{9}) to 38 for the process under consideration. 

As a consequence of CP invariance and Bose symmetry
only the sum of each SME and the one 
with  $(\varepsilon_1,k_1,\varepsilon_3,k_3)$ and
$(\varepsilon_2,k_2,\varepsilon_4,k_4)$ interchanged occurs.
For instance, $\M_{0401}$ only appears in the
combination $\M_{0401}+\M_{3020}$ in the expansion of $\M$ in
\refeq{decom}. This leaves 22 independent SME. 

In four dimensions, the matrix elements ${\cal M}^{(t)}_{0000}$ and
${\cal M}^{(u)}_{0000}$ are not linearly independent from the set of all
$\M_{ijkl}$ and 
can be reduced to linear combinations of the other matrix
elements using the identities 
\begin{eqnarray}
\delta^{\varepsilon_1\varepsilon_3^*k_1k_2k_3}
      _{\varepsilon_2\varepsilon_4^*k_1k_2k_3} =
\delta^{\varepsilon_1\varepsilon_4^*k_1k_2k_3}
      _{\varepsilon_2\varepsilon_3^*k_1k_2k_3} = 0
\end{eqnarray}
involving the Gram determinant
\begin{eqnarray}
\delta^{p_1 \ldots p_n}_{q_1 \ldots q_n} =
\left| \matrix{p_1\cdot q_1&\ldots&p_1\cdot q_n\cr
                   \vdots&\ddots&\vdots\cr
                   p_n\cdot q_1&\ldots&p_n\cdot q_n\cr}\right| \;.
\end{eqnarray}
Nevertheless, we keep ${\cal M}^{(t)}_{0000}$ and
${\cal M}^{(u)}_{0000}$ for convenience.

Bose symmetry implies that the amplitude $\M$ is invariant under 
the interchange $(k_1,\varepsilon_1) \leftrightarrow (k_2,\varepsilon_2)$.
Since many diagrams can be related to others by this transformation,
it is useful to introduce a second set of SME which is obtained
from \refeqs{5}--\refeqf{9} by this interchange.
Of course, this second set of SME can be expressed by the original set.

Besides Bose symmetry also CP is an exact symmetry,
since we use a unit quark-mixing 
matrix.%
\footnote{For a non-trivial quark-mixing matrix, CP would be violated
in the considered process first at two-loop level.} 
The helicity amplitudes for fixed 
polarization configurations are related as follows
\beq
\begin{array}[b]{ll}
{\cal M}_{\lambda_1 \lambda_2 \lambda_3 \lambda_4}(s,t,u) = 
{\cal M}_{\lambda_2 \lambda_1 \lambda_3 \lambda_4}(s,u,t) 
& \hbox{(Bose)} \\
{\cal M}_{\lambda_1 \lambda_2 \lambda_3 \lambda_4}(s,t,u) = 
{\cal M}_{-\lambda_1 -\lambda_2 -\lambda_4 -\lambda_3}(s,u,t) 
\qquad & \hbox{(CP)} \\
{\cal M}_{\lambda_1 \lambda_2 \lambda_3 \lambda_4}(s,t,u) =
{\cal M}_{-\lambda_2 -\lambda_1 -\lambda_4 -\lambda_3}(s,t,u)
\qquad & \hbox{(Bose+CP)}.
\earr
\eeq

In the following, we only consider the sum of the two transverse 
\PW~polarizations.
Therefore we indicate the polarizations of the external particles 
by four labels, the first pair corresponding to the photons, and the 
second pair to
the \PW~bosons. The labels $+$,$-$ represent right-handed and left-handed
photons, respectively, $\rL$ stands for longitudinal, and
$\rT$ for the sum of the two transverse \PW~polarizations.

The combination of Bose and CP symmetry leads to the following relations
between the differential \css\ with equal photon helicities
\beqar
\left(\frac{\rd\sigma}{\rd\Omega}\right)_{--{\rm TT}} &=&
\left(\frac{\rd\sigma}{\rd\Omega}\right)_{++{\rm TT}}\; ,
\nonumber \\
\left(\frac{\rd\sigma}{\rd\Omega}\right)_{--{\rm LL}} &=&
\left(\frac{\rd\sigma}{\rd\Omega}\right)_{++{\rm LL}}\; ,
\nonumber \\
\left(\frac{\rd\sigma}{\rd\Omega}\right)_{--{\rm (LT+TL)}} &=&
\left(\frac{\rd\sigma}{\rd\Omega}\right)_{++{\rm (LT+TL)}} \; .
\label{eq:CPBds}
\eeqar
Moreover, Bose symmetry implies that all \css\ in (\ref{eq:CPBds})
are forward--backward symmetric. 
For different photon helicities Bose symmetry leads to
\beqar
\left(\frac{\rd\sigma}{\rd\Omega}\right)_{-+{\rm TT}}(s,t,u) &=&
\left(\frac{\rd\sigma}{\rd\Omega}\right)_{+-{\rm TT}}(s,u,t)\; ,
\nonumber \\
\left(\frac{\rd\sigma}{\rd\Omega}\right)_{-+{\rm LL}}(s,t,u) &=&
\left(\frac{\rd\sigma}{\rd\Omega}\right)_{+-{\rm LL}}(s,u,t)\; ,
\nonumber \\
\left(\frac{\rd\sigma}{\rd\Omega}\right)_{-+{\rm (LT+TL)}}(s,t,u) &=&
\left(\frac{\rd\sigma}{\rd\Omega}\right)_{+-{\rm (LT+TL)}}(s,u,t) \; ,
\label{eq:Bds}
\eeqar
whereas Bose+CP does not yield further relations.

C and P symmetry are only violated by the fermionic loop corrections,
but hold in lowest order and for the bosonic loop corrections. We
indicate these restricted symmetries by a modified equality sign
\beqar
{\cal M}_{\lambda_1 \lambda_2 \lambda_3 \lambda_4}(s,t,u) &\Nequal{P}&
{\cal M}_{-\lambda_1 -\lambda_2 -\lambda_3 -\lambda_4}(s,t,u) \; ,
\nn\\
{\cal M}_{\lambda_1 \lambda_2 \lambda_3 \lambda_4}(s,t,u) &\Nequal{C}&
{\cal M}_{\lambda_1 \lambda_2 \lambda_4 \lambda_3}(s,u,t) \; .
\eeqar
P invariance then implies for the differential \css\
\beqar
\left(\frac{\rd\sigma}{\rd\Omega}\right)_{-+{\rm TT}} &\Nequal{P}&
\left(\frac{\rd\sigma}{\rd\Omega}\right)_{+-{\rm TT}}\; ,
\nonumber \\
\left(\frac{\rd\sigma}{\rd\Omega}\right)_{-+{\rm LL}} &\Nequal{P}&
\left(\frac{\rd\sigma}{\rd\Omega}\right)_{+-{\rm LL}}\; ,
\nonumber \\
\left(\frac{\rd\sigma}{\rd\Omega}\right)_{-+{\rm (LT+TL)}} &\Nequal{P}&
\left(\frac{\rd\sigma}{\rd\Omega}\right)_{+-{\rm (LT+TL)}}\; .
\label{eq:Pds}
\eeqar
In combination with (\ref{eq:Bds}) this means that the forward-backward
asymmetries of the differential \css\ for unequal photon helicities 
are entirely due to fermionic corrections.

We perform the calculation both in 't~Hooft--Feynman (tHF) gauge and in a
gauge with the following non-linear (NL) gauge-fixing term \cite{gavela}
\begin{eqnarray}
{\cal L}_{\rm GF} &=& -\left| \partial^\mu W^+_\mu + i e
                (A^\mu - \frac{\cw}{\sw} Z^\mu) W^+_\mu 
               -i \MW \phi^+ \right|^2
\nn \\
            & &{}-  \frac{1}{2} (\partial^\mu Z_\mu - \MZ \chi)^2
              - \frac{1}{2} (\partial^\mu A_\mu)^2 \;,
\label{nl}
\end{eqnarray}
with the conventions of \citere{ad&mex} for the fields.
In particular, $\phi^\pm$ and $\chi$ denote the charged and neutral 
would-be-Goldstone fields, respectively.
In this NL gauge the $\phi^\pm W^\mp A$ vertices
vanish. This reduces the number of Feynman graphs in comparison to the 
tHF gauge considerably.

\section{Lowest-order \cs}
\label{se:born}
In NL gauge, only the three diagrams of \reffi{fi:borndia} 
contribute to the lowest-order amplitude. 
\bfi
\setlength{\unitlength}{1mm}
\begin{picture}(160,35)
\put(-13,-154){\includegraphics{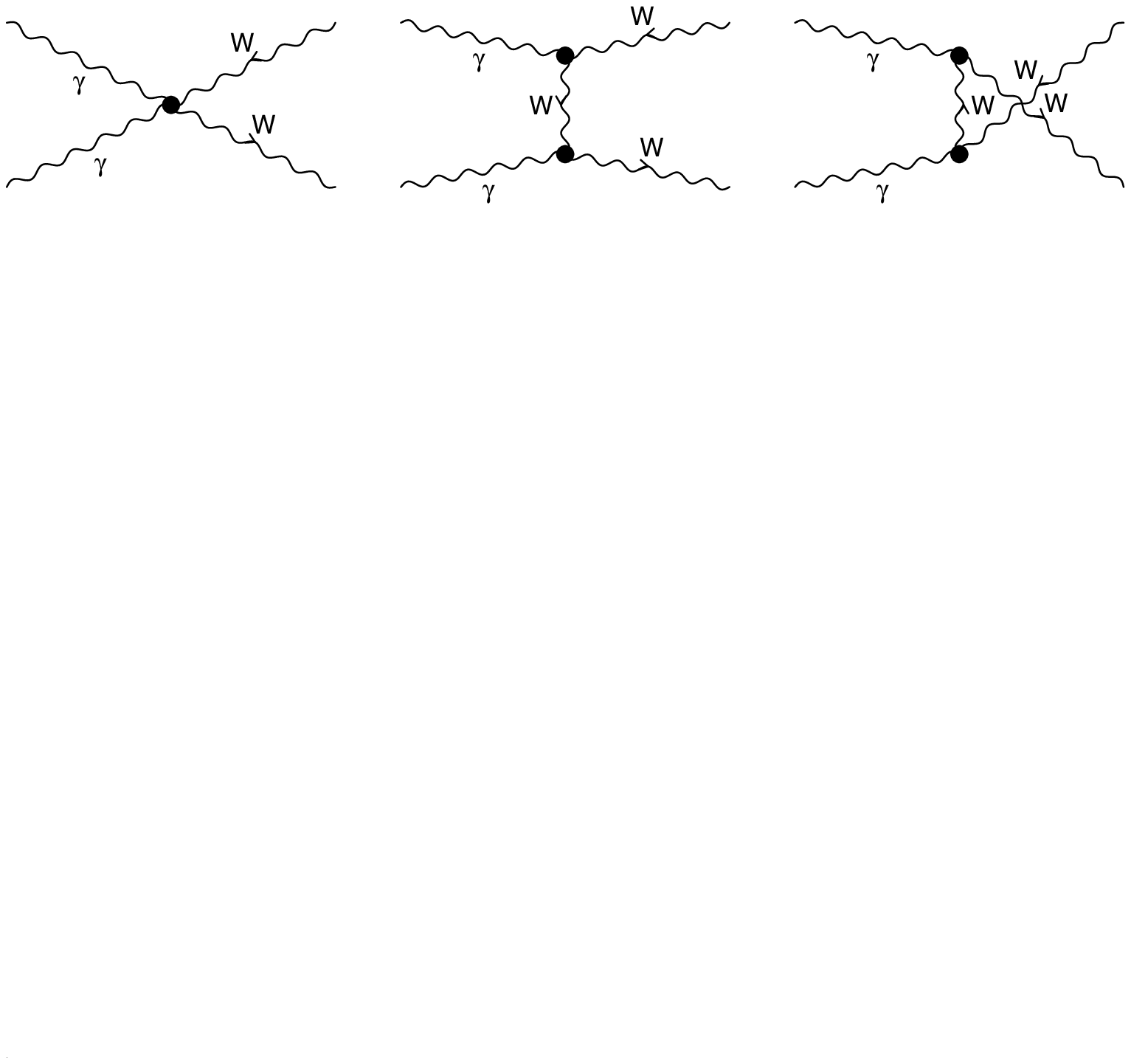}}
\end{picture}
\caption{Lowest-order diagrams for $\AAWW$ in NL gauge}
\label{fi:borndia}
\efi
In tHF gauge two additional diagrams exist which 
involve internal $\phi$ fields.
Evaluation of the tree diagrams in either gauge yields the Born amplitude
\beq
\M_{\Born} = 8\pi\alpha\Biggl\{
\frac{s}{\MW^2-t}\M_{0,t} + \frac{s}{\MW^2-u}\M_{0,u} - \M_{0000} 
\Biggr\},
\eeq
where
\beqar
\M_{0,t} &=& 
2 \M_{0012} + 2 \M_{3400} - 2 \M_{0401} - 2 \M_{3020} + 2 \M_{0410} 
+ 2 \M_{3002} + \M_{0000}^{(t)},  \nl
\M_{0,u} &=& 
2 \M_{0021} + 2 \M_{4300} - 2 \M_{0310} - 2 \M_{4002} + 2 \M_{0301} 
+ 2 \M_{4020} + \M_{0000}^{(u)}.
\eeqar
The lowest-order matrix element vanishes for the helicities 
$(\lambda_1,\lambda_2,\lambda_3,\lambda_4) = 
(\pm,\pm,0,\pm)$, $(\pm,\pm,\pm,0)$, $(\pm,\pm,0,\mp)$, 
$(\pm,\pm,\mp,0)$, $(\pm,\pm,\pm,\mp)$, $(\pm,\pm,\mp,\pm)$.

The differential Born \cs\ \cite{Ye91,Be92} is obtained as
\begin{eqnarray}
\left(\frac{\rd\sigma}{\rd\Omega}\right)^{\Born} &=&
\frac{\be}{64 \pi^2 s}
\sum_{\lambda_1 \lambda_2 \lambda_3 \lambda_4}
\frac{1}{4} \left(1+ \lambda_1 P^\gamma_1\right)
            \left(1+ \lambda_2 P^\gamma_2\right)
\left| {\cal M}_{{\Born}} \right|^2 \; ,
\end{eqnarray} 
where $P^\gamma_{1,2}$ denote the degrees of photon-beam polarization and
the sum over $\lambda_3$, $\lambda_4$ include the desired W polarizations.

We list the differential \css\ for several helicity configurations:
\begin{eqnarray}
\left(\frac{\rd\sigma}{\rd\Omega}\right)^{\Born}_{\pm\pm\rT\rT} &=& 
\frac{\alpha^2 \beta s (2 \MW^4 -4 \MW^2 s +s^2)}{(\MW^2 -t)^2(\MW^2 -u)^2} \; ,
\nonumber \\
\left(\frac{\rd\sigma}{\rd\Omega}\right)^{\Born}_{\pm\pm\rL\rL} &=& 
 \frac{\alpha^2 \beta \MW^4 s}{(\MW^2 -t)^2(\MW^2 -u)^2} \; ,
\nonumber \\
\left(\frac{\rd\sigma}{\rd\Omega}\right)^{\Born}_{\pm\pm(\rL\rT+\rT\rL)} 
&=& 0 \; ,  \nl
\left(\frac{\rd\sigma}{\rd\Omega}\right)^{\Born}_{\pm\mp\rT\rT} &=& 
 \frac{\alpha^2 \beta s^3}{(\MW^2 -t)^2(\MW^2 -u)^2} 
       \biggl\{2 \frac{(16 \MW^4 +s^2) (ut-\MW^4)^2}{s^6\be^4}
           +\frac{(t-u)^2}{s^2\beta^2} \biggr\} \; , \nl 
\left(\frac{\rd\sigma}{\rd\Omega}\right)^{\Born}_{\pm\mp\rL\rL} &=& 
 \frac{\alpha^2 (4 \MW^2+s)^2 (\MW^4 - t u)^2}
      {\beta^3 s^3  (\MW^2 -t)^2(\MW^2 -u)^2} \; ,
\nonumber \\
\left(\frac{\rd\sigma}{\rd\Omega}\right)^{\Born}_{\pm\mp(\rL\rT+\rT\rL)} &=& 
 \frac{16 \alpha^2 \MW^2 (2 \MW^4 -t^2-u^2) (\MW^4 - t u)}{
      \beta^3 s^2  (\MW^2 -t)^2(\MW^2 -u)^2}  \; .
\end{eqnarray}
These results can be reconstructed from equation (5) in \citere{Be92}
or from equation (4.5) in \citere{Ye91}.

Adding up the single contributions,
we get for the unpolarized differential \cs
\footnote{The second term in equation (6) of \citere{Be92} should be 
multiplied by 2.}
\begin{eqnarray}
\left(\frac{\rd\sigma}{\rd\Omega}\right)^{\Born}_{\unpol} &=&
 \frac{3 \alpha^2 \beta}{2s} 
\Biggl\{1 - \frac{2s^2}{(\MW^2 -t)(\MW^2 -u)} \left(
                    \frac{2}{3} + \frac{\MW^2}{s} \right) 
\nonumber \\ 
& &{}+ \frac{2s^4}{(\MW^2 -t)^2(\MW^2 -u)^2} \left(
                    \frac{1}{3} + \frac{\MW^4}{s^2} \right)
   \Biggr\} \; .
\label{unpol}
\end{eqnarray}

Integration over $\theta_{\cut} \leq \theta \leq \pi - \theta_{\cut}$
yields:
\begin{eqnarray}
\sigma^{\Born}_{\pm\pm\rT\rT} &=& \frac{16 \pi \alpha^2}{s}\, 
\frac{s^2 - 4 \MW^2s + 2\MW^4}{s^2} \left\{
\log\left(\frac{1 + \beta \cos\theta_{\cut}}{1 - \beta\cos\theta_{\cut}}\right)
   + \frac{2\beta \cos\theta_{\cut}}{1 - \beta^2 \cos^2\theta_{\cut}}
       \right\} \; ,
\nonumber \\
\sigma^{\Born}_{\pm\pm\rL\rL} &=& \frac{16 \pi \alpha^2}{s}\frac{\MW^4}{s^2} 
\left\{
\log\left(\frac{1 + \beta\cos\theta_{\cut}}{1 - \beta\cos\theta_{\cut}}\right)
   + \frac{2\beta \cos\theta_{\cut}}{1 - \beta^2 \cos^2\theta_{\cut}}
       \right\} \; ,
\nonumber \\
\sigma^{\Born}_{\pm\mp\rT\rT} &=& \frac{8 \pi \alpha^2 }{s \beta^4} \Biggl\{
 \frac{s^2 + 16\MW^4}{s^2} \beta \cos\theta_{\cut}\nl
& & {} -2\frac{s^4 - 2 \MW^2 s^3 - 2 \MW^4 s^2 + 32 \MW^6 s - 32 \MW^8} {s^4}
\log\left(\frac{1 + \beta \cos\theta_{\cut}}{1 - \beta\cos\theta_{\cut}}\right) 
\nl
& &{}+ 4 \frac{s^4 - 4 \MW^2 s^3 + 2 \MW^4 s^2 + 32 \MW^8}{s^4}\,
   \frac{\beta \cos\theta_{{\cut}}} {1 - \beta^2 \cos^2\theta_{{\cut}}} 
\Biggr\} \; , \nl
\sigma^{\Born}_{\pm\mp\rL\rL} &=& \frac{4 \pi \alpha^2}{s\be^4}
\frac{ (4 \MW^2 + s)^2}{s^2}
\Biggl\{\beta \cos\theta_{{\cut}}
- 4 \frac{\MW^2 (s-\MW^2)}{s^2}
\log\left(\frac{1 + \beta\cos\theta_{\cut}}{1 - \beta\cos\theta_{\cut}}\right) 
\nl
& &{} + \frac{8\MW^4}{s^2}
\frac{\beta \cos\theta_{{\cut}}}{1 - \beta^2 \cos^2\theta_{{\cut}}} 
       \Biggr\} \; ,
\nonumber \\
\sigma^{\Born}_{\pm\mp(\rL\rT+\rT\rL)} &=&  \frac{128 \pi \alpha^2}{s\be^4}
\frac{\MW^2}{s}
\Biggl\{
- \beta \cos\theta_{{\cut}}
+ \frac{s^2 - 2 \MW^2s + 4\MW^4}{s^2} 
\log\left(\frac{1 + \beta\cos\theta_{{\cut}}} {1 - \beta\cos\theta_{{\cut}}}
\right) \nl
& &{}
- \frac{4 \MW^2 (s - 2\MW^2)}{s^2} \,
\frac{\beta \cos\theta_{{\cut}}}{1 - \beta^2 \cos^2\theta_{\cut}}  
       \Biggr\} \; ,
\end{eqnarray}
and for the unpolarized \cs\
\begin{eqnarray}
\sigma^{\Born}_{{\unpol}} &=&  \frac{6\pi \alpha^2 }{s}
\Biggl\{
\beta \cos\theta_{{\cut}} - 
4 \frac{\MW^2}{s}\left(1 - \frac{2\MW^2}{s}\right)
\log\left(\frac{1 + \beta \cos\theta_{{\cut}}}
{1 - \beta \cos\theta_{{\cut}}}\right)
\nonumber \\
& &{}+ \left(\frac{1}{3} + \frac{\MW^4}{s^2} \right)
\frac{16 \beta \cos\theta_{\cut}}{1 - \beta^2\cos^2\theta_{\cut}} \Biggr\} \;.
\end{eqnarray}

In \reffis{fi:intcs10} and \ref{fi:intcs20} 
we show the lowest-order \css\ for various polarizations
and two different angular cuts $\theta_{\cut} = 10^\circ, 20^\circ$.
For $\theta_{{\cut}} = 0$, the \css\ for
transverse W~bosons approach a constant at high energies, $s \gg \MW^2$, 
owing to the massive $t$-channel exchange 
\beq
\sigma^{\Born}_{\pm\pm\rT\rT}, \sigma^{\Born}_{\pm\mp\rT\rT}
\;\Nlim{s\rightarrow\infty}\;
\frac{8 \pi \alpha^2}{\MW^2} = 80.8 \pb.
\eeq
For a finite cut, $\sigma^{\Born}_{\pm\pm\rT\rT}$  and
$\sigma^{\Born}_{\pm\mp\rT\rT}$ behave as $1/s$ for large $s$.
The \css\ $\sigma^{\Born}_{\pm\mp\rL\rL}$ and
$\sigma^{\Born}_{\pm\mp(\rT\rL+\rL\rT)}$ are
proportional to $1/s$ and $1/s^2$, respectively, independently
 of the cut-off.
The \cs\ $\sigma^{\Born}_{\pm\pm\rL\rL}$ goes 
like $1/s^2$ at high energies for $\theta_{\cut} = 0$
and like $1/s^3$ for a finite cut-off. 
It is suppressed by about a factor of $10^3$ at $E_{\CMS} = 500$ GeV.
Note that the latter \cs\ can be
enhanced drastically by non-standard physics \cite{HH}.
At high energies, the unpolarized \cs\ $\si^{\Born}_{\unpol}$
is dominated by transverse \PW~bosons,
and all polarized \css\ involving two transverse \PW~bosons are of the
same order-of-magnitude.
Close to threshold the differential and integrated 
\css\ for all polarization configurations vanish like $\beta$.
Numerical values for the lowest-order \css\ can be found in 
\refta{table_born}.
\begin{table}
\footnotesize
\begin{center}
\arraycolsep 6pt
$$\begin{array}{|c|c||c|c|c|c|c|c|}
\hline
\sqrt{s}/\mathrm{GeV}
 & \theta &
{\mathrm{unpol}} & {{\pm\pm}\mathrm{TT}} & {{\pm\pm}\mathrm{LL}} &
{{\pm\mp}\mathrm{TT}} & {{\pm\mp}\mathrm{LL}} & {{\pm\mp}\mathrm{(LT+TL)}} \\
\hline\hline
\phantom{0}500 & \phantom{0}0^\circ<\theta<180^\circ & 
77.6 &  82.2 & 6.10\times 10^{-2} & 70.2 & 9.99 \times 10^{-1} & 
1.69 \phantom{{}\times 10^{-1}} \\
\cline{2-8}
& 20^\circ<\theta<160^\circ & 
36.7 & 42.7 & 3.17\times 10^{-2}& 28.2 & 9.89\times 10^{-1}& 
1.49\phantom{{}\times 10^{-1}} \\
\hline\hline
1000 & \phantom{0}0^\circ<\theta<180^\circ & 
80.1 & 82.8 & 3.54 \times 10^{-3} & 76.9 & 2.52\times 10^{-1} & 
1.70\times 10^{-1}\\
\cline{2-8}
& 20^\circ<\theta<160^\circ & 
14.2 & 16.8 &7.18 \times 10^{-4}& 11.2 & 2.44\times 10^{-1} & 
1.21\times 10^{-1}\\
\hline\hline
2000 & \phantom{0}0^\circ<\theta<180^\circ & 
80.6 & 81.6 & 2.14\times 10^{-4}& 79.5 & 6.41\times 10^{-2}& 
1.50\times 10^{-2}\\
\cline{2-8}
& 20^\circ<\theta<160^\circ & 
4.07& 4.84 & 1.27\times 10^{-5}& 3.23 & 6.11\times 10^{-2}& 
8.26\times 10^{-3}  \\
\hline
\end{array}$$
\caption{Lowest-order integrated \css\ in pb for several polarizations}
\label{table_born}
\end{center}
\end{table}

Figures \ref{fi:difcspp} and \ref{fi:difcsmm} show 
the angular distributions of the differential
 lowest-order \css\ for various polarizations at 
 $E_{\CMS} = 500$, $1000$ and $2000$ GeV.
The \css\ involving transverse W bosons are characterized by the 
$t$- and $u$-channel poles in the forward and backward directions,
respectively. With increasing energy they increase in the very forward and
backward direction proportional to $s$ but decrease in the central angular
region proportional to $1/s$.
The respective behavior of 
$({{\rd}\sigma}/{{\rd}\Omega})^{\Born}_{\pm\pm\rL\rL}$
is $1/s$ and $1/s^3$.
The \css\ $({{\rd}\sigma}/{{\rd}\Omega})^{\Born}_{\pm\mp{\rL\rL}}$
and $({{\rd}\sigma}/{{\rd}\Omega})^{\Born}_{\pm\mp{(\rL\rT+\rT\rL)}}$
vanish in the forward and backward direction. 
While $({{\rd}\sigma}/{{\rd}\Omega})^{\Born}_{\pm\mp{\rL\rL}}$
reach their maxima at $90^\circ$ and decrease proportional to $1/s$
for all angles,
$({{\rd}\sigma}/{{\rd}\Omega})^{\Born}_{\pm\mp{(\rL\rT+\rT\rL)}}$
possess maxima at $|\!\cos\theta| = \beta$ decreasing
proportional to $1/s$ and relative minima at
$\theta = 90^\circ$ decreasing proportional to $1/s^2$.
   
\section{Radiative corrections}
\label{se:RC}

\subsection{Non-linear gauge fixing}

We have performed the calculation of the radiative corrections
in tHF gauge
and a NL gauge with the gauge-fixing term given in equation
\refeq{nl} applying the complete on-shell renormalization
scheme in both cases \cite{ad&mex}.
As pointed out in Sect.~2 the $\phi^\pm W^\mp A$ vertices
vanish in NL gauge. As a consequence 
the $\phi$ \se\ and the $\phi W$ mixing energy do not contribute, and 
the number of vertex and box diagrams is reduced from 441 in 
tHF gauge to 268 in NL gauge (for one fermion generation).

Furthermore, the analytical expressions for the
$W^\pm \overline{u}^\pm u^{A,Z}$
vertices (with $\bar u$, $u$ denoting the Fadeev--Popov ghost fields)
are proportional to the W-boson momentum
in NL gauge and thus vanish for on-shell W~bosons.
For this reason most of the box
and vertex diagrams with internal ghost fields vanish.
As the corrections to the $AAA$ and $AAZ$ vertices vanish in both gauges,
the number of non-vanishing vertex and box diagrams reduces to 365
in tHF gauge and to 168 in NL gauge.
Moreover, many diagrams have a simpler structure in NL gauge.

In order to determine the \cts\ necessary for renormalization
one has to calculate the \ses.  Here we list the differences
$\Delta\Sigma = \Sigma^{\NL} -\Sigma^{\tHF}$ between the \ses\ in
NL gauge and the ones in tHF gauge.
The transverse parts of the latter can be found e.g.~in Ref.~\cite{ad&mex}.
For the transverse part of the W \se\ we find
\beq
\Delta\Sigma^{WW}_{\rT} = 
 \frac{\alpha}{2 \pi} \bigl(k^2 - \MW^2\bigr)
       \biggl[B_0\bigl(k^2,0,\MW\bigr) + 
             \frac{\cw^2}{\sw^2} B_0\bigl(k^2,\MZ,\MW\bigr)
       \biggr] \; ,
\eeq
and for its longitudinal part
\beqar
\Delta\Sigma^{WW}_{\rL} &=& 
- \frac{\alpha}{4 \pi} \biggl\{
      \frac{\cw^2}{\sw^2} \bigl(5 k^2 + 5 \MZ^2 - 3 \MW^2\bigr) 
                                 B_0\bigl(k^2,\MW,\MZ\bigr) +
          \bigl(5 k^2 - 3 \MW^2\bigr) B_0\bigl(k^2,\MW,0\bigr) 
\nn \\
&&\phantom{\frac{- \alpha}{4 \pi} \biggl\{}
+ \frac{5}{\sw^2}\MW^2 \bigl[
             B_0\bigl(0,0,\MW\bigr) - B_0\bigl(0,0,\MZ\bigr)\bigr]
 - \frac{2}{\sw^2}k^2
       \biggr\}\;,
\eeqar
where $B_0$ is the scalar one-loop two-point function \cite{ad&mex,scalar}.
The differences for the \ses\ involving neutral gauge bosons can be given 
in a compact way
\beqar
\Delta\Sigma^{BB'}_{\rT} &=& 
- \frac{\alpha}{2 \pi} f^{ B B'} 
             \bigl(2 k^2 - M^2_{ B} - M^2_{ B'}\bigr)
               B_0\bigl(k^2,\MW,\MW\bigr) \; ,
\\
\Delta\Sigma^{BB'}_{\rL} &=&
\frac{\alpha}{2 \pi} f^{B B'}
             \bigl(M^2_{B} + M^2_{B'}\bigr)
               B_0\bigl(k^2,\MW,\MW\bigr) \; ,
\eeqar
with ${B}^{(\prime)} = A,Z$ and
\beq
f^{A A} = 1, \qquad
f^{A Z} = -\frac{\cw}{\sw}, \qquad
f^{ZZ} = \frac{\cw^2}{\sw^2} \;.
\eeq

Note that the differences for the transverse parts of the W, Z, and $A$ 
\ses\ are proportional to $(k^2 - M_{\PW,\PZ}^2)$ and $k^2$,
respectively.

\subsection{Inventory of ${\cal O}(\alpha)$ corrections}

In the following we list the virtual corrections,
i.e.\ the contributions to $\delta\cal M$, in NL gauge.
We adopt the conventions of
Ref.~\cite{ad&mex},
where the necessary explicit results for the transverse parts of the 
\ses\ and the renormalization constants can be found.
Because of the length of our results we do not explicitly write down the
analytic expressions.

Owing to our renormalization scheme,
we have to deal with \se\ insertions only into the internal
lines of the tree diagrams of Fig.{} 1.
These result in the following contribution to the invariant matrix
element
\beqar
\delta {\M}_{\self} &=& 4\pi\alpha\Biggl\{
\frac{2s}{(\MW^2-t)^2} \Sigma^{WW}_{\rT}(t) \M_{0,t} 
+ \frac{2s}{(\MW^2-u)^2} \Sigma^{WW}_{\rT}(u)\M_{0,u} \\
&& \phantom{4\pi\alpha\biggl(}
{}+\frac{1}{t} 
\Bigl(\Sigma^{WW}_{\rT}(t) - \Sigma^{WW}_{\rL}(t)\Bigr)\M_{0000}^{(t)} 
+\frac{1}{u} 
\Bigl(\Sigma^{WW}_{\rT}(u) - \Sigma^{WW}_{\rL}(u)\Bigr)\M_{0000}^{(u)} 
\Biggr\} .  \nn
\eeqar

\bfi
\setlength{\unitlength}{1mm}
\begin{picture}(160,168)
\put(-2,-55){\includegraphics{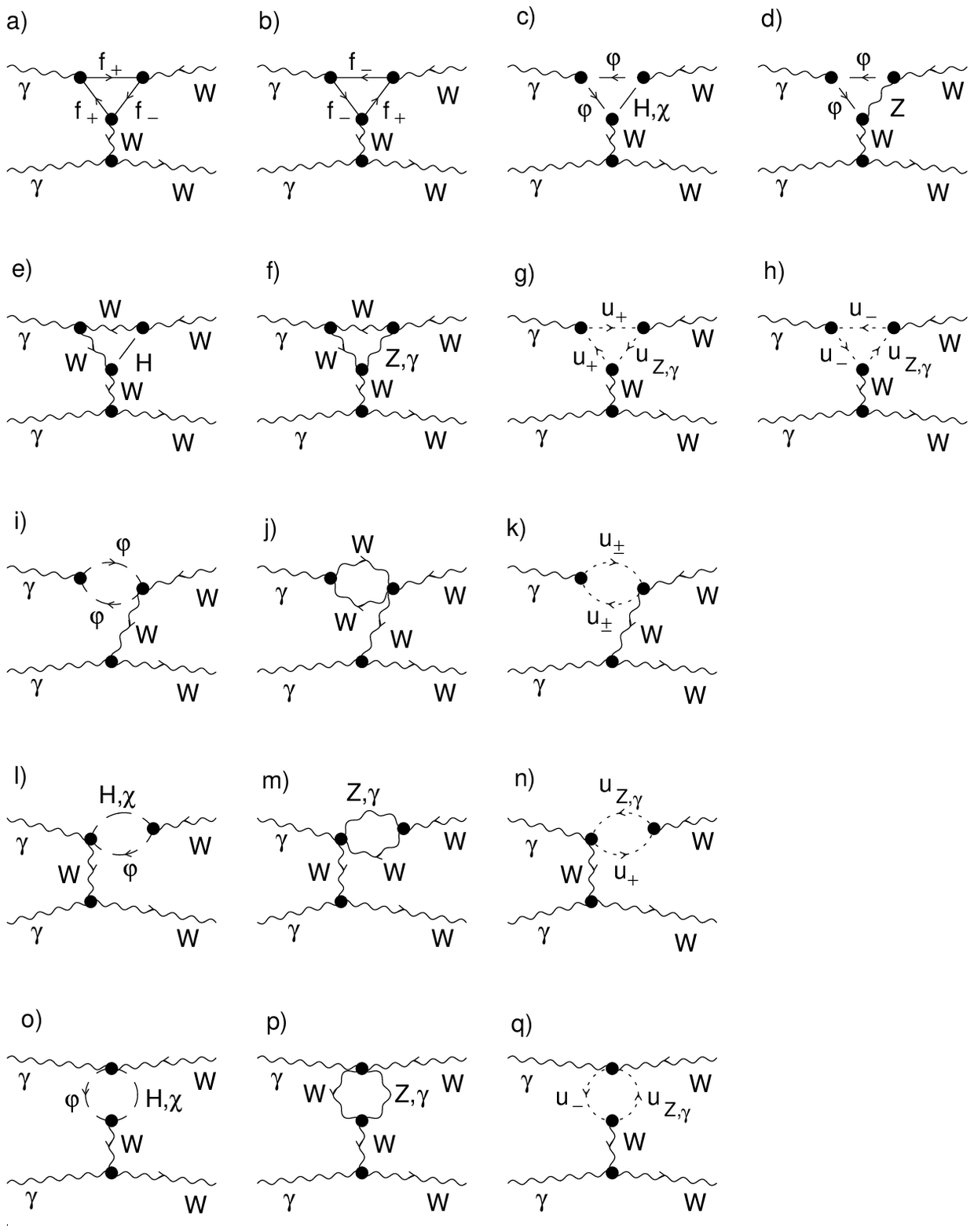}}
\end{picture}
\caption{The $t$-channel diagrams for the upper $AWW^*$ vertex }
\label{fi:AWW}
\efi

Figure \ref{fi:AWW} shows the $t$-channel graphs for the upper
$A W W^*$ vertex (asterics denote off-shell fields);
the diagrams (h), (i), (k), (l), and (n) vanish for on-shell 
external photons and \PW~bosons.
The diagrams for the lower $A W W^*$ vertex can be
constructed in an analogous way, and the $u$-channel diagrams are obtained
via crossing, i.e.\
the interchange of the two external photons.

The $AAA^*$ and $AAZ^*$ vertex corrections vanish according
to Yangs theorem \cite{ya49} and because the virtual $A$ and $Z$ are 
coupled to a conserved current.
Thus, the only $s$-channel vertex corrections, which contribute to 
$\delta{\cal M}$, are the Higgs-resonant 
$AAH^*$-vertex graphs shown in \reffi{fi:AAH}. For the graphs (a)--(d)
also crossed ones exist.  The $AAH^*$ corrections are discussed in the
next subsection.

The box diagrams are shown in \reffis{fi:box1} and \ref{fi:box2}.
While to each diagram in \reffi{fi:box1} a crossed partner diagram
corresponds, those in \reffi{fi:box2}  are symmetric under crossing.
For on-shell external bosons the graphs (h) and (i) of \reffi{fi:box1}
and the graphs (e) and (f) of \reffi{fi:box2} vanish.
\bfi
\setlength{\unitlength}{1mm}
\begin{picture}(160,165)(0,0)
\put(-34,-113){\includegraphics{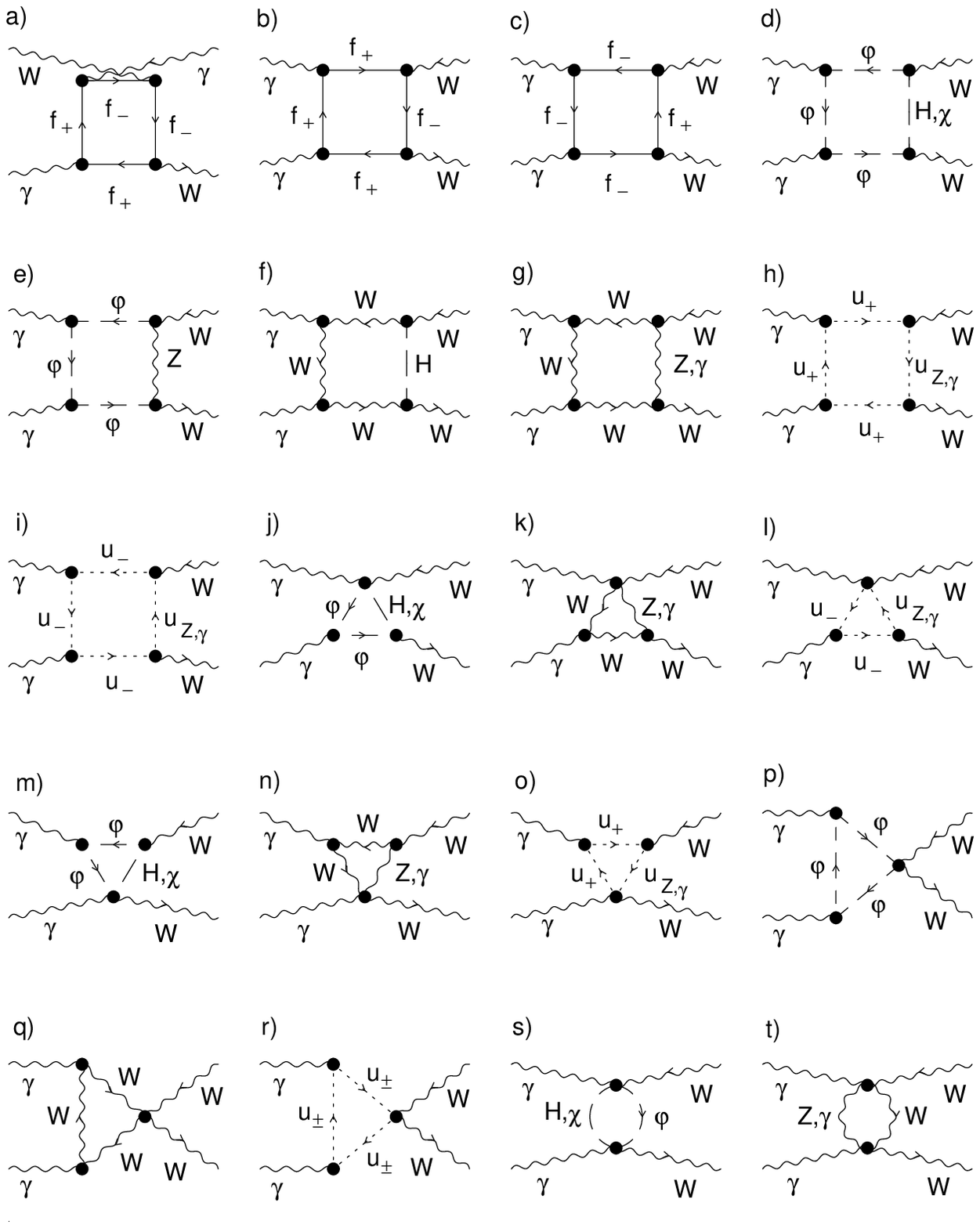}}
\end{picture}
\caption{Non-crossing-symmetric box diagrams}
\label{fi:box1}
\efi
\bfi
\setlength{\unitlength}{1mm}
\begin{picture}(160,98)(0,0)
\put(+3,-179){\includegraphics{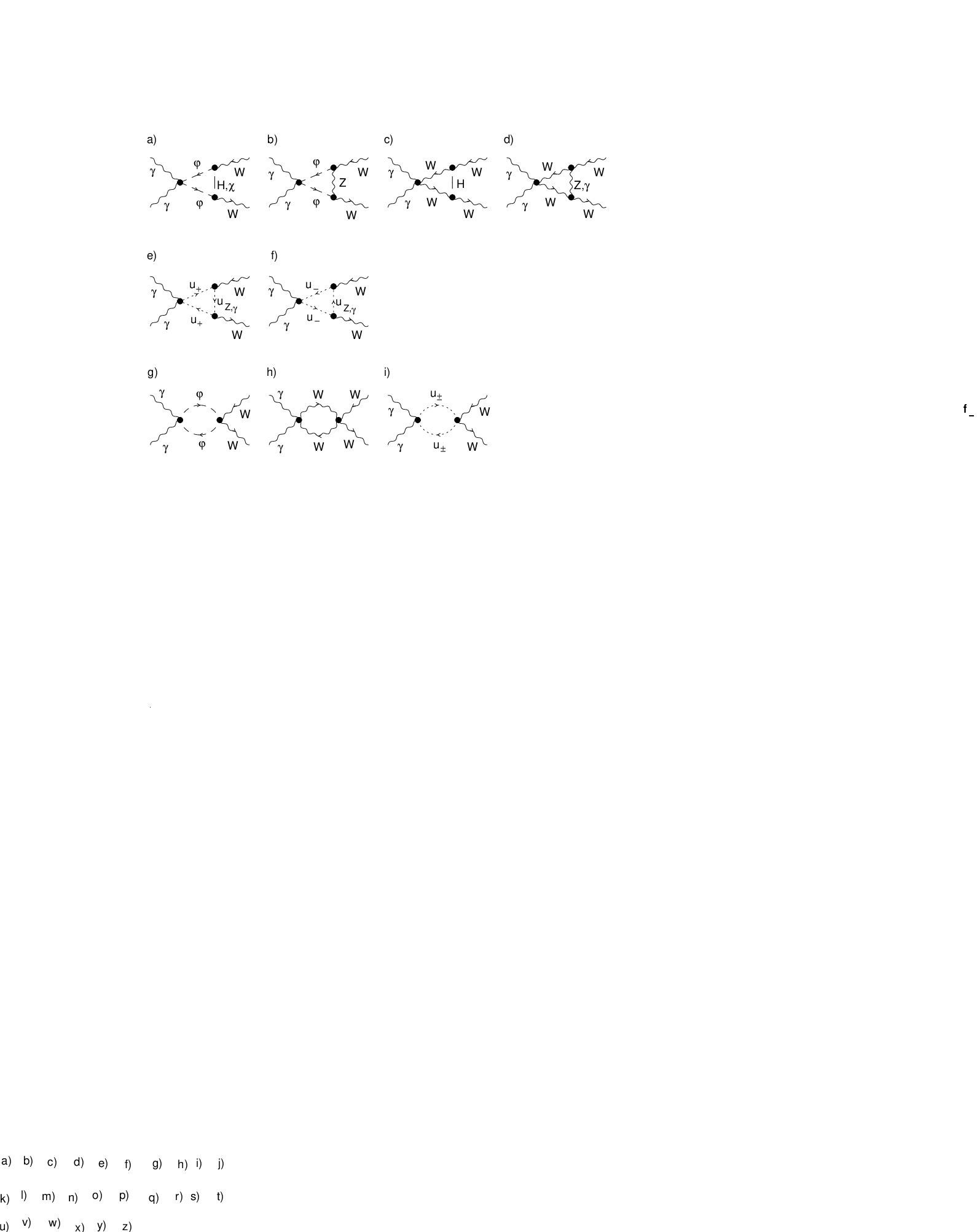}}
\end{picture}
\caption{Crossing-symmetric box diagrams}
\label{fi:box2}
\efi

The renormalization is performed in the on-shell renormalization
scheme.  Evaluation of the \cts\ diagrams yields
\beqar
\delta{\cal M}_{\counter} &=&
{\cal M}_{\Born} \biggl(
2 \delta { Z}_{ e} + \delta { Z}_{ W} +
\delta { Z}_{A A} - \frac{\cw}{\sw} \delta Z_{ZA} \biggr)
\nn \\ && {}-  8\pi\alpha   
\left(\frac{s\delta \MW^2}{(\MW^2-t)^2}\M_{0,t} 
+ \frac{s\delta \MW^2}{(\MW^2-u)^2}\M_{0,u}  \right) .
\eeqar
In this context we mention that the massive gauge-boson sector does not
break electromagnetic gauge invariance if the NL gauge fixing
\refeq{nl} is applied. As a consequence on-shell photons do not mix with
\PZ\ bosons rendering the \ct\ $\delta Z_{ZA}$ zero,
\beq
\delta Z_{ZA}=2\frac{\Si^{AZ}_{\rm T}(0)}{\MZ^2}=0.
\eeq
The charge renormalization constant $\de Z_e$ is then given by
\cite{ad&mex}
\beq
\de Z_e = -\frac{1}{2}\de Z_{AA} =
\left.\frac{1}{2}\frac{\partial\Si^{AA}_{\rm T}(k^2)}{\partial k^2}
\right|_{k^2=0},
\label{eq:dze}
\eeq
so that the complete \ct\ contribution to the matrix element $\M$
reduces to 
\beqar
\delta{\cal M}^{\NL}_{\counter} &=&
\M_{\Born} \delta { Z}_{ W} 
-  8\pi\alpha   
\left(\frac{s\delta \MW^2}{(\MW^2-t)^2}\M_{0,t} 
+ \frac{s\delta \MW^2}{(\MW^2-u)^2}\M_{0,u}  \right) .
\eeqar

\subsection{Higgs resonance}
The Higgs resonance in $\AAWW$
was discussed extensively in the literature;
see e.g.\ \citeres{Mo94,Ve94}.
So we restrict ourselves to the listing of our results for the 
Higgs-resonant graphs.

The Higgs-resonant part of the process is caused by the graphs
of \reffi{fi:AAH}.
\bfi
\setlength{\unitlength}{1mm}
\begin{picture}(160,63)(0,0)
\put(16,-133){\includegraphics{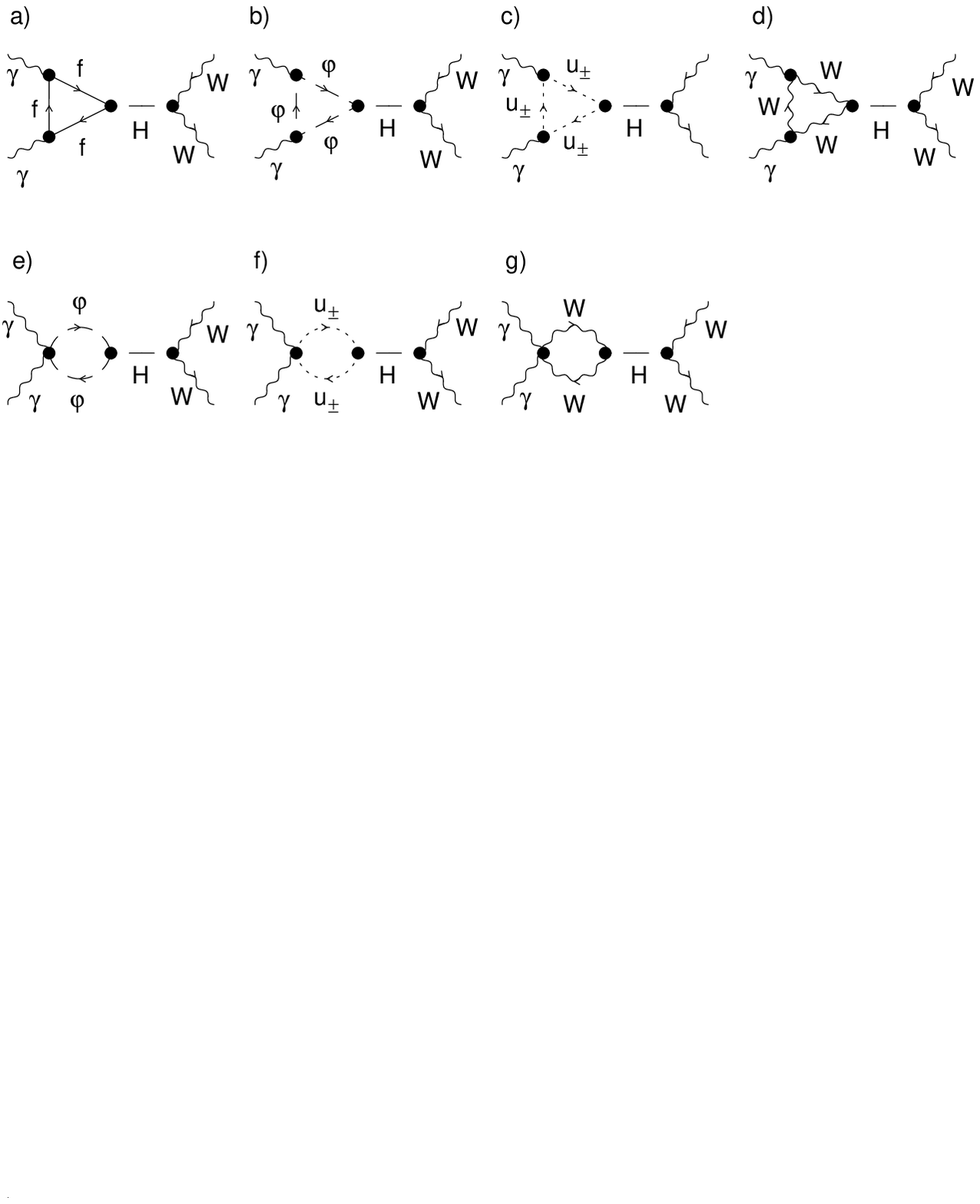}}
\end{picture}
\caption{The $AAH^*$ vertex diagrams}
\label{fi:AAH}
\efi
These yield a contribution of the form (compare \citere{Ve94})
\begin{eqnarray}
\delta \M_{AAH^*} = \frac{F^H(s)}{s-\MH^2}
\M_{0000} 
\label{eq_higgs}
\end{eqnarray}
with
\strut
\begin{eqnarray}
F^H(s) &=& -\frac{\alpha^2}{\sw^2} \biggl\{
  6 \MW^2 + \MH^2 + \MW^2  C^{(\gauge)}
C_0\bigl(s,0,0,\MW^2,\MW^2,\MW^2\bigr)
\nonumber \\
&&\phantom{\frac{e^2}{8 \pi \sw^2} \biggl\{}
-2 \sum_{f} N_f^c Q_{f}^2m_f^2 \left[ 2 + (4m_f^2 -s)
                 C_0\bigl(s,0,0,m_{f}^2,m_{f}^2,m_{f}^2\bigr) \right]  
\biggr\}.
\label{eq:FH}
\end{eqnarray}
\strut
The sum in (\ref{eq:FH}) extends over all massive fermions with charge
\pagebreak[2]
$Q_f$ and color factor $N_f^c$. 
The coefficient $C^{(\gauge)}$ is gauge-dependent and
reads
\begin{eqnarray}
C^{(\gauge)} = 12\MW^2 + 2\MH^2 - 8s
\end{eqnarray}
in NL gauge and
\begin{eqnarray}
C^{(\gauge)} = 12\MW^2 +  \MH^2 - 7s
\end{eqnarray}
in tHF gauge.
Note that $\delta \M_{AAH^*}$ vanishes for opposite helicities of the
incoming photons or outgoing \PW~bosons together with $\M_{0000}$. 
Hence the Higgs resonance is only present for 
photons and \PW~bosons with equal helicities.

In the literature \cite{Bo92,Mo94,Ve94}, the Higgs-boson width has been
introduced na{\"\i}vely by the replacement 
\beq
\frac{F^H(s)}{s-\MH^2} \rightarrow \frac{F^H(s)}{s-\MH^2 + i \MH\Gamma_{\PH}}
\eeq
in \refeq{eq_higgs}.
Owing to the gauge dependence of $F^H(s)$, this treatment 
destroys gauge invariance. The violation of gauge invariance occurs at
the level of the non-resonant $\Oa$ corrections, which were neglected in
\citeres{Bo92,Mo94,Ve94}. Since our main concern are exactly these
corrections we have to take care of gauge invariance.
To this end we 
decompose (\ref{eq_higgs}) into a gauge-invariant resonant part and a 
gauge-dependent non-invariant part and introduce $\Gamma_\PH$ only 
in the former. This results in the following replacement in (\ref{eq_higgs})
\beq
\frac{F^H(s)}{s-\MH^2} \rightarrow
\frac{F^H(\MH^2)}{s-\MH^2 + i \MH\Gamma_{\PH}} 
+ \frac{F^H(s) - F^H(\MH^2)}{s-\MH^2} \;.
\label{eq_higgs2}
\eeq
Equations (\ref{eq_higgs}) and (\ref{eq_higgs2}) yield a gauge-invariant
amplitude including the finite width in the resonant Higgs contributions.

Since the resonant Higgs contributions are large for $s\approx\MH^2$,
we take also the square of the resonant part of the
matrix element into account in the numerical analysis [compare
\refeq{eq:dsido}].

For a calculation with order $\Oa$ accuracy also near $s=\MH^2$, 
one should take into account the $\Oa$ corrections to the 
Higgs-boson width \cite{Fl81} and to $F^H(\MH^2)$ in the resonant
contribution. Since the
Higgs resonance is not our main concern, we only take into account the
lowest-order decay width determined from the imaginary part of the
one-loop Higgs-boson \se\ and \refeq{eq:FH} for $F^H(\MH^2)$. 

\subsection{Leading corrections}
\label{leadrcs}

The electroweak radiative corrections typically involve leading
contributions of universal origin such as the leading-logarithmic QED
corrections, corrections arising from the running of $\al$,
corrections associated with large top-quark or Higgs-boson 
masses, and the Coulomb
singularity at threshold for the production of a pair of charged
particles.

We first discuss the leading weak corrections:
\begin{itemize}
\item
For $\AAWW$, the running of $\al$ is not relevant, as the {\it two} 
external
photons are on mass shell, \ie the relevant effective coupling is the
one at zero-momentum transfer. Technically, the large logarithms present
in the renormalization constant $\de Z_e$ of the electron charge are 
canceled by the corresponding logarithms in the wave-function 
renormalization constant $\de Z_{AA}$ of the external photons, as can
be explicitly seen in \refeq{eq:dze}.
\item
The Higgs-mass-dependent corrections have been discussed in detail in
\citere{HH}. 
In the heavy-Higgs limit, $\MH\gg\sqrt{s}$, no corrections involving 
$\log(\MH/\MW)$ or $\MH^2/\MW^2$ arise. Consequently the Higgs-mass
dependence is small. However, for $\sqrt{s}\gg\MH\gg\MW$ corrections
proportional to $\MH^2/\MW^2$ appear for the \css\ involving
longitudinal gauge bosons as a remnant of the unitarity cancellations
(compare \citere{eeWWhe}). These give rise to large effects in
particular for $\si_{\pm\pm\rL\rL}$.
\item
The situation is similar for the top-dependent corrections.
As the lowest-order matrix element is independent of the weak mixing 
angle, no universal corrections proportional to $\Mt^2/\MW^2$ arise from
renormalization. It can be easily derived by power counting that such
terms do also not result from loop diagrams in the heavy top limit 
$\Mt\gg\sqrt{s}$. A more accurate analysis
reveals that even no terms involving $\log(\Mt/\MW)$ occur in this limit. 
On the other hand, for $\sqrt{s}\gg\Mt\gg\MW$ corrections
proportional to $\Mt^2/\MW^2$ appear for longitudinal gauge bosons 
(compare \citere{eeWWhe}).
\end{itemize}
All these statements hold in the on-shell renormalization scheme with
$\alpha$, \MW\ and \MZ\ as input parameters. 
If the \MW\ mass is determined from \GF, 
corrections involving logarithms of
the light fermion masses, \Mt, and \MH\ occur together with the
universal corrections proportional to $\Mt^2/\MW^2$ associated with the
$\rho$ parameter.

The leading corrections of electromagnetic origin are independent of 
the renormalization scheme and the input parameters:
\begin{itemize}
\item
As $\AAWW$ involves no light charged external particles,
no large logarithmic corrections associated with collinear
photons show up apart from the region of very high energies, $s\gg\MW^2$.
As a consequence, the photonic corrections are not enhanced with respect
to the weak corrections.
\item
Close to threshold, the Coulomb singularity gives rise to large effects 
as in any pair-production process of charged particles. 
These effects can be extracted on general grounds or directly
from the Feynman diagrams. To this end one has to consider all diagrams
resulting from the lowest-order diagrams with an additional photon
exchanged between the final state \PW~bosons (\reffi{fi:coul}). 
In the limit $\be\ll 1$ one obtains:
\beq
\de\si^{\Coul} = \frac{\al\pi}{2\be} \si^{\Born}.
\label{eq:coul}
\eeq
The $\beta^{-1}$ correction factor in \refeq{eq:coul} to the Born
\cs\ near threshold is typical for the pair production of
stable (on-shell) particles. The generalization to unstable (off-shell)
particles can be found in the literature \cite{coul}.
\end{itemize}

\bfi
\setlength{\unitlength}{1mm}
\begin{picture}(160,30)(0,0)
\put(14,-188){\includegraphics{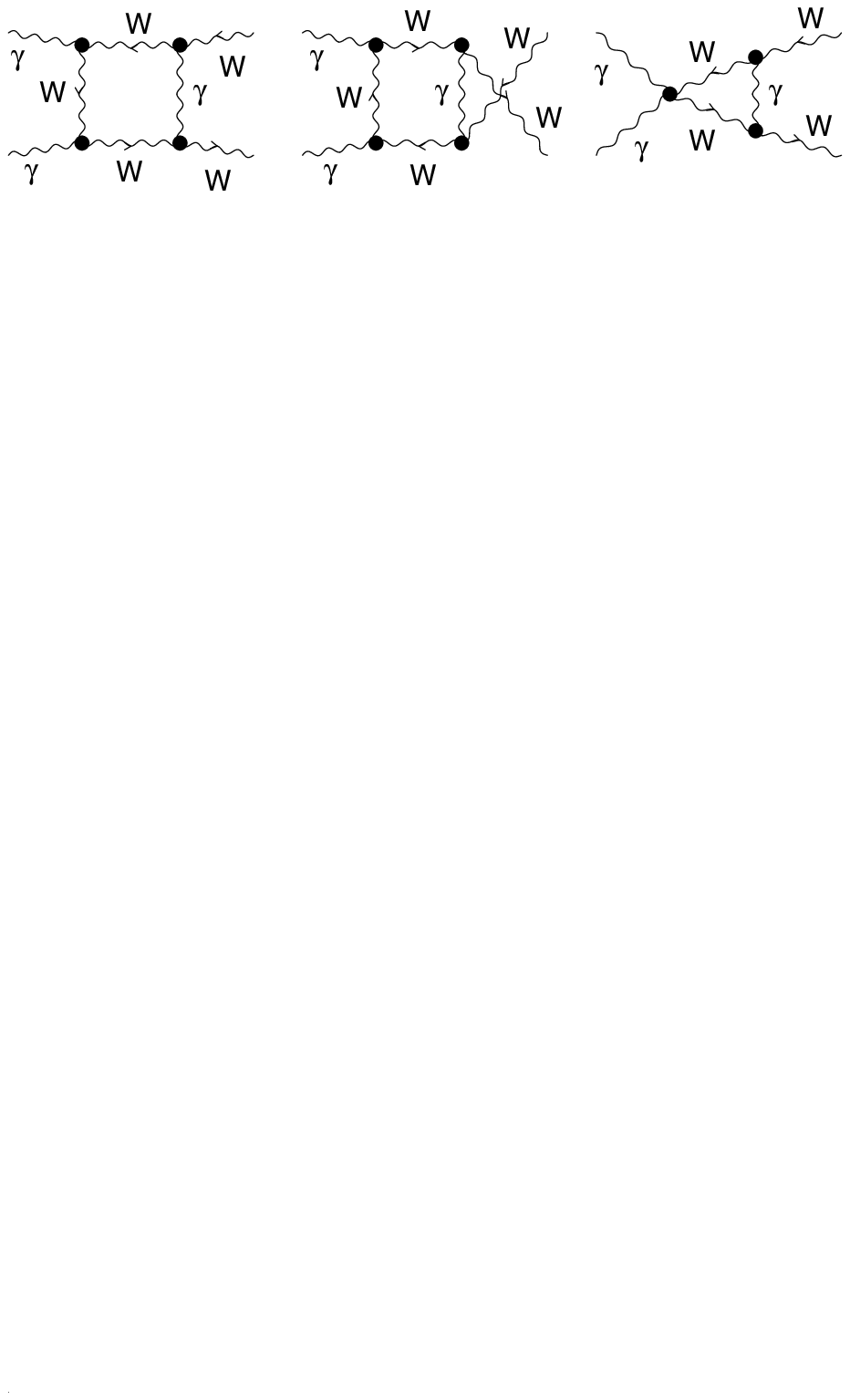}}
\end{picture}
\caption{The diagrams that contribute to the Coulomb singularity in
NL gauge}
\label{fi:coul}
\efi

At high energies, $s\gg\MW^2$, the radiative corrections are dominated 
by terms like
$(\al/\pi)\log^2(s/\MW^2)$, which arise from vertex and box diagrams
(comp.\ \citere{eeWWhe}).
At $1\TeV$ these are about 10\%, setting the scale for the (weak)
radiative corrections at this energy.

\subsection{Structure of the final result}

For a consistent treatment of the virtual one-loop radiative corrections
the squared transition matrix element has to be expanded in a power series
of the coupling constant $\alpha$
\beq
|{\cal M}|^2 = |{\cal M}_{\Born}|^2 
+ 2\Re\{\delta {\cal M} {\cal M}^*_{\Born}\} + \hbox{higher orders}.
\eeq
The ${\cal {O}}(\alpha)$ correction $\delta {\cal M}$ to the matrix element
${\cal M}$ is decomposed as in (\ref{decom}).
We do not consider those polarization configurations for which the
lowest-order matrix element vanishes.

The invariant functions $F_{ijkl}$ are calculated in terms
of standard tensor integrals, which are reduced to scalar integrals by the
procedure proposed in \citere{pas79}. The scalar one-loop integrals are
evaluated using the methods and general results of \citere{scalar}. Whereas
UV divergences are regularized dimensionally, we treat IR divergences
by introducing an infinitesimal photon mass $\lambda$. The
artificial $\lambda$ dependence drops out when soft-photon
bremsstrahlung is added. 

The \cs\ including full ${\cal {O}}(\alpha)$ corrections and the 
squared Higgs-resonant $\Oa$ contributions read
\begin{eqnarray}
\left(\frac{\rd\sigma}{\rd\Omega}\right) &=&
\frac{\be}{64\pi^2s} \sum_{\lambda_1 \lambda_2 \lambda_3 \lambda_4}
\biggl[
\left| {\cal M}_{\Born} \right|^2 (1 + \delta_{\SB}) 
+ 2\Re\{\delta {\cal M} \M_{{\Born}}^*\}
+ \frac{|F^H(\MH^2)\M_{0000}|^2}{(s-\MH^2)^2 + \Ga_H^2 \MH^2}\biggr]
\nonumber \\*
&=& \left(\frac{\rd\sigma}{\rd\Omega}\right)^{\Born}(1+\delta),
\label{eq:dsido}
\end{eqnarray}
where 
\beqar
\delta_{\SB} &=& -\frac{\alpha}{\pi}
\biggl\{2 \log \frac{2 \Delta E}{\lambda} +
\frac{1}{\beta} \log\biggl(\frac{1-\beta}{1+\beta}\biggr)
  +      \frac{s - 2 \MW^2}{s \beta} \biggl[2 \log \frac{2 \Delta E}{\lambda}
         \log\biggl(\frac{1-\beta}{1+\beta}\biggr)
\nn \\
\phantom{-\frac{\alpha}{\pi}\biggl\{}
&& 
       - 2 {\Li}_2 \biggl(\frac{1-\beta}{1+\beta}\biggr) + \frac{1}{2}
       \log^2\biggl(\frac{1-\beta}{1+\beta}\biggr) + \frac{\pi^2}{3}
      - 2 \log\biggl(\frac{1-\beta}{1+\beta}\biggr)
      \log\biggl(\frac{2 \beta}{1+\beta}\biggr)\biggr]\biggr\}
\eeqar
denotes the soft-photon correction factor, 
$\De E$ is the maximal energy of the emitted photon, 
$F^H$ is given in \refeq{eq:FH},
and $\delta$ is the relative correction.

For the integrated \cs\ $\si$, the relative correction is defined
analogously
\beq
\si = \int_{\theta_{\min}}^{\theta_{\max}} \rd\!\cos\theta \int_0^{2\pi}
\rd\phi\,\left(\frac{\rd\si}{\rd\Omega}\right) = \si^{\Born}(1+\delta).
\eeq

In order to ensure the correctness of our results we have performed 
three different calculations. The corrections were calculated with 
{\it FeynCalc\/} \cite{fc} both in tHF gauge and NL
gauge \refeq{nl}. A further calculation was performed independently with 
{\it Mathematica\/} without using {\it FeynCalc\/} in NL gauge.
The results of these various calculations agree numerically within
8--9 digits for the corrected \cs.
Moreover, we have checked that all 
UV and IR singularities cancel, and that the symmetries discussed in
\refse{se:notcon}
hold. Finally, the leading corrections discussed in \refse{leadrcs} have
been deduced analytically and checked numerically.
 
\section{Numerical results}
\label{se:num}

For the numerical evaluation we used the following set of parameters
\cite{PDG94}
\beq
\begin{array}[b]{lcllcllcl}
\alpha &=& 1/137.0359895 &
\GF & = & 1.166390 \times 10^{-5} \GeV^{-2} \\[.3em]
\MZ & = & 91.187\GeV, &
\MH & = & 250\GeV, &&& \\[.3em]
\Me & = & 0.51099906\MeV,  \hspace{1.5em} &
m_{\mu} & = & 105.65839\MeV,  \hspace{1.5em} &
m_{\tau} & = & 1.777\;\GeV, \\[.3em]
\Mu & = & 46.0\;\MeV, &
\Mc & = & 1.50\;\GeV, &
\Mt & = & 170\;\GeV, \\[.3em]
\Md & = & 46.0\;\MeV, &
\Ms & = & 150\;\MeV, &
\Mb & = & 4.50\;\GeV.
\end{array}
\label{eq:par}
\eeq
The masses of the light quarks are adjusted such that the
experimentally measured hadronic vacuum polarization is reproduced
\cite{Ei95}. As discussed in the previous section, 
no large logarithms associated with fermion masses enter the
$\Oa$ corrections for $\AAWW$ in the on-shell renormalization scheme,
and the fermion mass contributions are only of the order $\alpha
m_f^2/\MW^2$.
However, as the Fermi-constant $\GF$ is empirically much better
known than the \PW~mass, \MW\ is usually calculated from all the other 
parameters using the muon decay width including radiative corrections. 
In this calculation of \MW\ all parameters given above enter sensibly.
If not stated otherwise, $\MW$ is determined in the following using
formulae (2.56) and (2.57) of \citere{wwrev}. The above set of
parameters yields
$$\MW=80.333\GeV.$$

As discussed above, no leading collinear logarithms occur in 
$\AAWW$. Thus, the only source of enhanced photonic corrections are
the soft-photon-cut-off-dependent terms which yield the following
relative correction
\beq
\de_{\cut} = -\frac{2\al}{\pi} \log\frac{\De E}{E} 
\left(1 - \frac{s-2\MW^2}{s\be}\log\frac{1+\be}{1-\be}\right).
\label{eq:cut}
\eeq
While these cut-off-dependent terms are 
definitely of electromagnetic origin, the complete electroweak \Oa\ 
corrections cannot be
separated on the basis of Feynman diagrams in a gauge-invariant way.
Since we are mainly interested in the weak corrections we discard the 
cut-off-dependent terms \refeq{eq:cut} 
and consider the rest as a suitable measure of
the weak corrections for the process at hand. The elimination of the 
cut-off-dependent terms can be achieved simply by setting the soft-photon
cut-off energy equal to the beam energy. If not stated otherwise, the
correction $\de$ stands in the following for
the complete soft-photonic and virtual electroweak corrections
as defined in \refeq{eq:dsido} for $\De E = E$.

Figure \ref{fi:intcs10} shows the corrections to the total \css\ integrated 
over $10^\circ \leq \theta \leq 170^\circ$ for different boson polarizations.
\begin{figure}
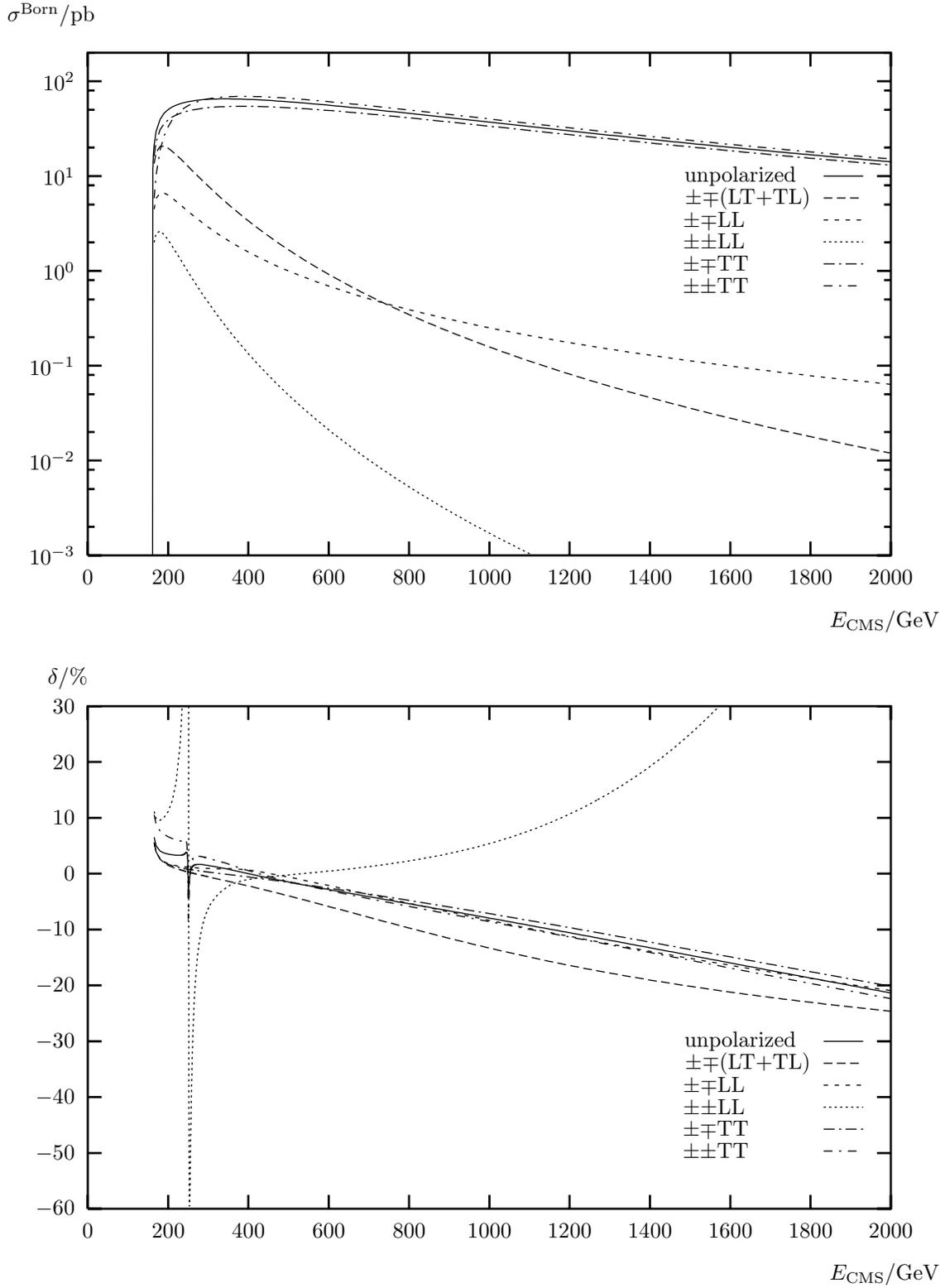

\setlength{\unitlength}{1mm}
\begin{picture}(160,210)(0,0)
\put(0,110){\input{aaww0.tot10.tex}}
\put(0,05){\input{aaww.tot10.tex}}
\end{picture}
\caption{Integrated lowest-order \css\ and corresponding relative 
corrections for several polarizations with an angular cut 
         $10^\circ \leq \theta \leq 170^\circ$}
\label{fi:intcs10}
\end{figure}
The dominating channels involving transverse \PW~bosons get 
corrections which almost coincide with each other as well as
the unpolarized case and reach roughly $-20\%$ at $\sqrt{s}=2\TeV$.
For $\theta_{\cut} = 10^\circ$ the corrections to $\si_{\pm\mp\rL\rL}$ are 
similar, and those to $\si_{\pm\mp(\rL\rT+\rT\rL)}$ are only slightly larger.
The corrections to $\si_{\pm\pm\rL\rL}$ are 
completely different. At low energies they are dominated by the Higgs
resonance, at high energies 
by corrections proportional to $\MH^2/\MW^2$ which are additionally
enhanced owing to the suppression of the corresponding lowest-order \cs.
This \cs, which is also most sensitive to a very heavy Higgs boson, has been
discussed in detail in \citere{HH}. 
Note that owing to helicity conservation only the \css\ with
equal photon and \PW~boson helicities are affected by the Higgs resonance.

Imposing a more stringent angular cut $20^\circ <\theta <160^\circ$ 
to the phase-space integration, the corrections become
larger at high energies for all polarizations 
involving $t$- and $u$-channel poles
and reach about $-35\%$ at $\sqrt{s}=2\TeV$  (\reffi{fi:intcs20}).
\begin{figure}
\setlength{\unitlength}{1mm}
\begin{picture}(160,210)(0,0)
\put(0,110){\input{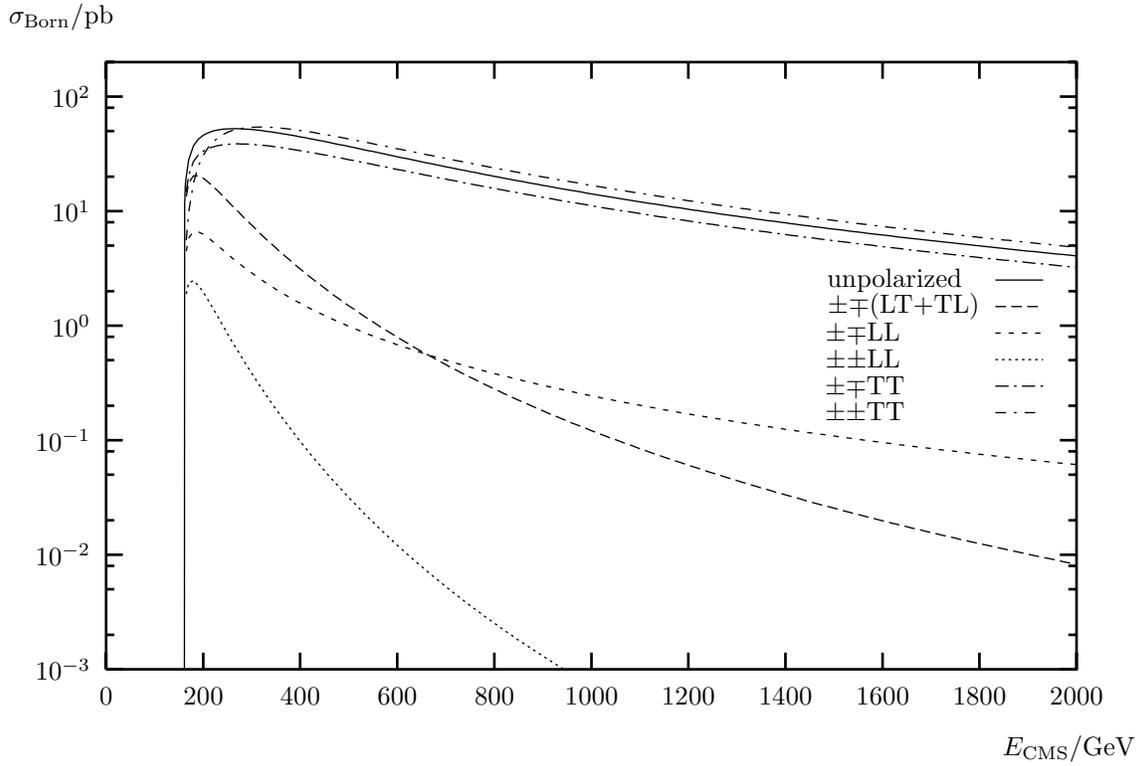}}
\put(0,05){\input{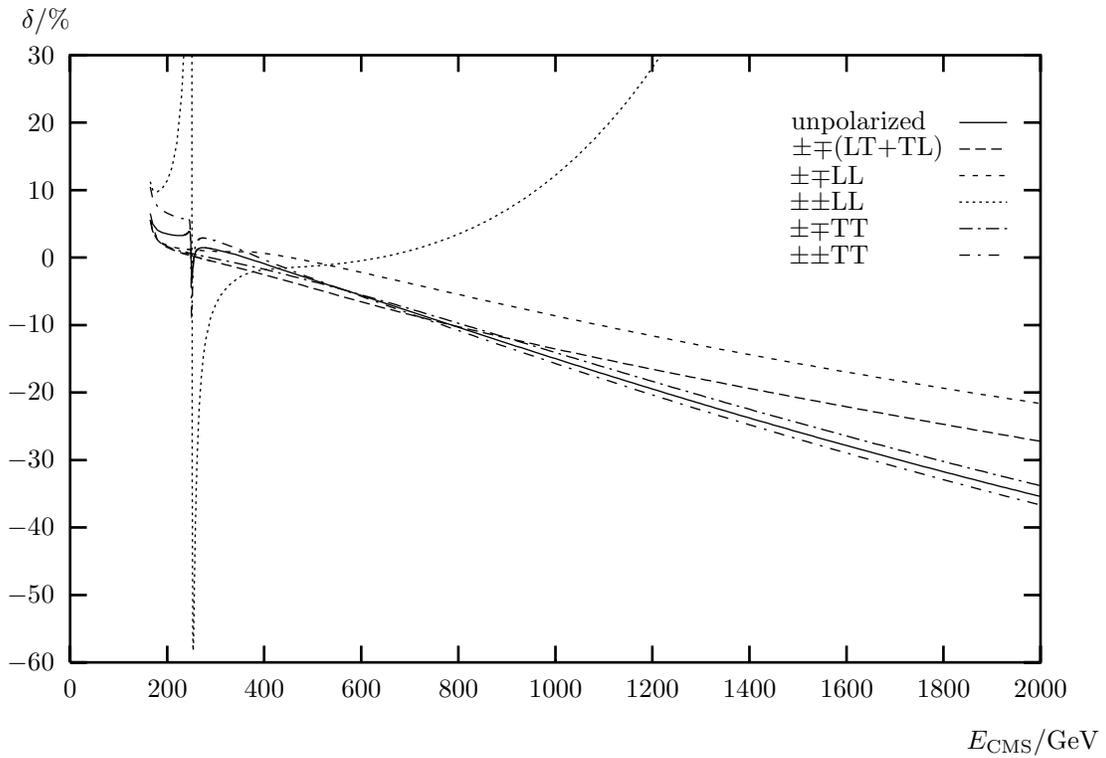}}
\end{picture}
\caption{Same as in \protect\reffi{fi:intcs10} but with an angular cut 
         $20^\circ \leq \theta \leq 160^\circ$}
\label{fi:intcs20}
\end{figure}
This is due to the fact that after cutting off the dominant forward 
and backward peaks we are left
with a region in phase space where the influence of the radiative
corrections becomes more important.
The corrections to the other \css, in particular to $\si_{\pm\mp\rL\rL}$,
are hardly affected.
 
In \reffis{fi:difcspp} and \ref{fi:difcsmm} we
show the corrections to the differential \css\ for $\sqrt{s} =
0.5$, 1 and $2\TeV$.
\begin{figure}
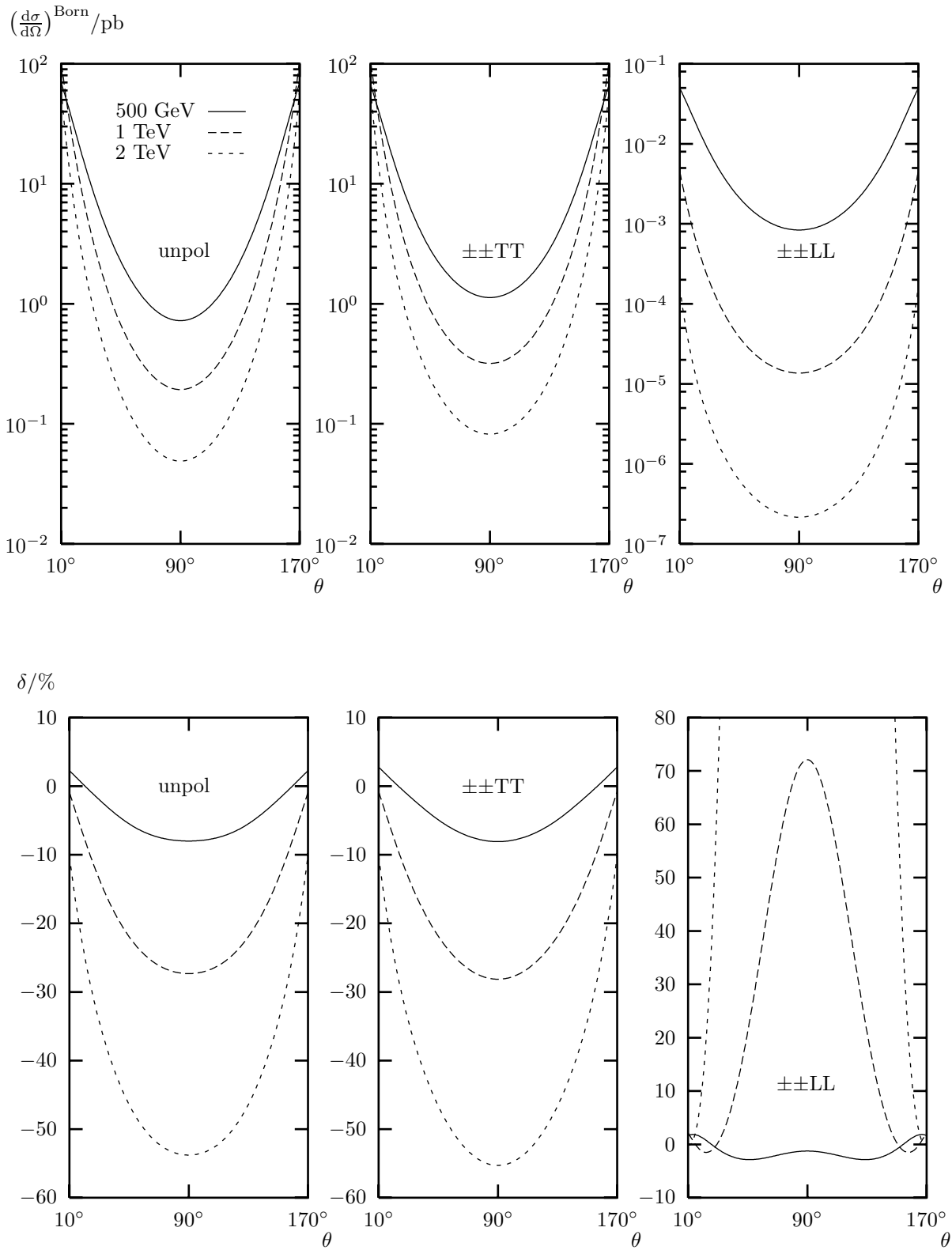

\setlength{\unitlength}{1mm}
\begin{picture}(160,210)(-1,0)
\put(26,168){\makebox(0,0)[l]{{\fs unpol}}}
\put(26,78){\makebox(0,0)[l]{{\fs unpol}}}
\put(77,168){\makebox(0,0)[l]{{\fs $\pm\pm$TT}}}
\put(77,78){\makebox(0,0)[l]{{\fs $\pm\pm$TT}}}
\put(130,168){\makebox(0,0)[l]{{\fs $\pm\pm$LL}}}
\put(130,28){\makebox(0,0)[l]{{\fs $\pm\pm$LL}}}
\put(-10,110){\input{aaww0.unpol}
\hspace*{-1.7cm}\input{aaww0.22TT}
\hspace*{-1.7cm}\input{aaww0.22LL}}
\put(-10,0){
\setlength{\unitlength}{0.1bp}
\special{!
/gnudict 40 dict def
gnudict begin
/Color false def
/Solid false def
/gnulinewidth 5.000 def
/vshift -33 def
/dl {10 mul} def
/hpt 31.5 def
/vpt 31.5 def
/M {moveto} bind def
/L {lineto} bind def
/R {rmoveto} bind def
/V {rlineto} bind def
/vpt2 vpt 2 mul def
/hpt2 hpt 2 mul def
/Lshow { currentpoint stroke M
  0 vshift R show } def
/Rshow { currentpoint stroke M
  dup stringwidth pop neg vshift R show } def
/Cshow { currentpoint stroke M
  dup stringwidth pop -2 div vshift R show } def
/DL { Color {setrgbcolor Solid {pop []} if 0 setdash }
 {pop pop pop Solid {pop []} if 0 setdash} ifelse } def
/BL { stroke gnulinewidth 2 mul setlinewidth } def
/AL { stroke gnulinewidth 2 div setlinewidth } def
/PL { stroke gnulinewidth setlinewidth } def
/LTb { BL [] 0 0 0 DL } def
/LTa { AL [1 dl 2 dl] 0 setdash 0 0 0 setrgbcolor } def
/LT0 { PL [] 0 1 0 DL } def
/LT1 { PL [4 dl 2 dl] 0 0 1 DL } def
/LT2 { PL [2 dl 3 dl] 1 0 0 DL } def
/LT3 { PL [1 dl 1.5 dl] 1 0 1 DL } def
/LT4 { PL [5 dl 2 dl 1 dl 2 dl] 0 1 1 DL } def
/LT5 { PL [4 dl 3 dl 1 dl 3 dl] 1 1 0 DL } def
/LT6 { PL [2 dl 2 dl 2 dl 4 dl] 0 0 0 DL } def
/LT7 { PL [2 dl 2 dl 2 dl 2 dl 2 dl 4 dl] 1 0.3 0 DL } def
/LT8 { PL [2 dl 2 dl 2 dl 2 dl 2 dl 2 dl 2 dl 4 dl] 0.5 0.5 0.5 DL } def
/P { stroke [] 0 setdash
  currentlinewidth 2 div sub M
  0 currentlinewidth V stroke } def
/D { stroke [] 0 setdash 2 copy vpt add M
  hpt neg vpt neg V hpt vpt neg V
  hpt vpt V hpt neg vpt V closepath stroke
  P } def
/A { stroke [] 0 setdash vpt sub M 0 vpt2 V
  currentpoint stroke M
  hpt neg vpt neg R hpt2 0 V stroke
  } def
/B { stroke [] 0 setdash 2 copy exch hpt sub exch vpt add M
  0 vpt2 neg V hpt2 0 V 0 vpt2 V
  hpt2 neg 0 V closepath stroke
  P } def
/C { stroke [] 0 setdash exch hpt sub exch vpt add M
  hpt2 vpt2 neg V currentpoint stroke M
  hpt2 neg 0 R hpt2 vpt2 V stroke } def
/T { stroke [] 0 setdash 2 copy vpt 1.12 mul add M
  hpt neg vpt -1.62 mul V
  hpt 2 mul 0 V
  hpt neg vpt 1.62 mul V closepath stroke
  P  } def
/S { 2 copy A C} def
end
}
\begin{picture}(1800,2592)(0,0)
\special{"
gnudict begin
gsave
50 50 translate
0.100 0.100 scale
0 setgray
/Helvetica findfont 100 scalefont setfont
newpath
-500.000000 -500.000000 translate
LTa
LTb
480 251 M
63 0 V
1074 0 R
-63 0 V
480 578 M
63 0 V
1074 0 R
-63 0 V
480 905 M
63 0 V
1074 0 R
-63 0 V
480 1232 M
63 0 V
1074 0 R
-63 0 V
480 1560 M
63 0 V
1074 0 R
-63 0 V
480 1887 M
63 0 V
1074 0 R
-63 0 V
480 2214 M
63 0 V
1074 0 R
-63 0 V
480 2541 M
63 0 V
1074 0 R
-63 0 V
480 251 M
0 63 V
0 2227 R
0 -63 V
1049 251 M
0 63 V
0 2227 R
0 -63 V
1617 251 M
0 63 V
0 2227 R
0 -63 V
480 251 M
1137 0 V
0 2290 V
-1137 0 V
480 251 L
LT0
480 2287 M
4 -3 V
3 -3 V
4 -4 V
3 -3 V
4 -3 V
3 -4 V
4 -3 V
3 -4 V
4 -4 V
4 -3 V
3 -4 V
4 -3 V
3 -4 V
4 -3 V
3 -4 V
4 -3 V
3 -4 V
4 -4 V
4 -3 V
3 -4 V
4 -3 V
3 -4 V
4 -3 V
3 -3 V
4 -4 V
3 -3 V
4 -4 V
3 -3 V
4 -4 V
4 -3 V
3 -3 V
4 -4 V
3 -3 V
4 -3 V
3 -4 V
4 -3 V
3 -3 V
4 -4 V
4 -3 V
3 -3 V
4 -4 V
3 -3 V
4 -3 V
3 -3 V
4 -4 V
3 -3 V
4 -3 V
4 -3 V
3 -3 V
4 -4 V
3 -3 V
4 -3 V
3 -3 V
4 -3 V
3 -3 V
4 -3 V
4 -3 V
3 -3 V
4 -3 V
3 -3 V
4 -3 V
3 -3 V
4 -3 V
3 -2 V
4 -3 V
4 -3 V
3 -3 V
4 -2 V
3 -3 V
4 -3 V
3 -2 V
4 -3 V
3 -2 V
4 -3 V
3 -3 V
4 -2 V
4 -2 V
3 -3 V
4 -2 V
3 -2 V
4 -3 V
3 -2 V
4 -2 V
3 -2 V
4 -3 V
4 -2 V
3 -2 V
4 -2 V
3 -2 V
4 -2 V
3 -2 V
4 -2 V
3 -2 V
4 -1 V
4 -2 V
3 -2 V
4 -2 V
3 -1 V
4 -2 V
3 -2 V
4 -1 V
3 -2 V
4 -1 V
4 -2 V
3 -1 V
4 -2 V
3 -1 V
4 -1 V
3 -2 V
4 -1 V
3 -1 V
4 -1 V
4 -2 V
3 -1 V
4 -1 V
3 -1 V
4 -1 V
3 -1 V
4 -1 V
3 -1 V
4 -1 V
3 -1 V
4 -1 V
4 -1 V
3 0 V
4 -1 V
3 -1 V
4 -1 V
3 0 V
4 -1 V
3 -1 V
4 0 V
4 -1 V
3 -1 V
4 0 V
3 -1 V
4 0 V
3 -1 V
4 0 V
3 -1 V
4 0 V
4 0 V
3 -1 V
4 0 V
3 0 V
4 -1 V
3 0 V
4 0 V
3 0 V
4 -1 V
4 0 V
3 0 V
4 0 V
3 0 V
4 0 V
3 -1 V
4 0 V
3 0 V
4 0 V
4 0 V
3 0 V
4 0 V
3 0 V
4 0 V
3 1 V
4 0 V
3 0 V
4 0 V
3 0 V
4 0 V
4 1 V
3 0 V
4 0 V
3 0 V
4 1 V
3 0 V
4 0 V
3 1 V
4 0 V
4 0 V
3 1 V
4 0 V
3 1 V
4 0 V
3 1 V
4 0 V
3 1 V
4 1 V
4 0 V
3 1 V
4 1 V
3 0 V
4 1 V
3 1 V
4 1 V
3 0 V
4 1 V
4 1 V
3 1 V
4 1 V
3 1 V
4 1 V
3 1 V
4 1 V
3 1 V
4 1 V
3 1 V
4 2 V
4 1 V
3 1 V
4 1 V
3 2 V
4 1 V
3 1 V
4 2 V
3 1 V
4 2 V
4 1 V
3 2 V
4 1 V
3 2 V
4 2 V
3 1 V
4 2 V
3 2 V
4 2 V
4 1 V
3 2 V
4 2 V
3 2 V
4 2 V
3 2 V
4 2 V
3 2 V
4 2 V
4 3 V
3 2 V
4 2 V
3 2 V
4 3 V
3 2 V
4 2 V
3 3 V
4 2 V
4 2 V
3 3 V
4 3 V
3 2 V
4 3 V
3 2 V
4 3 V
3 3 V
4 2 V
3 3 V
4 3 V
4 3 V
3 2 V
4 3 V
3 3 V
4 3 V
3 3 V
4 3 V
3 3 V
4 3 V
4 3 V
3 3 V
4 3 V
3 3 V
4 3 V
3 3 V
4 4 V
3 3 V
4 3 V
4 3 V
3 3 V
4 4 V
3 3 V
4 3 V
3 3 V
4 4 V
3 3 V
4 3 V
4 4 V
3 3 V
4 3 V
3 4 V
4 3 V
3 3 V
4 4 V
3 3 V
4 3 V
4 4 V
3 3 V
4 4 V
3 3 V
4 4 V
3 3 V
4 3 V
3 4 V
4 3 V
3 4 V
4 3 V
4 4 V
3 4 V
4 3 V
3 4 V
4 3 V
3 4 V
4 3 V
3 4 V
4 3 V
4 4 V
3 4 V
4 3 V
3 4 V
4 3 V
3 3 V
4 4 V
3 3 V
4 3 V
LT1
480 2183 M
4 -10 V
3 -11 V
4 -11 V
3 -11 V
4 -11 V
3 -11 V
4 -11 V
3 -11 V
4 -11 V
4 -11 V
3 -11 V
4 -11 V
3 -11 V
4 -11 V
3 -11 V
4 -10 V
3 -11 V
4 -11 V
4 -11 V
3 -10 V
4 -11 V
3 -10 V
4 -11 V
3 -10 V
4 -11 V
3 -10 V
4 -10 V
3 -10 V
4 -10 V
4 -10 V
3 -9 V
4 -10 V
3 -10 V
4 -9 V
3 -9 V
4 -10 V
3 -9 V
4 -9 V
4 -9 V
3 -9 V
4 -8 V
3 -9 V
4 -9 V
3 -8 V
4 -8 V
3 -9 V
4 -8 V
4 -8 V
3 -8 V
4 -8 V
3 -7 V
4 -8 V
3 -8 V
4 -7 V
3 -7 V
4 -8 V
4 -7 V
3 -7 V
4 -7 V
3 -7 V
4 -7 V
3 -6 V
4 -7 V
3 -6 V
4 -7 V
4 -6 V
3 -6 V
4 -6 V
3 -6 V
4 -6 V
3 -6 V
4 -6 V
3 -6 V
4 -5 V
3 -6 V
4 -5 V
4 -5 V
3 -6 V
4 -5 V
3 -5 V
4 -5 V
3 -5 V
4 -4 V
3 -5 V
4 -5 V
4 -4 V
3 -5 V
4 -4 V
3 -4 V
4 -5 V
3 -4 V
4 -4 V
3 -4 V
4 -4 V
4 -3 V
3 -4 V
4 -4 V
3 -3 V
4 -4 V
3 -3 V
4 -4 V
3 -3 V
4 -3 V
4 -3 V
3 -4 V
4 -3 V
3 -2 V
4 -3 V
3 -3 V
4 -3 V
3 -3 V
4 -2 V
4 -3 V
3 -2 V
4 -3 V
3 -2 V
4 -2 V
3 -2 V
4 -2 V
3 -3 V
4 -2 V
3 -1 V
4 -2 V
4 -2 V
3 -2 V
4 -2 V
3 -1 V
4 -2 V
3 -2 V
4 -1 V
3 -1 V
4 -2 V
4 -1 V
3 -1 V
4 -2 V
3 -1 V
4 -1 V
3 -1 V
4 -1 V
3 -1 V
4 -1 V
4 -1 V
3 -1 V
4 0 V
3 -1 V
4 -1 V
3 0 V
4 -1 V
3 0 V
4 -1 V
4 0 V
3 -1 V
4 0 V
3 0 V
4 -1 V
3 0 V
4 0 V
3 0 V
4 0 V
4 0 V
3 0 V
4 0 V
3 0 V
4 0 V
3 0 V
4 1 V
3 0 V
4 0 V
3 1 V
4 0 V
4 1 V
3 0 V
4 1 V
3 0 V
4 1 V
3 1 V
4 0 V
3 1 V
4 1 V
4 1 V
3 1 V
4 1 V
3 1 V
4 1 V
3 1 V
4 2 V
3 1 V
4 1 V
4 2 V
3 1 V
4 1 V
3 2 V
4 2 V
3 1 V
4 2 V
3 2 V
4 2 V
4 2 V
3 1 V
4 2 V
3 3 V
4 2 V
3 2 V
4 2 V
3 2 V
4 3 V
3 2 V
4 3 V
4 2 V
3 3 V
4 3 V
3 3 V
4 3 V
3 2 V
4 3 V
3 4 V
4 3 V
4 3 V
3 3 V
4 4 V
3 3 V
4 4 V
3 3 V
4 4 V
3 4 V
4 3 V
4 4 V
3 4 V
4 4 V
3 4 V
4 5 V
3 4 V
4 4 V
3 5 V
4 4 V
4 5 V
3 5 V
4 4 V
3 5 V
4 5 V
3 5 V
4 5 V
3 6 V
4 5 V
4 5 V
3 6 V
4 5 V
3 6 V
4 6 V
3 6 V
4 6 V
3 6 V
4 6 V
3 6 V
4 6 V
4 7 V
3 6 V
4 7 V
3 6 V
4 7 V
3 7 V
4 7 V
3 7 V
4 7 V
4 8 V
3 7 V
4 7 V
3 8 V
4 8 V
3 7 V
4 8 V
3 8 V
4 8 V
4 8 V
3 9 V
4 8 V
3 8 V
4 9 V
3 9 V
4 8 V
3 9 V
4 9 V
4 9 V
3 9 V
4 10 V
3 9 V
4 9 V
3 10 V
4 10 V
3 9 V
4 10 V
4 10 V
3 10 V
4 10 V
3 10 V
4 11 V
3 10 V
4 11 V
3 10 V
4 11 V
3 10 V
4 11 V
4 11 V
3 11 V
4 10 V
3 11 V
4 11 V
3 11 V
4 11 V
3 11 V
4 11 V
4 11 V
3 11 V
4 11 V
3 11 V
4 11 V
3 11 V
4 11 V
3 11 V
4 10 V
LT2
480 1882 M
4 -28 V
3 -27 V
4 -27 V
3 -26 V
4 -26 V
3 -26 V
4 -25 V
3 -25 V
4 -24 V
4 -24 V
3 -23 V
4 -23 V
3 -22 V
4 -22 V
3 -21 V
4 -21 V
3 -21 V
4 -20 V
4 -20 V
3 -19 V
4 -19 V
3 -19 V
4 -18 V
3 -18 V
4 -17 V
3 -18 V
4 -17 V
3 -16 V
4 -17 V
4 -16 V
3 -16 V
4 -15 V
3 -15 V
4 -15 V
3 -15 V
4 -15 V
3 -14 V
4 -14 V
4 -14 V
3 -13 V
4 -13 V
3 -14 V
4 -12 V
3 -13 V
4 -13 V
3 -12 V
4 -12 V
4 -12 V
3 -12 V
4 -11 V
3 -11 V
4 -12 V
3 -11 V
4 -10 V
3 -11 V
4 -10 V
4 -11 V
3 -10 V
4 -10 V
3 -10 V
4 -9 V
3 -10 V
4 -9 V
3 -9 V
4 -9 V
4 -9 V
3 -9 V
4 -9 V
3 -8 V
4 -9 V
3 -8 V
4 -8 V
3 -8 V
4 -8 V
3 -7 V
4 -8 V
4 -7 V
3 -7 V
4 -8 V
3 -7 V
4 -6 V
3 -7 V
4 -7 V
3 -6 V
4 -7 V
4 -6 V
3 -6 V
4 -6 V
3 -6 V
4 -6 V
3 -6 V
4 -5 V
3 -6 V
4 -5 V
4 -5 V
3 -5 V
4 -5 V
3 -5 V
4 -5 V
3 -5 V
4 -4 V
3 -5 V
4 -4 V
4 -5 V
3 -4 V
4 -4 V
3 -4 V
4 -4 V
3 -4 V
4 -3 V
3 -4 V
4 -3 V
4 -4 V
3 -3 V
4 -3 V
3 -4 V
4 -3 V
3 -3 V
4 -3 V
3 -2 V
4 -3 V
3 -3 V
4 -2 V
4 -3 V
3 -2 V
4 -3 V
3 -2 V
4 -2 V
3 -2 V
4 -2 V
3 -2 V
4 -2 V
4 -2 V
3 -1 V
4 -2 V
3 -2 V
4 -1 V
3 -2 V
4 -1 V
3 -1 V
4 -1 V
4 -2 V
3 -1 V
4 -1 V
3 -1 V
4 -1 V
3 0 V
4 -1 V
3 -1 V
4 -1 V
4 0 V
3 -1 V
4 0 V
3 -1 V
4 0 V
3 0 V
4 0 V
3 -1 V
4 0 V
4 0 V
3 0 V
4 0 V
3 1 V
4 0 V
3 0 V
4 0 V
3 1 V
4 0 V
3 1 V
4 0 V
4 1 V
3 1 V
4 1 V
3 0 V
4 1 V
3 1 V
4 1 V
3 1 V
4 2 V
4 1 V
3 1 V
4 1 V
3 2 V
4 1 V
3 2 V
4 2 V
3 1 V
4 2 V
4 2 V
3 2 V
4 2 V
3 2 V
4 2 V
3 2 V
4 3 V
3 2 V
4 3 V
4 2 V
3 3 V
4 3 V
3 2 V
4 3 V
3 3 V
4 3 V
3 4 V
4 3 V
3 3 V
4 4 V
4 3 V
3 4 V
4 3 V
3 4 V
4 4 V
3 4 V
4 4 V
3 4 V
4 5 V
4 4 V
3 5 V
4 4 V
3 5 V
4 5 V
3 5 V
4 5 V
3 5 V
4 5 V
4 5 V
3 6 V
4 5 V
3 6 V
4 6 V
3 6 V
4 6 V
3 6 V
4 6 V
4 7 V
3 6 V
4 7 V
3 7 V
4 6 V
3 7 V
4 8 V
3 7 V
4 7 V
4 8 V
3 7 V
4 8 V
3 8 V
4 8 V
3 8 V
4 9 V
3 8 V
4 9 V
3 9 V
4 9 V
4 9 V
3 9 V
4 9 V
3 10 V
4 9 V
3 10 V
4 10 V
3 10 V
4 11 V
4 10 V
3 11 V
4 10 V
3 11 V
4 12 V
3 11 V
4 11 V
3 12 V
4 12 V
4 12 V
3 12 V
4 13 V
3 13 V
4 12 V
3 14 V
4 13 V
3 13 V
4 14 V
4 14 V
3 14 V
4 15 V
3 15 V
4 15 V
3 15 V
4 15 V
3 16 V
4 16 V
4 17 V
3 16 V
4 17 V
3 18 V
4 17 V
3 18 V
4 18 V
3 19 V
4 19 V
3 19 V
4 20 V
4 20 V
3 21 V
4 21 V
3 21 V
4 22 V
3 22 V
4 23 V
3 23 V
4 24 V
4 24 V
3 25 V
4 25 V
3 26 V
4 26 V
3 26 V
4 27 V
3 27 V
4 28 V
stroke
grestore
end
showpage
}
\put(231,2705){\makebox(0,0)[l]{{\fs$\delta$/\%}}}
\put(1708,51){\makebox(0,0){{\fs $\theta$}}}
\put(1617,151){\makebox(0,0){{\fs $170^\circ$}}}
\put(1049,151){\makebox(0,0){{\fs $90^\circ$}}}
\put(480,151){\makebox(0,0){{\fs $10^\circ$}}}
\put(420,2541){\makebox(0,0)[r]{{\fs $10$}}}
\put(420,2214){\makebox(0,0)[r]{{\fs $0$}}}
\put(420,1887){\makebox(0,0)[r]{{\fs $-10$}}}
\put(420,1560){\makebox(0,0)[r]{{\fs $-20$}}}
\put(420,1232){\makebox(0,0)[r]{{\fs $-30$}}}
\put(420,905){\makebox(0,0)[r]{{\fs $-40$}}}
\put(420,578){\makebox(0,0)[r]{{\fs $-50$}}}
\put(420,251){\makebox(0,0)[r]{{\fs $-60$}}}
\end{picture}
\hspace*{-1.7cm}
\setlength{\unitlength}{0.1bp}
\special{!
/gnudict 40 dict def
gnudict begin
/Color false def
/Solid false def
/gnulinewidth 5.000 def
/vshift -33 def
/dl {10 mul} def
/hpt 31.5 def
/vpt 31.5 def
/M {moveto} bind def
/L {lineto} bind def
/R {rmoveto} bind def
/V {rlineto} bind def
/vpt2 vpt 2 mul def
/hpt2 hpt 2 mul def
/Lshow { currentpoint stroke M
  0 vshift R show } def
/Rshow { currentpoint stroke M
  dup stringwidth pop neg vshift R show } def
/Cshow { currentpoint stroke M
  dup stringwidth pop -2 div vshift R show } def
/DL { Color {setrgbcolor Solid {pop []} if 0 setdash }
 {pop pop pop Solid {pop []} if 0 setdash} ifelse } def
/BL { stroke gnulinewidth 2 mul setlinewidth } def
/AL { stroke gnulinewidth 2 div setlinewidth } def
/PL { stroke gnulinewidth setlinewidth } def
/LTb { BL [] 0 0 0 DL } def
/LTa { AL [1 dl 2 dl] 0 setdash 0 0 0 setrgbcolor } def
/LT0 { PL [] 0 1 0 DL } def
/LT1 { PL [4 dl 2 dl] 0 0 1 DL } def
/LT2 { PL [2 dl 3 dl] 1 0 0 DL } def
/LT3 { PL [1 dl 1.5 dl] 1 0 1 DL } def
/LT4 { PL [5 dl 2 dl 1 dl 2 dl] 0 1 1 DL } def
/LT5 { PL [4 dl 3 dl 1 dl 3 dl] 1 1 0 DL } def
/LT6 { PL [2 dl 2 dl 2 dl 4 dl] 0 0 0 DL } def
/LT7 { PL [2 dl 2 dl 2 dl 2 dl 2 dl 4 dl] 1 0.3 0 DL } def
/LT8 { PL [2 dl 2 dl 2 dl 2 dl 2 dl 2 dl 2 dl 4 dl] 0.5 0.5 0.5 DL } def
/P { stroke [] 0 setdash
  currentlinewidth 2 div sub M
  0 currentlinewidth V stroke } def
/D { stroke [] 0 setdash 2 copy vpt add M
  hpt neg vpt neg V hpt vpt neg V
  hpt vpt V hpt neg vpt V closepath stroke
  P } def
/A { stroke [] 0 setdash vpt sub M 0 vpt2 V
  currentpoint stroke M
  hpt neg vpt neg R hpt2 0 V stroke
  } def
/B { stroke [] 0 setdash 2 copy exch hpt sub exch vpt add M
  0 vpt2 neg V hpt2 0 V 0 vpt2 V
  hpt2 neg 0 V closepath stroke
  P } def
/C { stroke [] 0 setdash exch hpt sub exch vpt add M
  hpt2 vpt2 neg V currentpoint stroke M
  hpt2 neg 0 R hpt2 vpt2 V stroke } def
/T { stroke [] 0 setdash 2 copy vpt 1.12 mul add M
  hpt neg vpt -1.62 mul V
  hpt 2 mul 0 V
  hpt neg vpt 1.62 mul V closepath stroke
  P  } def
/S { 2 copy A C} def
end
}
\begin{picture}(1800,2592)(0,0)
\special{"
gnudict begin
gsave
50 50 translate
0.100 0.100 scale
0 setgray
/Helvetica findfont 100 scalefont setfont
newpath
-500.000000 -500.000000 translate
LTa
LTb
480 251 M
63 0 V
1074 0 R
-63 0 V
480 578 M
63 0 V
1074 0 R
-63 0 V
480 905 M
63 0 V
1074 0 R
-63 0 V
480 1232 M
63 0 V
1074 0 R
-63 0 V
480 1560 M
63 0 V
1074 0 R
-63 0 V
480 1887 M
63 0 V
1074 0 R
-63 0 V
480 2214 M
63 0 V
1074 0 R
-63 0 V
480 2541 M
63 0 V
1074 0 R
-63 0 V
480 251 M
0 63 V
0 2227 R
0 -63 V
1049 251 M
0 63 V
0 2227 R
0 -63 V
1617 251 M
0 63 V
0 2227 R
0 -63 V
480 251 M
1137 0 V
0 2290 V
-1137 0 V
480 251 L
LT0
480 2305 M
4 -4 V
3 -3 V
4 -3 V
3 -4 V
4 -3 V
3 -4 V
4 -3 V
3 -4 V
4 -3 V
4 -4 V
3 -3 V
4 -4 V
3 -3 V
4 -4 V
3 -3 V
4 -4 V
3 -3 V
4 -4 V
4 -3 V
3 -3 V
4 -4 V
3 -3 V
4 -3 V
3 -4 V
4 -3 V
3 -3 V
4 -4 V
3 -3 V
4 -3 V
4 -3 V
3 -4 V
4 -3 V
3 -3 V
4 -3 V
3 -4 V
4 -3 V
3 -3 V
4 -3 V
4 -3 V
3 -4 V
4 -3 V
3 -3 V
4 -3 V
3 -3 V
4 -3 V
3 -3 V
4 -4 V
4 -3 V
3 -3 V
4 -3 V
3 -3 V
4 -3 V
3 -3 V
4 -3 V
3 -3 V
4 -3 V
4 -3 V
3 -3 V
4 -3 V
3 -3 V
4 -3 V
3 -3 V
4 -3 V
3 -3 V
4 -3 V
4 -3 V
3 -2 V
4 -3 V
3 -3 V
4 -3 V
3 -3 V
4 -2 V
3 -3 V
4 -3 V
3 -2 V
4 -3 V
4 -3 V
3 -2 V
4 -3 V
3 -2 V
4 -3 V
3 -2 V
4 -3 V
3 -2 V
4 -3 V
4 -2 V
3 -2 V
4 -3 V
3 -2 V
4 -2 V
3 -2 V
4 -3 V
3 -2 V
4 -2 V
4 -2 V
3 -2 V
4 -2 V
3 -2 V
4 -2 V
3 -2 V
4 -2 V
3 -2 V
4 -2 V
4 -2 V
3 -1 V
4 -2 V
3 -2 V
4 -2 V
3 -1 V
4 -2 V
3 -2 V
4 -1 V
4 -2 V
3 -1 V
4 -2 V
3 -1 V
4 -2 V
3 -1 V
4 -2 V
3 -1 V
4 -1 V
3 -1 V
4 -2 V
4 -1 V
3 -1 V
4 -1 V
3 -1 V
4 -1 V
3 -1 V
4 -1 V
3 -1 V
4 -1 V
4 -1 V
3 -1 V
4 -1 V
3 -1 V
4 -1 V
3 0 V
4 -1 V
3 -1 V
4 0 V
4 -1 V
3 -1 V
4 0 V
3 -1 V
4 0 V
3 -1 V
4 0 V
3 0 V
4 -1 V
4 0 V
3 0 V
4 -1 V
3 0 V
4 0 V
3 0 V
4 0 V
3 0 V
4 0 V
4 0 V
3 0 V
4 0 V
3 0 V
4 0 V
3 0 V
4 0 V
3 0 V
4 1 V
3 0 V
4 0 V
4 1 V
3 0 V
4 0 V
3 1 V
4 0 V
3 1 V
4 0 V
3 1 V
4 1 V
4 0 V
3 1 V
4 1 V
3 0 V
4 1 V
3 1 V
4 1 V
3 1 V
4 1 V
4 1 V
3 1 V
4 1 V
3 1 V
4 1 V
3 1 V
4 1 V
3 1 V
4 1 V
4 2 V
3 1 V
4 1 V
3 1 V
4 2 V
3 1 V
4 2 V
3 1 V
4 2 V
3 1 V
4 2 V
4 1 V
3 2 V
4 2 V
3 1 V
4 2 V
3 2 V
4 2 V
3 1 V
4 2 V
4 2 V
3 2 V
4 2 V
3 2 V
4 2 V
3 2 V
4 2 V
3 2 V
4 2 V
4 2 V
3 2 V
4 3 V
3 2 V
4 2 V
3 2 V
4 3 V
3 2 V
4 2 V
4 3 V
3 2 V
4 3 V
3 2 V
4 3 V
3 2 V
4 3 V
3 2 V
4 3 V
4 3 V
3 2 V
4 3 V
3 3 V
4 2 V
3 3 V
4 3 V
3 3 V
4 3 V
3 2 V
4 3 V
4 3 V
3 3 V
4 3 V
3 3 V
4 3 V
3 3 V
4 3 V
3 3 V
4 3 V
4 3 V
3 3 V
4 3 V
3 3 V
4 3 V
3 3 V
4 3 V
3 3 V
4 3 V
4 4 V
3 3 V
4 3 V
3 3 V
4 3 V
3 3 V
4 3 V
3 4 V
4 3 V
4 3 V
3 3 V
4 3 V
3 4 V
4 3 V
3 3 V
4 3 V
3 4 V
4 3 V
4 3 V
3 3 V
4 4 V
3 3 V
4 3 V
3 4 V
4 3 V
3 3 V
4 4 V
3 3 V
4 3 V
4 4 V
3 3 V
4 4 V
3 3 V
4 4 V
3 3 V
4 4 V
3 3 V
4 4 V
4 3 V
3 4 V
4 3 V
3 4 V
4 3 V
3 4 V
4 3 V
3 3 V
4 4 V
LT1
480 2188 M
4 -11 V
3 -11 V
4 -11 V
3 -10 V
4 -11 V
3 -11 V
4 -11 V
3 -11 V
4 -11 V
4 -11 V
3 -11 V
4 -10 V
3 -11 V
4 -11 V
3 -11 V
4 -11 V
3 -10 V
4 -11 V
4 -11 V
3 -10 V
4 -11 V
3 -10 V
4 -11 V
3 -10 V
4 -10 V
3 -10 V
4 -10 V
3 -10 V
4 -10 V
4 -10 V
3 -10 V
4 -9 V
3 -10 V
4 -9 V
3 -9 V
4 -10 V
3 -9 V
4 -9 V
4 -9 V
3 -9 V
4 -8 V
3 -9 V
4 -8 V
3 -9 V
4 -8 V
3 -9 V
4 -8 V
4 -8 V
3 -8 V
4 -8 V
3 -7 V
4 -8 V
3 -8 V
4 -7 V
3 -8 V
4 -7 V
4 -7 V
3 -7 V
4 -7 V
3 -7 V
4 -7 V
3 -7 V
4 -7 V
3 -6 V
4 -7 V
4 -6 V
3 -7 V
4 -6 V
3 -6 V
4 -7 V
3 -6 V
4 -6 V
3 -5 V
4 -6 V
3 -6 V
4 -6 V
4 -5 V
3 -6 V
4 -5 V
3 -5 V
4 -6 V
3 -5 V
4 -5 V
3 -5 V
4 -5 V
4 -5 V
3 -5 V
4 -5 V
3 -4 V
4 -5 V
3 -4 V
4 -5 V
3 -4 V
4 -5 V
4 -4 V
3 -4 V
4 -4 V
3 -4 V
4 -4 V
3 -4 V
4 -4 V
3 -4 V
4 -3 V
4 -4 V
3 -3 V
4 -4 V
3 -3 V
4 -4 V
3 -3 V
4 -3 V
3 -3 V
4 -3 V
4 -3 V
3 -3 V
4 -3 V
3 -3 V
4 -3 V
3 -3 V
4 -2 V
3 -3 V
4 -2 V
3 -3 V
4 -2 V
4 -2 V
3 -3 V
4 -2 V
3 -2 V
4 -2 V
3 -2 V
4 -2 V
3 -2 V
4 -1 V
4 -2 V
3 -2 V
4 -1 V
3 -2 V
4 -1 V
3 -2 V
4 -1 V
3 -1 V
4 -2 V
4 -1 V
3 -1 V
4 -1 V
3 -1 V
4 -1 V
3 -1 V
4 0 V
3 -1 V
4 -1 V
4 0 V
3 -1 V
4 0 V
3 -1 V
4 0 V
3 0 V
4 -1 V
3 0 V
4 0 V
4 0 V
3 0 V
4 0 V
3 0 V
4 1 V
3 0 V
4 0 V
3 1 V
4 0 V
3 1 V
4 0 V
4 1 V
3 1 V
4 0 V
3 1 V
4 1 V
3 1 V
4 1 V
3 1 V
4 1 V
4 2 V
3 1 V
4 1 V
3 2 V
4 1 V
3 2 V
4 1 V
3 2 V
4 2 V
4 1 V
3 2 V
4 2 V
3 2 V
4 2 V
3 2 V
4 2 V
3 3 V
4 2 V
4 2 V
3 3 V
4 2 V
3 3 V
4 2 V
3 3 V
4 3 V
3 3 V
4 3 V
3 3 V
4 3 V
4 3 V
3 3 V
4 3 V
3 3 V
4 4 V
3 3 V
4 4 V
3 3 V
4 4 V
4 3 V
3 4 V
4 4 V
3 4 V
4 4 V
3 4 V
4 4 V
3 4 V
4 4 V
4 5 V
3 4 V
4 5 V
3 4 V
4 5 V
3 4 V
4 5 V
3 5 V
4 5 V
4 5 V
3 5 V
4 5 V
3 5 V
4 6 V
3 5 V
4 5 V
3 6 V
4 5 V
4 6 V
3 6 V
4 6 V
3 5 V
4 6 V
3 6 V
4 7 V
3 6 V
4 6 V
3 7 V
4 6 V
4 7 V
3 6 V
4 7 V
3 7 V
4 7 V
3 7 V
4 7 V
3 7 V
4 7 V
4 7 V
3 8 V
4 7 V
3 8 V
4 8 V
3 7 V
4 8 V
3 8 V
4 8 V
4 8 V
3 9 V
4 8 V
3 9 V
4 8 V
3 9 V
4 8 V
3 9 V
4 9 V
4 9 V
3 9 V
4 10 V
3 9 V
4 9 V
3 10 V
4 9 V
3 10 V
4 10 V
4 10 V
3 10 V
4 10 V
3 10 V
4 10 V
3 10 V
4 11 V
3 10 V
4 11 V
3 10 V
4 11 V
4 11 V
3 10 V
4 11 V
3 11 V
4 11 V
3 11 V
4 10 V
3 11 V
4 11 V
4 11 V
3 11 V
4 11 V
3 11 V
4 11 V
3 10 V
4 11 V
3 11 V
4 11 V
LT2
480 1883 M
4 -27 V
3 -28 V
4 -27 V
3 -26 V
4 -26 V
3 -26 V
4 -25 V
3 -24 V
4 -25 V
4 -23 V
3 -23 V
4 -23 V
3 -22 V
4 -22 V
3 -22 V
4 -20 V
3 -21 V
4 -20 V
4 -20 V
3 -19 V
4 -19 V
3 -19 V
4 -18 V
3 -18 V
4 -18 V
3 -17 V
4 -17 V
3 -16 V
4 -17 V
4 -16 V
3 -16 V
4 -15 V
3 -15 V
4 -15 V
3 -15 V
4 -15 V
3 -14 V
4 -14 V
4 -14 V
3 -13 V
4 -14 V
3 -13 V
4 -13 V
3 -13 V
4 -12 V
3 -12 V
4 -13 V
4 -12 V
3 -11 V
4 -12 V
3 -11 V
4 -12 V
3 -11 V
4 -11 V
3 -11 V
4 -10 V
4 -11 V
3 -10 V
4 -10 V
3 -10 V
4 -10 V
3 -10 V
4 -9 V
3 -10 V
4 -9 V
4 -9 V
3 -9 V
4 -9 V
3 -9 V
4 -9 V
3 -8 V
4 -9 V
3 -8 V
4 -8 V
3 -8 V
4 -8 V
4 -8 V
3 -8 V
4 -7 V
3 -8 V
4 -7 V
3 -7 V
4 -7 V
3 -7 V
4 -7 V
4 -7 V
3 -7 V
4 -6 V
3 -7 V
4 -6 V
3 -7 V
4 -6 V
3 -6 V
4 -6 V
4 -6 V
3 -5 V
4 -6 V
3 -6 V
4 -5 V
3 -6 V
4 -5 V
3 -5 V
4 -5 V
4 -5 V
3 -5 V
4 -5 V
3 -4 V
4 -5 V
3 -5 V
4 -4 V
3 -4 V
4 -5 V
4 -4 V
3 -4 V
4 -4 V
3 -4 V
4 -3 V
3 -4 V
4 -4 V
3 -3 V
4 -4 V
3 -3 V
4 -3 V
4 -3 V
3 -3 V
4 -3 V
3 -3 V
4 -3 V
3 -3 V
4 -2 V
3 -3 V
4 -2 V
4 -3 V
3 -2 V
4 -2 V
3 -2 V
4 -2 V
3 -2 V
4 -2 V
3 -2 V
4 -1 V
4 -2 V
3 -2 V
4 -1 V
3 -1 V
4 -2 V
3 -1 V
4 -1 V
3 -1 V
4 -1 V
4 0 V
3 -1 V
4 -1 V
3 0 V
4 -1 V
3 0 V
4 0 V
3 -1 V
4 0 V
4 0 V
3 0 V
4 0 V
3 1 V
4 0 V
3 0 V
4 1 V
3 0 V
4 1 V
3 1 V
4 0 V
4 1 V
3 1 V
4 1 V
3 1 V
4 2 V
3 1 V
4 1 V
3 2 V
4 2 V
4 1 V
3 2 V
4 2 V
3 2 V
4 2 V
3 2 V
4 2 V
3 2 V
4 3 V
4 2 V
3 3 V
4 2 V
3 3 V
4 3 V
3 3 V
4 3 V
3 3 V
4 3 V
4 3 V
3 3 V
4 4 V
3 3 V
4 4 V
3 4 V
4 3 V
3 4 V
4 4 V
3 4 V
4 4 V
4 5 V
3 4 V
4 4 V
3 5 V
4 5 V
3 4 V
4 5 V
3 5 V
4 5 V
4 5 V
3 5 V
4 5 V
3 6 V
4 5 V
3 6 V
4 6 V
3 5 V
4 6 V
4 6 V
3 6 V
4 6 V
3 7 V
4 6 V
3 7 V
4 6 V
3 7 V
4 7 V
4 7 V
3 7 V
4 7 V
3 7 V
4 7 V
3 8 V
4 7 V
3 8 V
4 8 V
4 8 V
3 8 V
4 8 V
3 8 V
4 9 V
3 8 V
4 9 V
3 9 V
4 9 V
3 9 V
4 9 V
4 9 V
3 10 V
4 9 V
3 10 V
4 10 V
3 10 V
4 10 V
3 10 V
4 11 V
4 10 V
3 11 V
4 11 V
3 11 V
4 12 V
3 11 V
4 12 V
3 11 V
4 12 V
4 13 V
3 12 V
4 12 V
3 13 V
4 13 V
3 13 V
4 14 V
3 13 V
4 14 V
4 14 V
3 14 V
4 15 V
3 15 V
4 15 V
3 15 V
4 15 V
3 16 V
4 16 V
4 17 V
3 16 V
4 17 V
3 17 V
4 18 V
3 18 V
4 18 V
3 19 V
4 19 V
3 19 V
4 20 V
4 20 V
3 21 V
4 20 V
3 22 V
4 22 V
3 22 V
4 23 V
3 23 V
4 23 V
4 25 V
3 24 V
4 25 V
3 26 V
4 26 V
3 26 V
4 27 V
3 28 V
4 27 V
stroke
grestore
end
showpage
}
\put(1708,51){\makebox(0,0){{\fs $\theta$}}}
\put(1617,151){\makebox(0,0){{\fs $170^\circ$}}}
\put(1049,151){\makebox(0,0){{\fs $90^\circ$}}}
\put(480,151){\makebox(0,0){{\fs $10^\circ$}}}
\put(420,2541){\makebox(0,0)[r]{{\fs $10$}}}
\put(420,2214){\makebox(0,0)[r]{{\fs $0$}}}
\put(420,1887){\makebox(0,0)[r]{{\fs $-10$}}}
\put(420,1560){\makebox(0,0)[r]{{\fs $-20$}}}
\put(420,1232){\makebox(0,0)[r]{{\fs $-30$}}}
\put(420,905){\makebox(0,0)[r]{{\fs $-40$}}}
\put(420,578){\makebox(0,0)[r]{{\fs $-50$}}}
\put(420,251){\makebox(0,0)[r]{{\fs $-60$}}}
\end{picture}
\hspace*{-1.7cm}
\setlength{\unitlength}{0.1bp}
\special{!
/gnudict 40 dict def
gnudict begin
/Color false def
/Solid false def
/gnulinewidth 5.000 def
/vshift -33 def
/dl {10 mul} def
/hpt 31.5 def
/vpt 31.5 def
/M {moveto} bind def
/L {lineto} bind def
/R {rmoveto} bind def
/V {rlineto} bind def
/vpt2 vpt 2 mul def
/hpt2 hpt 2 mul def
/Lshow { currentpoint stroke M
  0 vshift R show } def
/Rshow { currentpoint stroke M
  dup stringwidth pop neg vshift R show } def
/Cshow { currentpoint stroke M
  dup stringwidth pop -2 div vshift R show } def
/DL { Color {setrgbcolor Solid {pop []} if 0 setdash }
 {pop pop pop Solid {pop []} if 0 setdash} ifelse } def
/BL { stroke gnulinewidth 2 mul setlinewidth } def
/AL { stroke gnulinewidth 2 div setlinewidth } def
/PL { stroke gnulinewidth setlinewidth } def
/LTb { BL [] 0 0 0 DL } def
/LTa { AL [1 dl 2 dl] 0 setdash 0 0 0 setrgbcolor } def
/LT0 { PL [] 0 1 0 DL } def
/LT1 { PL [4 dl 2 dl] 0 0 1 DL } def
/LT2 { PL [2 dl 3 dl] 1 0 0 DL } def
/LT3 { PL [1 dl 1.5 dl] 1 0 1 DL } def
/LT4 { PL [5 dl 2 dl 1 dl 2 dl] 0 1 1 DL } def
/LT5 { PL [4 dl 3 dl 1 dl 3 dl] 1 1 0 DL } def
/LT6 { PL [2 dl 2 dl 2 dl 4 dl] 0 0 0 DL } def
/LT7 { PL [2 dl 2 dl 2 dl 2 dl 2 dl 4 dl] 1 0.3 0 DL } def
/LT8 { PL [2 dl 2 dl 2 dl 2 dl 2 dl 2 dl 2 dl 4 dl] 0.5 0.5 0.5 DL } def
/P { stroke [] 0 setdash
  currentlinewidth 2 div sub M
  0 currentlinewidth V stroke } def
/D { stroke [] 0 setdash 2 copy vpt add M
  hpt neg vpt neg V hpt vpt neg V
  hpt vpt V hpt neg vpt V closepath stroke
  P } def
/A { stroke [] 0 setdash vpt sub M 0 vpt2 V
  currentpoint stroke M
  hpt neg vpt neg R hpt2 0 V stroke
  } def
/B { stroke [] 0 setdash 2 copy exch hpt sub exch vpt add M
  0 vpt2 neg V hpt2 0 V 0 vpt2 V
  hpt2 neg 0 V closepath stroke
  P } def
/C { stroke [] 0 setdash exch hpt sub exch vpt add M
  hpt2 vpt2 neg V currentpoint stroke M
  hpt2 neg 0 R hpt2 vpt2 V stroke } def
/T { stroke [] 0 setdash 2 copy vpt 1.12 mul add M
  hpt neg vpt -1.62 mul V
  hpt 2 mul 0 V
  hpt neg vpt 1.62 mul V closepath stroke
  P  } def
/S { 2 copy A C} def
end
}
\begin{picture}(1800,2592)(0,0)
\special{"
gnudict begin
gsave
50 50 translate
0.100 0.100 scale
0 setgray
/Helvetica findfont 100 scalefont setfont
newpath
-500.000000 -500.000000 translate
LTa
LTb
480 251 M
63 0 V
1074 0 R
-63 0 V
480 505 M
63 0 V
1074 0 R
-63 0 V
480 760 M
63 0 V
1074 0 R
-63 0 V
480 1014 M
63 0 V
1074 0 R
-63 0 V
480 1269 M
63 0 V
1074 0 R
-63 0 V
480 1523 M
63 0 V
1074 0 R
-63 0 V
480 1778 M
63 0 V
1074 0 R
-63 0 V
480 2032 M
63 0 V
1074 0 R
-63 0 V
480 2287 M
63 0 V
1074 0 R
-63 0 V
480 2541 M
63 0 V
1074 0 R
-63 0 V
480 251 M
0 63 V
0 2227 R
0 -63 V
1049 251 M
0 63 V
0 2227 R
0 -63 V
1617 251 M
0 63 V
0 2227 R
0 -63 V
480 251 M
1137 0 V
0 2290 V
-1137 0 V
480 251 L
LT0
480 548 M
4 1 V
3 1 V
4 1 V
3 1 V
4 0 V
3 1 V
4 0 V
3 -1 V
4 0 V
4 -1 V
3 -1 V
4 -1 V
3 -1 V
4 -1 V
3 -2 V
4 -2 V
3 -2 V
4 -2 V
4 -2 V
3 -2 V
4 -2 V
3 -3 V
4 -2 V
3 -3 V
4 -2 V
3 -3 V
4 -3 V
3 -3 V
4 -2 V
4 -3 V
3 -3 V
4 -2 V
3 -3 V
4 -3 V
3 -2 V
4 -3 V
3 -3 V
4 -2 V
4 -3 V
3 -2 V
4 -3 V
3 -2 V
4 -2 V
3 -2 V
4 -3 V
3 -2 V
4 -2 V
4 -2 V
3 -2 V
4 -2 V
3 -2 V
4 -1 V
3 -2 V
4 -2 V
3 -1 V
4 -2 V
4 -1 V
3 -2 V
4 -1 V
3 -1 V
4 -1 V
3 -2 V
4 -1 V
3 -1 V
4 -1 V
4 -1 V
3 0 V
4 -1 V
3 -1 V
4 -1 V
3 0 V
4 -1 V
3 0 V
4 -1 V
3 0 V
4 0 V
4 -1 V
3 0 V
4 0 V
3 0 V
4 0 V
3 0 V
4 0 V
3 0 V
4 0 V
4 0 V
3 0 V
4 0 V
3 1 V
4 0 V
3 0 V
4 1 V
3 0 V
4 0 V
4 1 V
3 0 V
4 1 V
3 0 V
4 1 V
3 1 V
4 0 V
3 1 V
4 0 V
4 1 V
3 1 V
4 1 V
3 0 V
4 1 V
3 1 V
4 1 V
3 0 V
4 1 V
4 1 V
3 1 V
4 1 V
3 0 V
4 1 V
3 1 V
4 1 V
3 1 V
4 0 V
3 1 V
4 1 V
4 1 V
3 1 V
4 0 V
3 1 V
4 1 V
3 1 V
4 0 V
3 1 V
4 1 V
4 0 V
3 1 V
4 1 V
3 0 V
4 1 V
3 1 V
4 0 V
3 1 V
4 0 V
4 1 V
3 0 V
4 1 V
3 0 V
4 1 V
3 0 V
4 0 V
3 1 V
4 0 V
4 0 V
3 1 V
4 0 V
3 0 V
4 0 V
3 0 V
4 0 V
3 1 V
4 0 V
4 0 V
3 0 V
4 0 V
3 -1 V
4 0 V
3 0 V
4 0 V
3 0 V
4 0 V
3 -1 V
4 0 V
4 0 V
3 -1 V
4 0 V
3 0 V
4 -1 V
3 0 V
4 -1 V
3 0 V
4 -1 V
4 0 V
3 -1 V
4 0 V
3 -1 V
4 -1 V
3 0 V
4 -1 V
3 -1 V
4 0 V
4 -1 V
3 -1 V
4 0 V
3 -1 V
4 -1 V
3 -1 V
4 0 V
3 -1 V
4 -1 V
4 -1 V
3 -1 V
4 0 V
3 -1 V
4 -1 V
3 -1 V
4 -1 V
3 0 V
4 -1 V
3 -1 V
4 -1 V
4 -1 V
3 0 V
4 -1 V
3 -1 V
4 -1 V
3 0 V
4 -1 V
3 -1 V
4 -1 V
4 0 V
3 -1 V
4 0 V
3 -1 V
4 -1 V
3 0 V
4 -1 V
3 0 V
4 -1 V
4 0 V
3 0 V
4 -1 V
3 0 V
4 0 V
3 -1 V
4 0 V
3 0 V
4 0 V
4 0 V
3 0 V
4 0 V
3 0 V
4 0 V
3 0 V
4 0 V
3 0 V
4 1 V
4 0 V
3 0 V
4 1 V
3 0 V
4 1 V
3 0 V
4 1 V
3 1 V
4 1 V
3 0 V
4 1 V
4 1 V
3 1 V
4 1 V
3 2 V
4 1 V
3 1 V
4 1 V
3 2 V
4 1 V
4 2 V
3 1 V
4 2 V
3 2 V
4 1 V
3 2 V
4 2 V
3 2 V
4 2 V
4 2 V
3 2 V
4 3 V
3 2 V
4 2 V
3 2 V
4 3 V
3 2 V
4 3 V
4 2 V
3 3 V
4 3 V
3 2 V
4 3 V
3 3 V
4 2 V
3 3 V
4 3 V
4 2 V
3 3 V
4 3 V
3 3 V
4 2 V
3 3 V
4 2 V
3 3 V
4 2 V
3 2 V
4 2 V
4 2 V
3 2 V
4 2 V
3 2 V
4 1 V
3 1 V
4 1 V
3 1 V
4 1 V
4 0 V
3 1 V
4 0 V
3 -1 V
4 0 V
3 -1 V
4 -1 V
3 -1 V
4 -1 V
LT1
480 560 M
4 -5 V
3 -6 V
4 -6 V
3 -6 V
4 -5 V
3 -6 V
4 -6 V
3 -5 V
4 -6 V
4 -5 V
3 -5 V
4 -4 V
3 -4 V
4 -4 V
3 -4 V
4 -3 V
3 -3 V
4 -2 V
4 -3 V
3 -1 V
4 -2 V
3 -1 V
4 -1 V
3 0 V
4 0 V
3 1 V
4 0 V
3 2 V
4 1 V
4 2 V
3 2 V
4 3 V
3 3 V
4 3 V
3 4 V
4 4 V
3 5 V
4 5 V
4 5 V
3 6 V
4 6 V
3 6 V
4 7 V
3 7 V
4 8 V
3 8 V
4 8 V
4 9 V
3 9 V
4 9 V
3 10 V
4 10 V
3 11 V
4 11 V
3 11 V
4 12 V
4 12 V
3 13 V
4 13 V
3 13 V
4 14 V
3 14 V
4 14 V
3 15 V
4 15 V
4 15 V
3 16 V
4 16 V
3 16 V
4 17 V
3 17 V
4 17 V
3 18 V
4 18 V
3 18 V
4 19 V
4 19 V
3 19 V
4 19 V
3 20 V
4 19 V
3 21 V
4 20 V
3 20 V
4 21 V
4 21 V
3 21 V
4 21 V
3 22 V
4 21 V
3 22 V
4 22 V
3 22 V
4 22 V
4 22 V
3 22 V
4 23 V
3 22 V
4 22 V
3 23 V
4 22 V
3 23 V
4 22 V
4 22 V
3 23 V
4 22 V
3 22 V
4 22 V
3 22 V
4 22 V
3 22 V
4 21 V
4 22 V
3 21 V
4 21 V
3 21 V
4 20 V
3 21 V
4 20 V
3 20 V
4 19 V
3 19 V
4 19 V
4 19 V
3 18 V
4 18 V
3 18 V
4 17 V
3 17 V
4 16 V
3 16 V
4 16 V
4 15 V
3 15 V
4 14 V
3 14 V
4 13 V
3 13 V
4 12 V
3 12 V
4 11 V
4 11 V
3 10 V
4 10 V
3 9 V
4 8 V
3 8 V
4 8 V
3 7 V
4 6 V
4 6 V
3 5 V
4 4 V
3 4 V
4 4 V
3 2 V
4 3 V
3 1 V
4 1 V
4 0 V
3 0 V
4 -1 V
3 -1 V
4 -3 V
3 -2 V
4 -4 V
3 -4 V
4 -4 V
3 -5 V
4 -6 V
4 -6 V
3 -7 V
4 -8 V
3 -8 V
4 -8 V
3 -9 V
4 -10 V
3 -10 V
4 -11 V
4 -11 V
3 -12 V
4 -12 V
3 -13 V
4 -13 V
3 -14 V
4 -14 V
3 -15 V
4 -15 V
4 -16 V
3 -16 V
4 -16 V
3 -17 V
4 -17 V
3 -18 V
4 -18 V
3 -18 V
4 -19 V
4 -19 V
3 -19 V
4 -19 V
3 -20 V
4 -20 V
3 -21 V
4 -20 V
3 -21 V
4 -21 V
3 -21 V
4 -22 V
4 -21 V
3 -22 V
4 -22 V
3 -22 V
4 -22 V
3 -22 V
4 -22 V
3 -23 V
4 -22 V
4 -22 V
3 -23 V
4 -22 V
3 -23 V
4 -22 V
3 -22 V
4 -23 V
3 -22 V
4 -22 V
4 -22 V
3 -22 V
4 -22 V
3 -22 V
4 -21 V
3 -22 V
4 -21 V
3 -21 V
4 -21 V
4 -21 V
3 -20 V
4 -20 V
3 -21 V
4 -19 V
3 -20 V
4 -19 V
3 -19 V
4 -19 V
4 -19 V
3 -18 V
4 -18 V
3 -18 V
4 -17 V
3 -17 V
4 -17 V
3 -16 V
4 -16 V
3 -16 V
4 -15 V
4 -15 V
3 -15 V
4 -14 V
3 -14 V
4 -14 V
3 -13 V
4 -13 V
3 -13 V
4 -12 V
4 -12 V
3 -11 V
4 -11 V
3 -11 V
4 -10 V
3 -10 V
4 -9 V
3 -9 V
4 -9 V
4 -8 V
3 -8 V
4 -8 V
3 -7 V
4 -7 V
3 -6 V
4 -6 V
3 -6 V
4 -5 V
4 -5 V
3 -5 V
4 -4 V
3 -4 V
4 -3 V
3 -3 V
4 -3 V
3 -2 V
4 -2 V
4 -1 V
3 -2 V
4 0 V
3 -1 V
4 0 V
3 0 V
4 1 V
3 1 V
4 2 V
3 1 V
4 3 V
4 2 V
3 3 V
4 3 V
3 4 V
4 4 V
3 4 V
4 4 V
3 5 V
4 5 V
4 6 V
3 5 V
4 6 V
3 6 V
4 5 V
3 6 V
4 6 V
3 6 V
4 5 V
LT2
480 505 M
4 -2 V
3 0 V
4 1 V
3 3 V
4 5 V
3 7 V
4 8 V
3 10 V
4 12 V
4 14 V
3 16 V
4 18 V
3 20 V
4 22 V
3 24 V
4 26 V
3 29 V
4 31 V
4 34 V
3 37 V
4 39 V
3 42 V
4 45 V
3 48 V
4 51 V
3 54 V
4 58 V
3 61 V
4 65 V
4 68 V
3 72 V
4 76 V
3 80 V
4 84 V
3 88 V
4 92 V
3 97 V
4 101 V
4 105 V
3 110 V
4 115 V
3 120 V
2 50 V
835 0 R
2 -50 V
3 -120 V
4 -115 V
3 -110 V
4 -105 V
4 -101 V
3 -97 V
4 -92 V
3 -88 V
4 -84 V
3 -80 V
4 -76 V
3 -72 V
4 -68 V
4 -65 V
3 -61 V
4 -58 V
3 -54 V
4 -51 V
3 -48 V
4 -45 V
3 -42 V
4 -39 V
3 -37 V
4 -34 V
4 -31 V
3 -29 V
4 -26 V
3 -24 V
4 -22 V
3 -20 V
4 -18 V
3 -16 V
4 -14 V
4 -12 V
3 -10 V
4 -8 V
3 -7 V
4 -5 V
3 -3 V
4 -1 V
3 0 V
4 2 V
stroke
grestore
end
showpage
}
\put(1708,51){\makebox(0,0){{\fs $\theta$}}}
\put(1617,151){\makebox(0,0){{\fs $170^\circ$}}}
\put(1049,151){\makebox(0,0){{\fs $90^\circ$}}}
\put(480,151){\makebox(0,0){{\fs $10^\circ$}}}
\put(420,2541){\makebox(0,0)[r]{{\fs $80$}}}
\put(420,2287){\makebox(0,0)[r]{{\fs $70$}}}
\put(420,2032){\makebox(0,0)[r]{{\fs $60$}}}
\put(420,1778){\makebox(0,0)[r]{{\fs $50$}}}
\put(420,1523){\makebox(0,0)[r]{{\fs $40$}}}
\put(420,1269){\makebox(0,0)[r]{{\fs $30$}}}
\put(420,1014){\makebox(0,0)[r]{{\fs $20$}}}
\put(420,760){\makebox(0,0)[r]{{\fs $10$}}}
\put(420,505){\makebox(0,0)[r]{{\fs $0$}}}
\put(420,251){\makebox(0,0)[r]{{\fs $-10$}}}
\end{picture}}
\end{picture}
\caption{Differential lowest-order \css\ and relative 
corrections for the unpolarized \cs\ and the \css\ with equal photon helicities}
\label{fi:difcspp}
\end{figure}%
\begin{figure}
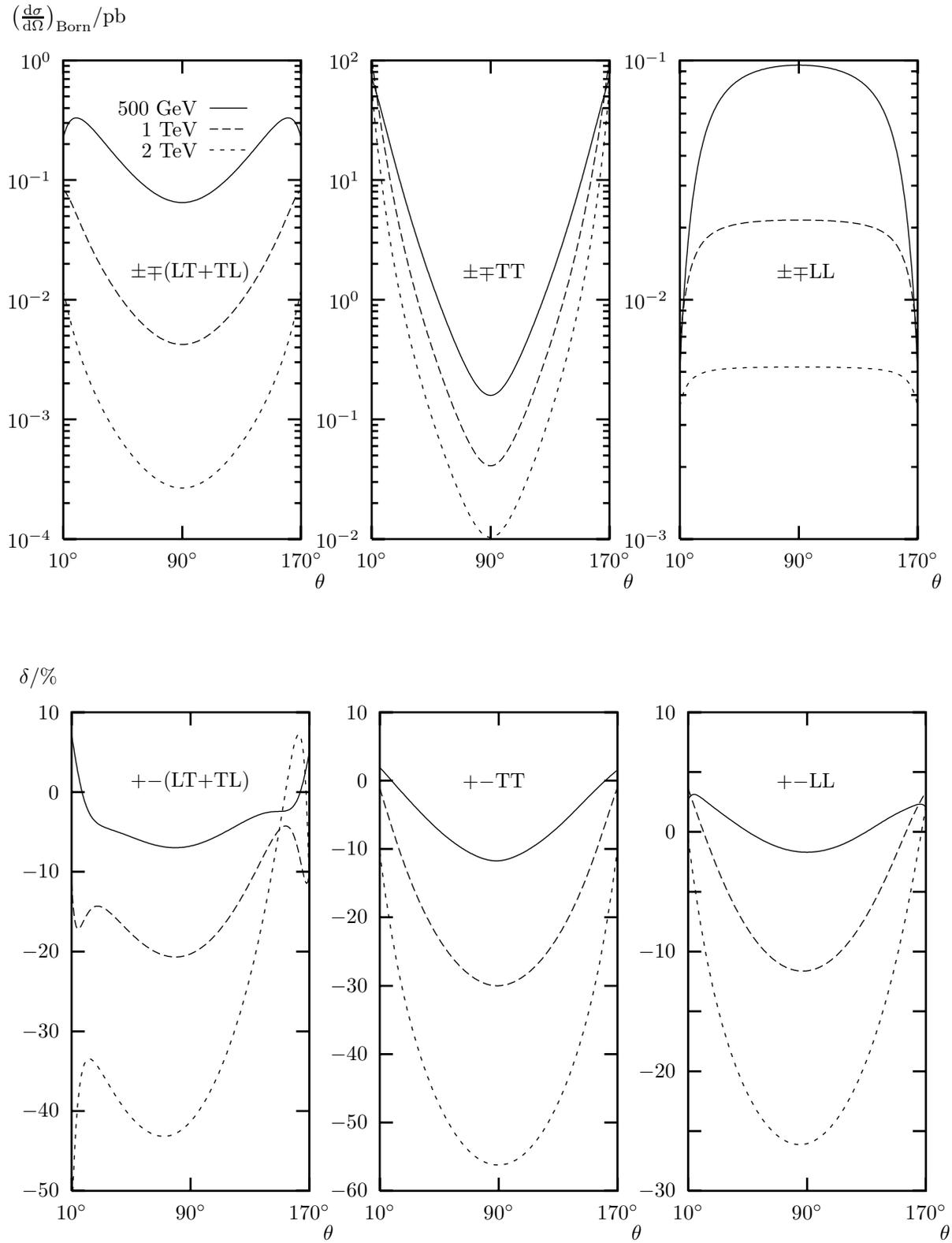

\setlength{\unitlength}{1mm}
\begin{picture}(160,210)(-1,0)
\put(21,164){\makebox(0,0)[l]{{\fs $\pm$$\mp$(LT+TL)}}}
\put(21,78){\makebox(0,0)[l]{{\fs $+$$-$(LT+TL)}}}
\put(77,164){\makebox(0,0)[l]{{\fs $\pm\mp$TT}}}
\put(77,78){\makebox(0,0)[l]{{\fs $+$$-$TT}}}
\put(130,164){\makebox(0,0)[l]{{\fs $\pm\mp$LL}}}
\put(130,78){\makebox(0,0)[l]{{\fs $+$$-$LL}}}
\put(-10,110){\input{aaww0.21_LT_TL}
\hspace*{-1.7cm}\input{aaww0.21TT}
\hspace*{-1.7cm}
\setlength{\unitlength}{0.1bp}
\special{!
/gnudict 40 dict def
gnudict begin
/Color false def
/Solid false def
/gnulinewidth 5.000 def
/vshift -33 def
/dl {10 mul} def
/hpt 31.5 def
/vpt 31.5 def
/M {moveto} bind def
/L {lineto} bind def
/R {rmoveto} bind def
/V {rlineto} bind def
/vpt2 vpt 2 mul def
/hpt2 hpt 2 mul def
/Lshow { currentpoint stroke M
  0 vshift R show } def
/Rshow { currentpoint stroke M
  dup stringwidth pop neg vshift R show } def
/Cshow { currentpoint stroke M
  dup stringwidth pop -2 div vshift R show } def
/DL { Color {setrgbcolor Solid {pop []} if 0 setdash }
 {pop pop pop Solid {pop []} if 0 setdash} ifelse } def
/BL { stroke gnulinewidth 2 mul setlinewidth } def
/AL { stroke gnulinewidth 2 div setlinewidth } def
/PL { stroke gnulinewidth setlinewidth } def
/LTb { BL [] 0 0 0 DL } def
/LTa { AL [1 dl 2 dl] 0 setdash 0 0 0 setrgbcolor } def
/LT0 { PL [] 0 1 0 DL } def
/LT1 { PL [4 dl 2 dl] 0 0 1 DL } def
/LT2 { PL [2 dl 3 dl] 1 0 0 DL } def
/LT3 { PL [1 dl 1.5 dl] 1 0 1 DL } def
/LT4 { PL [5 dl 2 dl 1 dl 2 dl] 0 1 1 DL } def
/LT5 { PL [4 dl 3 dl 1 dl 3 dl] 1 1 0 DL } def
/LT6 { PL [2 dl 2 dl 2 dl 4 dl] 0 0 0 DL } def
/LT7 { PL [2 dl 2 dl 2 dl 2 dl 2 dl 4 dl] 1 0.3 0 DL } def
/LT8 { PL [2 dl 2 dl 2 dl 2 dl 2 dl 2 dl 2 dl 4 dl] 0.5 0.5 0.5 DL } def
/P { stroke [] 0 setdash
  currentlinewidth 2 div sub M
  0 currentlinewidth V stroke } def
/D { stroke [] 0 setdash 2 copy vpt add M
  hpt neg vpt neg V hpt vpt neg V
  hpt vpt V hpt neg vpt V closepath stroke
  P } def
/A { stroke [] 0 setdash vpt sub M 0 vpt2 V
  currentpoint stroke M
  hpt neg vpt neg R hpt2 0 V stroke
  } def
/B { stroke [] 0 setdash 2 copy exch hpt sub exch vpt add M
  0 vpt2 neg V hpt2 0 V 0 vpt2 V
  hpt2 neg 0 V closepath stroke
  P } def
/C { stroke [] 0 setdash exch hpt sub exch vpt add M
  hpt2 vpt2 neg V currentpoint stroke M
  hpt2 neg 0 R hpt2 vpt2 V stroke } def
/T { stroke [] 0 setdash 2 copy vpt 1.12 mul add M
  hpt neg vpt -1.62 mul V
  hpt 2 mul 0 V
  hpt neg vpt 1.62 mul V closepath stroke
  P  } def
/S { 2 copy A C} def
end
}
\begin{picture}(1800,2592)(0,0)
\special{"
gnudict begin
gsave
50 50 translate
0.100 0.100 scale
0 setgray
/Helvetica findfont 100 scalefont setfont
newpath
-500.000000 -500.000000 translate
LTa
LTb
480 251 M
63 0 V
1074 0 R
-63 0 V
480 596 M
31 0 V
1106 0 R
-31 0 V
480 797 M
31 0 V
1106 0 R
-31 0 V
480 940 M
31 0 V
1106 0 R
-31 0 V
480 1051 M
31 0 V
1106 0 R
-31 0 V
480 1142 M
31 0 V
1106 0 R
-31 0 V
480 1219 M
31 0 V
1106 0 R
-31 0 V
480 1285 M
31 0 V
1106 0 R
-31 0 V
480 1344 M
31 0 V
1106 0 R
-31 0 V
480 1396 M
63 0 V
1074 0 R
-63 0 V
480 1741 M
31 0 V
1106 0 R
-31 0 V
480 1942 M
31 0 V
1106 0 R
-31 0 V
480 2085 M
31 0 V
1106 0 R
-31 0 V
480 2196 M
31 0 V
1106 0 R
-31 0 V
480 2287 M
31 0 V
1106 0 R
-31 0 V
480 2364 M
31 0 V
1106 0 R
-31 0 V
480 2430 M
31 0 V
1106 0 R
-31 0 V
480 2489 M
31 0 V
1106 0 R
-31 0 V
480 2541 M
63 0 V
1074 0 R
-63 0 V
480 251 M
0 63 V
0 2227 R
0 -63 V
1049 251 M
0 63 V
0 2227 R
0 -63 V
1617 251 M
0 63 V
0 2227 R
0 -63 V
480 251 M
1137 0 V
0 2290 V
-1137 0 V
480 251 L
LT0
480 1063 M
4 75 V
3 71 V
4 65 V
3 62 V
4 57 V
3 54 V
4 51 V
3 47 V
4 45 V
4 42 V
3 40 V
4 38 V
3 35 V
4 34 V
3 31 V
4 30 V
3 29 V
4 27 V
4 25 V
3 24 V
4 24 V
3 21 V
4 21 V
3 20 V
4 19 V
3 18 V
4 17 V
3 17 V
4 15 V
4 15 V
3 14 V
4 14 V
3 13 V
4 13 V
3 12 V
4 11 V
3 11 V
4 10 V
4 11 V
3 9 V
4 10 V
3 8 V
4 9 V
3 8 V
4 8 V
3 8 V
4 7 V
4 7 V
3 7 V
4 7 V
3 6 V
4 6 V
3 6 V
4 5 V
3 6 V
4 5 V
4 5 V
3 5 V
4 5 V
3 5 V
4 4 V
3 4 V
4 4 V
3 4 V
4 4 V
4 4 V
3 4 V
4 3 V
3 4 V
4 3 V
3 3 V
4 3 V
3 3 V
4 3 V
3 3 V
4 3 V
4 2 V
3 3 V
4 2 V
3 3 V
4 2 V
3 2 V
4 3 V
3 2 V
4 2 V
4 2 V
3 2 V
4 2 V
3 2 V
4 2 V
3 1 V
4 2 V
3 2 V
4 1 V
4 2 V
3 1 V
4 2 V
3 1 V
4 2 V
3 1 V
4 1 V
3 2 V
4 1 V
4 1 V
3 1 V
4 1 V
3 1 V
4 1 V
3 2 V
4 1 V
3 1 V
4 0 V
4 1 V
3 1 V
4 1 V
3 1 V
4 1 V
3 1 V
4 0 V
3 1 V
4 1 V
3 1 V
4 0 V
4 1 V
3 0 V
4 1 V
3 1 V
4 0 V
3 1 V
4 0 V
3 1 V
4 0 V
4 1 V
3 0 V
4 1 V
3 0 V
4 0 V
3 1 V
4 0 V
3 0 V
4 1 V
4 0 V
3 0 V
4 1 V
3 0 V
4 0 V
3 0 V
4 0 V
3 1 V
4 0 V
4 0 V
3 0 V
4 0 V
3 0 V
4 0 V
3 0 V
4 1 V
3 0 V
4 0 V
4 0 V
3 0 V
4 0 V
3 0 V
4 -1 V
3 0 V
4 0 V
3 0 V
4 0 V
3 0 V
4 0 V
4 0 V
3 -1 V
4 0 V
3 0 V
4 0 V
3 0 V
4 -1 V
3 0 V
4 0 V
4 -1 V
3 0 V
4 0 V
3 -1 V
4 0 V
3 0 V
4 -1 V
3 0 V
4 -1 V
4 0 V
3 -1 V
4 0 V
3 -1 V
4 0 V
3 -1 V
4 -1 V
3 0 V
4 -1 V
4 0 V
3 -1 V
4 -1 V
3 -1 V
4 0 V
3 -1 V
4 -1 V
3 -1 V
4 -1 V
3 -1 V
4 -1 V
4 0 V
3 -1 V
4 -1 V
3 -2 V
4 -1 V
3 -1 V
4 -1 V
3 -1 V
4 -1 V
4 -1 V
3 -2 V
4 -1 V
3 -1 V
4 -2 V
3 -1 V
4 -2 V
3 -1 V
4 -2 V
4 -1 V
3 -2 V
4 -2 V
3 -1 V
4 -2 V
3 -2 V
4 -2 V
3 -2 V
4 -2 V
4 -2 V
3 -2 V
4 -3 V
3 -2 V
4 -2 V
3 -3 V
4 -2 V
3 -3 V
4 -2 V
4 -3 V
3 -3 V
4 -3 V
3 -3 V
4 -3 V
3 -3 V
4 -3 V
3 -4 V
4 -3 V
3 -4 V
4 -4 V
4 -4 V
3 -4 V
4 -4 V
3 -4 V
4 -4 V
3 -5 V
4 -5 V
3 -5 V
4 -5 V
4 -5 V
3 -6 V
4 -5 V
3 -6 V
4 -6 V
3 -6 V
4 -7 V
3 -7 V
4 -7 V
4 -7 V
3 -8 V
4 -8 V
3 -8 V
4 -9 V
3 -8 V
4 -10 V
3 -9 V
4 -11 V
4 -10 V
3 -11 V
4 -11 V
3 -12 V
4 -13 V
3 -13 V
4 -14 V
3 -14 V
4 -15 V
4 -15 V
3 -17 V
4 -17 V
3 -18 V
4 -19 V
3 -20 V
4 -21 V
3 -21 V
4 -24 V
3 -24 V
4 -25 V
4 -27 V
3 -29 V
4 -30 V
3 -31 V
4 -34 V
3 -35 V
4 -38 V
3 -40 V
4 -42 V
4 -45 V
3 -47 V
4 -51 V
3 -54 V
4 -57 V
3 -62 V
4 -65 V
3 -71 V
4 -75 V
LT1
480 1176 M
4 44 V
3 40 V
4 35 V
3 33 V
4 30 V
3 27 V
4 24 V
3 23 V
4 21 V
4 19 V
3 18 V
4 16 V
3 15 V
4 14 V
3 13 V
4 12 V
3 11 V
4 11 V
4 9 V
3 9 V
4 9 V
3 8 V
4 7 V
3 7 V
4 7 V
3 6 V
4 6 V
3 6 V
4 5 V
4 5 V
3 5 V
4 4 V
3 4 V
4 4 V
3 4 V
4 4 V
3 3 V
4 3 V
4 3 V
3 3 V
4 3 V
3 3 V
4 2 V
3 3 V
4 2 V
3 2 V
4 3 V
4 2 V
3 2 V
4 1 V
3 2 V
4 2 V
3 2 V
4 1 V
3 2 V
4 1 V
4 2 V
3 1 V
4 1 V
3 2 V
4 1 V
3 1 V
4 1 V
3 1 V
4 1 V
4 1 V
3 1 V
4 1 V
3 1 V
4 1 V
3 1 V
4 1 V
3 1 V
4 0 V
3 1 V
4 1 V
4 1 V
3 0 V
4 1 V
3 1 V
4 0 V
3 1 V
4 0 V
3 1 V
4 0 V
4 1 V
3 1 V
4 0 V
3 1 V
4 0 V
3 0 V
4 1 V
3 0 V
4 1 V
4 0 V
3 1 V
4 0 V
3 0 V
4 1 V
3 0 V
4 0 V
3 1 V
4 0 V
4 0 V
3 1 V
4 0 V
3 0 V
4 0 V
3 1 V
4 0 V
3 0 V
4 0 V
4 1 V
3 0 V
4 0 V
3 0 V
4 1 V
3 0 V
4 0 V
3 0 V
4 0 V
3 0 V
4 1 V
4 0 V
3 0 V
4 0 V
3 0 V
4 0 V
3 1 V
4 0 V
3 0 V
4 0 V
4 0 V
3 0 V
4 0 V
3 0 V
4 0 V
3 1 V
4 0 V
3 0 V
4 0 V
4 0 V
3 0 V
4 0 V
3 0 V
4 0 V
3 0 V
4 0 V
3 0 V
4 0 V
4 0 V
3 0 V
4 0 V
3 0 V
4 0 V
3 0 V
4 0 V
3 0 V
4 0 V
4 0 V
3 0 V
4 0 V
3 0 V
4 0 V
3 0 V
4 0 V
3 0 V
4 0 V
3 0 V
4 0 V
4 0 V
3 0 V
4 0 V
3 0 V
4 0 V
3 0 V
4 0 V
3 0 V
4 0 V
4 0 V
3 0 V
4 0 V
3 -1 V
4 0 V
3 0 V
4 0 V
3 0 V
4 0 V
4 0 V
3 0 V
4 0 V
3 -1 V
4 0 V
3 0 V
4 0 V
3 0 V
4 0 V
4 -1 V
3 0 V
4 0 V
3 0 V
4 0 V
3 0 V
4 -1 V
3 0 V
4 0 V
3 0 V
4 -1 V
4 0 V
3 0 V
4 0 V
3 -1 V
4 0 V
3 0 V
4 0 V
3 -1 V
4 0 V
4 0 V
3 -1 V
4 0 V
3 0 V
4 -1 V
3 0 V
4 0 V
3 -1 V
4 0 V
4 -1 V
3 0 V
4 -1 V
3 0 V
4 0 V
3 -1 V
4 0 V
3 -1 V
4 -1 V
4 0 V
3 -1 V
4 0 V
3 -1 V
4 0 V
3 -1 V
4 -1 V
3 0 V
4 -1 V
4 -1 V
3 -1 V
4 0 V
3 -1 V
4 -1 V
3 -1 V
4 -1 V
3 -1 V
4 -1 V
3 -1 V
4 -1 V
4 -1 V
3 -1 V
4 -1 V
3 -1 V
4 -1 V
3 -2 V
4 -1 V
3 -1 V
4 -2 V
4 -1 V
3 -2 V
4 -1 V
3 -2 V
4 -2 V
3 -2 V
4 -1 V
3 -2 V
4 -2 V
4 -3 V
3 -2 V
4 -2 V
3 -3 V
4 -2 V
3 -3 V
4 -3 V
3 -3 V
4 -3 V
4 -3 V
3 -3 V
4 -4 V
3 -4 V
4 -4 V
3 -4 V
4 -4 V
3 -5 V
4 -5 V
4 -5 V
3 -6 V
4 -6 V
3 -6 V
4 -7 V
3 -7 V
4 -7 V
3 -8 V
4 -9 V
3 -9 V
4 -9 V
4 -11 V
3 -11 V
4 -12 V
3 -13 V
4 -14 V
3 -15 V
4 -16 V
3 -18 V
4 -19 V
4 -21 V
3 -23 V
4 -24 V
3 -27 V
4 -30 V
3 -33 V
4 -35 V
3 -40 V
4 -44 V
LT2
480 886 M
4 17 V
3 14 V
4 13 V
3 11 V
4 10 V
3 9 V
4 9 V
3 7 V
4 6 V
4 6 V
3 6 V
4 5 V
3 4 V
4 5 V
3 4 V
4 3 V
3 3 V
4 3 V
4 3 V
3 3 V
4 2 V
3 2 V
4 3 V
3 2 V
4 1 V
3 2 V
4 2 V
3 1 V
4 2 V
4 1 V
3 1 V
4 1 V
3 2 V
4 1 V
3 1 V
4 1 V
3 1 V
4 0 V
4 1 V
3 1 V
4 1 V
3 0 V
4 1 V
3 1 V
4 0 V
3 1 V
4 1 V
4 0 V
3 1 V
4 0 V
3 1 V
4 0 V
3 0 V
4 1 V
3 0 V
4 1 V
4 0 V
3 0 V
4 1 V
3 0 V
4 0 V
3 1 V
4 0 V
3 0 V
4 0 V
4 1 V
3 0 V
4 0 V
3 0 V
4 1 V
3 0 V
4 0 V
3 0 V
4 0 V
3 1 V
4 0 V
4 0 V
3 0 V
4 0 V
3 1 V
4 0 V
3 0 V
4 0 V
3 0 V
4 0 V
4 0 V
3 1 V
4 0 V
3 0 V
4 0 V
3 0 V
4 0 V
3 0 V
4 0 V
4 0 V
3 1 V
4 0 V
3 0 V
4 0 V
3 0 V
4 0 V
3 0 V
4 0 V
4 0 V
3 0 V
4 0 V
3 0 V
4 1 V
3 0 V
4 0 V
3 0 V
4 0 V
4 0 V
3 0 V
4 0 V
3 0 V
4 0 V
3 0 V
4 0 V
3 0 V
4 0 V
3 0 V
4 0 V
4 0 V
3 0 V
4 0 V
3 0 V
4 1 V
3 0 V
4 0 V
3 0 V
4 0 V
4 0 V
3 0 V
4 0 V
3 0 V
4 0 V
3 0 V
4 0 V
3 0 V
4 0 V
4 0 V
3 0 V
4 0 V
3 0 V
4 0 V
3 0 V
4 0 V
3 0 V
4 0 V
4 0 V
3 0 V
4 0 V
3 0 V
4 0 V
3 0 V
4 0 V
3 0 V
4 0 V
4 0 V
3 0 V
4 0 V
3 0 V
4 0 V
3 0 V
4 0 V
3 0 V
4 0 V
3 0 V
4 0 V
4 0 V
3 0 V
4 0 V
3 0 V
4 0 V
3 0 V
4 0 V
3 0 V
4 0 V
4 0 V
3 0 V
4 0 V
3 0 V
4 0 V
3 0 V
4 0 V
3 0 V
4 0 V
4 0 V
3 0 V
4 0 V
3 0 V
4 -1 V
3 0 V
4 0 V
3 0 V
4 0 V
4 0 V
3 0 V
4 0 V
3 0 V
4 0 V
3 0 V
4 0 V
3 0 V
4 0 V
3 0 V
4 0 V
4 0 V
3 0 V
4 0 V
3 0 V
4 -1 V
3 0 V
4 0 V
3 0 V
4 0 V
4 0 V
3 0 V
4 0 V
3 0 V
4 0 V
3 0 V
4 0 V
3 -1 V
4 0 V
4 0 V
3 0 V
4 0 V
3 0 V
4 0 V
3 0 V
4 0 V
3 -1 V
4 0 V
4 0 V
3 0 V
4 0 V
3 0 V
4 0 V
3 -1 V
4 0 V
3 0 V
4 0 V
4 0 V
3 -1 V
4 0 V
3 0 V
4 0 V
3 0 V
4 -1 V
3 0 V
4 0 V
3 0 V
4 -1 V
4 0 V
3 0 V
4 0 V
3 -1 V
4 0 V
3 0 V
4 -1 V
3 0 V
4 0 V
4 -1 V
3 0 V
4 -1 V
3 0 V
4 0 V
3 -1 V
4 0 V
3 -1 V
4 0 V
4 -1 V
3 -1 V
4 0 V
3 -1 V
4 -1 V
3 0 V
4 -1 V
3 -1 V
4 -1 V
4 0 V
3 -1 V
4 -1 V
3 -1 V
4 -1 V
3 -2 V
4 -1 V
3 -1 V
4 -1 V
4 -2 V
3 -1 V
4 -2 V
3 -2 V
4 -1 V
3 -2 V
4 -3 V
3 -2 V
4 -2 V
3 -3 V
4 -3 V
4 -3 V
3 -3 V
4 -3 V
3 -4 V
4 -5 V
3 -4 V
4 -5 V
3 -6 V
4 -6 V
4 -6 V
3 -7 V
4 -9 V
3 -9 V
4 -10 V
3 -11 V
4 -13 V
3 -14 V
4 -17 V
stroke
grestore
end
showpage
}
\put(1708,51){\makebox(0,0){{\fs $\theta$}}}
\put(1617,151){\makebox(0,0){{\fs $170^\circ$}}}
\put(1049,151){\makebox(0,0){{\fs $90^\circ$}}}
\put(480,151){\makebox(0,0){{\fs $10^\circ$}}}
\put(420,2541){\makebox(0,0)[r]{{\fs $10^{-1}$}}}
\put(420,1396){\makebox(0,0)[r]{{\fs $10^{-2}$}}}
\put(420,251){\makebox(0,0)[r]{{\fs $10^{-3}$}}}
\end{picture}}
\put(-10,0){
\setlength{\unitlength}{0.1bp}
\special{!
/gnudict 40 dict def
gnudict begin
/Color false def
/Solid false def
/gnulinewidth 5.000 def
/vshift -33 def
/dl {10 mul} def
/hpt 31.5 def
/vpt 31.5 def
/M {moveto} bind def
/L {lineto} bind def
/R {rmoveto} bind def
/V {rlineto} bind def
/vpt2 vpt 2 mul def
/hpt2 hpt 2 mul def
/Lshow { currentpoint stroke M
  0 vshift R show } def
/Rshow { currentpoint stroke M
  dup stringwidth pop neg vshift R show } def
/Cshow { currentpoint stroke M
  dup stringwidth pop -2 div vshift R show } def
/DL { Color {setrgbcolor Solid {pop []} if 0 setdash }
 {pop pop pop Solid {pop []} if 0 setdash} ifelse } def
/BL { stroke gnulinewidth 2 mul setlinewidth } def
/AL { stroke gnulinewidth 2 div setlinewidth } def
/PL { stroke gnulinewidth setlinewidth } def
/LTb { BL [] 0 0 0 DL } def
/LTa { AL [1 dl 2 dl] 0 setdash 0 0 0 setrgbcolor } def
/LT0 { PL [] 0 1 0 DL } def
/LT1 { PL [4 dl 2 dl] 0 0 1 DL } def
/LT2 { PL [2 dl 3 dl] 1 0 0 DL } def
/LT3 { PL [1 dl 1.5 dl] 1 0 1 DL } def
/LT4 { PL [5 dl 2 dl 1 dl 2 dl] 0 1 1 DL } def
/LT5 { PL [4 dl 3 dl 1 dl 3 dl] 1 1 0 DL } def
/LT6 { PL [2 dl 2 dl 2 dl 4 dl] 0 0 0 DL } def
/LT7 { PL [2 dl 2 dl 2 dl 2 dl 2 dl 4 dl] 1 0.3 0 DL } def
/LT8 { PL [2 dl 2 dl 2 dl 2 dl 2 dl 2 dl 2 dl 4 dl] 0.5 0.5 0.5 DL } def
/P { stroke [] 0 setdash
  currentlinewidth 2 div sub M
  0 currentlinewidth V stroke } def
/D { stroke [] 0 setdash 2 copy vpt add M
  hpt neg vpt neg V hpt vpt neg V
  hpt vpt V hpt neg vpt V closepath stroke
  P } def
/A { stroke [] 0 setdash vpt sub M 0 vpt2 V
  currentpoint stroke M
  hpt neg vpt neg R hpt2 0 V stroke
  } def
/B { stroke [] 0 setdash 2 copy exch hpt sub exch vpt add M
  0 vpt2 neg V hpt2 0 V 0 vpt2 V
  hpt2 neg 0 V closepath stroke
  P } def
/C { stroke [] 0 setdash exch hpt sub exch vpt add M
  hpt2 vpt2 neg V currentpoint stroke M
  hpt2 neg 0 R hpt2 vpt2 V stroke } def
/T { stroke [] 0 setdash 2 copy vpt 1.12 mul add M
  hpt neg vpt -1.62 mul V
  hpt 2 mul 0 V
  hpt neg vpt 1.62 mul V closepath stroke
  P  } def
/S { 2 copy A C} def
end
}
\begin{picture}(1800,2592)(0,0)
\special{"
gnudict begin
gsave
50 50 translate
0.100 0.100 scale
0 setgray
/Helvetica findfont 100 scalefont setfont
newpath
-500.000000 -500.000000 translate
LTa
LTb
480 251 M
63 0 V
1074 0 R
-63 0 V
480 633 M
63 0 V
1074 0 R
-63 0 V
480 1014 M
63 0 V
1074 0 R
-63 0 V
480 1396 M
63 0 V
1074 0 R
-63 0 V
480 1778 M
63 0 V
1074 0 R
-63 0 V
480 2159 M
63 0 V
1074 0 R
-63 0 V
480 2541 M
63 0 V
1074 0 R
-63 0 V
480 251 M
0 63 V
0 2227 R
0 -63 V
1049 251 M
0 63 V
0 2227 R
0 -63 V
1617 251 M
0 63 V
0 2227 R
0 -63 V
480 251 M
1137 0 V
0 2290 V
-1137 0 V
480 251 L
LT0
480 2429 M
4 -20 V
3 -21 V
4 -21 V
3 -22 V
4 -21 V
3 -21 V
4 -20 V
3 -20 V
4 -19 V
4 -18 V
3 -18 V
4 -16 V
3 -16 V
4 -15 V
3 -14 V
4 -13 V
3 -12 V
4 -12 V
4 -10 V
3 -10 V
4 -9 V
3 -8 V
4 -8 V
3 -7 V
4 -7 V
3 -6 V
4 -5 V
3 -5 V
4 -5 V
4 -4 V
3 -4 V
4 -3 V
3 -4 V
4 -3 V
3 -3 V
4 -2 V
3 -3 V
4 -2 V
4 -2 V
3 -2 V
4 -2 V
3 -1 V
4 -2 V
3 -2 V
4 -1 V
3 -1 V
4 -2 V
4 -1 V
3 -1 V
4 -2 V
3 -1 V
4 -1 V
3 -2 V
4 -1 V
3 -1 V
4 -1 V
4 -1 V
3 -2 V
4 -1 V
3 -1 V
4 -1 V
3 -2 V
4 -1 V
3 -1 V
4 -1 V
4 -2 V
3 -1 V
4 -1 V
3 -1 V
4 -2 V
3 -1 V
4 -1 V
3 -2 V
4 -1 V
3 -1 V
4 -2 V
4 -1 V
3 -1 V
4 -2 V
3 -1 V
4 -2 V
3 -1 V
4 -1 V
3 -2 V
4 -1 V
4 -1 V
3 -2 V
4 -1 V
3 -1 V
4 -2 V
3 -1 V
4 -1 V
3 -1 V
4 -2 V
4 -1 V
3 -1 V
4 -2 V
3 -1 V
4 -1 V
3 -1 V
4 -1 V
3 -1 V
4 -2 V
4 -1 V
3 -1 V
4 -1 V
3 -1 V
4 -1 V
3 -1 V
4 -1 V
3 -1 V
4 -1 V
4 -1 V
3 -1 V
4 0 V
3 -1 V
4 -1 V
3 -1 V
4 0 V
3 -1 V
4 -1 V
3 0 V
4 -1 V
4 -1 V
3 0 V
4 -1 V
3 0 V
4 -1 V
3 0 V
4 0 V
3 -1 V
4 0 V
4 0 V
3 0 V
4 -1 V
3 0 V
4 0 V
3 0 V
4 0 V
3 0 V
4 0 V
4 0 V
3 0 V
4 0 V
3 1 V
4 0 V
3 0 V
4 0 V
3 1 V
4 0 V
4 0 V
3 1 V
4 0 V
3 1 V
4 1 V
3 0 V
4 1 V
3 0 V
4 1 V
4 1 V
3 1 V
4 1 V
3 0 V
4 1 V
3 1 V
4 1 V
3 1 V
4 1 V
3 1 V
4 1 V
4 1 V
3 2 V
4 1 V
3 1 V
4 1 V
3 2 V
4 1 V
3 1 V
4 2 V
4 1 V
3 1 V
4 2 V
3 1 V
4 2 V
3 1 V
4 2 V
3 1 V
4 2 V
4 2 V
3 1 V
4 2 V
3 2 V
4 1 V
3 2 V
4 2 V
3 1 V
4 2 V
4 2 V
3 2 V
4 2 V
3 1 V
4 2 V
3 2 V
4 2 V
3 2 V
4 2 V
3 2 V
4 2 V
4 2 V
3 1 V
4 2 V
3 2 V
4 2 V
3 2 V
4 2 V
3 2 V
4 2 V
4 2 V
3 2 V
4 2 V
3 2 V
4 2 V
3 2 V
4 2 V
3 2 V
4 2 V
4 2 V
3 2 V
4 2 V
3 2 V
4 1 V
3 2 V
4 2 V
3 2 V
4 2 V
4 2 V
3 2 V
4 1 V
3 2 V
4 2 V
3 1 V
4 2 V
3 2 V
4 1 V
4 2 V
3 2 V
4 1 V
3 1 V
4 2 V
3 1 V
4 2 V
3 1 V
4 1 V
3 1 V
4 2 V
4 1 V
3 1 V
4 1 V
3 1 V
4 1 V
3 1 V
4 0 V
3 1 V
4 1 V
4 0 V
3 1 V
4 1 V
3 0 V
4 1 V
3 0 V
4 0 V
3 0 V
4 1 V
4 0 V
3 0 V
4 0 V
3 0 V
4 0 V
3 1 V
4 0 V
3 0 V
4 0 V
4 0 V
3 0 V
4 1 V
3 0 V
4 0 V
3 1 V
4 1 V
3 1 V
4 1 V
4 2 V
3 1 V
4 2 V
3 3 V
4 2 V
3 4 V
4 3 V
3 5 V
4 4 V
3 6 V
4 6 V
4 6 V
3 8 V
4 8 V
3 9 V
4 10 V
3 11 V
4 11 V
3 13 V
4 13 V
4 14 V
3 15 V
4 16 V
3 16 V
4 17 V
3 17 V
4 17 V
3 18 V
4 17 V
LT1
480 1726 M
4 -61 V
3 -47 V
4 -37 V
3 -28 V
4 -20 V
3 -14 V
4 -9 V
3 -4 V
4 -2 V
4 2 V
3 3 V
4 5 V
3 6 V
4 7 V
3 7 V
4 7 V
3 7 V
4 7 V
4 7 V
3 6 V
4 6 V
3 6 V
4 5 V
3 5 V
4 4 V
3 4 V
4 4 V
3 2 V
4 3 V
4 2 V
3 2 V
4 1 V
3 1 V
4 0 V
3 0 V
4 0 V
3 0 V
4 -1 V
4 -1 V
3 -1 V
4 -2 V
3 -1 V
4 -2 V
3 -2 V
4 -2 V
3 -3 V
4 -2 V
4 -3 V
3 -3 V
4 -3 V
3 -3 V
4 -3 V
3 -3 V
4 -3 V
3 -3 V
4 -4 V
4 -3 V
3 -4 V
4 -3 V
3 -4 V
4 -3 V
3 -4 V
4 -3 V
3 -4 V
4 -4 V
4 -3 V
3 -4 V
4 -4 V
3 -3 V
4 -4 V
3 -3 V
4 -4 V
3 -4 V
4 -3 V
3 -4 V
4 -3 V
4 -4 V
3 -3 V
4 -4 V
3 -3 V
4 -4 V
3 -3 V
4 -3 V
3 -4 V
4 -3 V
4 -3 V
3 -3 V
4 -3 V
3 -4 V
4 -3 V
3 -3 V
4 -3 V
3 -3 V
4 -2 V
4 -3 V
3 -3 V
4 -3 V
3 -2 V
4 -3 V
3 -3 V
4 -2 V
3 -2 V
4 -3 V
4 -2 V
3 -2 V
4 -3 V
3 -2 V
4 -2 V
3 -2 V
4 -2 V
3 -2 V
4 -2 V
4 -2 V
3 -1 V
4 -2 V
3 -2 V
4 -1 V
3 -2 V
4 -1 V
3 -2 V
4 -1 V
3 -1 V
4 -1 V
4 -1 V
3 -2 V
4 0 V
3 -1 V
4 -1 V
3 -1 V
4 -1 V
3 0 V
4 -1 V
4 0 V
3 -1 V
4 0 V
3 -1 V
4 0 V
3 0 V
4 0 V
3 0 V
4 0 V
4 0 V
3 0 V
4 1 V
3 0 V
4 0 V
3 1 V
4 0 V
3 1 V
4 1 V
4 1 V
3 0 V
4 1 V
3 1 V
4 1 V
3 2 V
4 1 V
3 1 V
4 1 V
4 2 V
3 1 V
4 2 V
3 2 V
4 1 V
3 2 V
4 2 V
3 2 V
4 2 V
3 2 V
4 2 V
4 3 V
3 2 V
4 2 V
3 3 V
4 2 V
3 3 V
4 3 V
3 2 V
4 3 V
4 3 V
3 3 V
4 3 V
3 3 V
4 3 V
3 4 V
4 3 V
3 3 V
4 4 V
4 4 V
3 3 V
4 4 V
3 4 V
4 3 V
3 4 V
4 4 V
3 4 V
4 4 V
4 5 V
3 4 V
4 4 V
3 5 V
4 4 V
3 5 V
4 4 V
3 5 V
4 5 V
3 4 V
4 5 V
4 5 V
3 5 V
4 5 V
3 5 V
4 5 V
3 6 V
4 5 V
3 5 V
4 6 V
4 5 V
3 6 V
4 6 V
3 5 V
4 6 V
3 6 V
4 6 V
3 6 V
4 5 V
4 7 V
3 6 V
4 6 V
3 6 V
4 6 V
3 7 V
4 6 V
3 6 V
4 7 V
4 6 V
3 7 V
4 6 V
3 7 V
4 7 V
3 6 V
4 7 V
3 7 V
4 7 V
4 7 V
3 6 V
4 7 V
3 7 V
4 7 V
3 7 V
4 7 V
3 7 V
4 7 V
3 7 V
4 7 V
4 7 V
3 7 V
4 7 V
3 7 V
4 7 V
3 6 V
4 7 V
3 7 V
4 7 V
4 6 V
3 7 V
4 6 V
3 6 V
4 7 V
3 6 V
4 6 V
3 5 V
4 6 V
4 5 V
3 6 V
4 5 V
3 4 V
4 5 V
3 4 V
4 4 V
3 4 V
4 3 V
4 3 V
3 2 V
4 2 V
3 2 V
4 1 V
3 0 V
4 0 V
3 -1 V
4 -2 V
4 -2 V
3 -3 V
4 -4 V
3 -4 V
4 -6 V
3 -6 V
4 -8 V
3 -8 V
4 -10 V
3 -10 V
4 -12 V
4 -12 V
3 -14 V
4 -14 V
3 -16 V
4 -16 V
3 -17 V
4 -17 V
3 -17 V
4 -17 V
4 -16 V
3 -15 V
4 -12 V
3 -9 V
4 -6 V
3 0 V
4 7 V
3 15 V
4 25 V
LT2
486 251 M
1 30 V
4 82 V
3 73 V
4 65 V
3 57 V
4 51 V
3 44 V
4 39 V
4 33 V
3 29 V
4 25 V
3 21 V
4 18 V
3 15 V
4 12 V
3 11 V
4 8 V
4 6 V
3 5 V
4 4 V
3 2 V
4 1 V
3 0 V
4 0 V
3 -1 V
4 -2 V
3 -3 V
4 -3 V
4 -3 V
3 -4 V
4 -5 V
3 -4 V
4 -5 V
3 -5 V
4 -5 V
3 -5 V
4 -6 V
4 -6 V
3 -5 V
4 -6 V
3 -6 V
4 -6 V
3 -6 V
4 -6 V
3 -6 V
4 -6 V
4 -6 V
3 -6 V
4 -6 V
3 -6 V
4 -6 V
3 -6 V
4 -6 V
3 -6 V
4 -5 V
4 -6 V
3 -6 V
4 -6 V
3 -5 V
4 -6 V
3 -5 V
4 -5 V
3 -6 V
4 -5 V
4 -5 V
3 -5 V
4 -5 V
3 -5 V
4 -5 V
3 -5 V
4 -5 V
3 -5 V
4 -4 V
3 -5 V
4 -4 V
4 -5 V
3 -4 V
4 -4 V
3 -4 V
4 -4 V
3 -4 V
4 -4 V
3 -4 V
4 -3 V
4 -4 V
3 -4 V
4 -3 V
3 -3 V
4 -4 V
3 -3 V
4 -3 V
3 -3 V
4 -3 V
4 -3 V
3 -3 V
4 -2 V
3 -3 V
4 -2 V
3 -3 V
4 -2 V
3 -2 V
4 -2 V
4 -2 V
3 -2 V
4 -2 V
3 -2 V
4 -1 V
3 -2 V
4 -1 V
3 -2 V
4 -1 V
4 -1 V
3 -1 V
4 -1 V
3 -1 V
4 -1 V
3 0 V
4 -1 V
3 0 V
4 -1 V
3 0 V
4 0 V
4 0 V
3 0 V
4 0 V
3 0 V
4 1 V
3 0 V
4 1 V
3 0 V
4 1 V
4 1 V
3 1 V
4 1 V
3 2 V
4 1 V
3 1 V
4 2 V
3 2 V
4 1 V
4 2 V
3 2 V
4 2 V
3 3 V
4 2 V
3 2 V
4 3 V
3 3 V
4 2 V
4 3 V
3 3 V
4 3 V
3 4 V
4 3 V
3 3 V
4 4 V
3 4 V
4 3 V
4 4 V
3 4 V
4 4 V
3 5 V
4 4 V
3 4 V
4 5 V
3 5 V
4 4 V
3 5 V
4 5 V
4 5 V
3 6 V
4 5 V
3 6 V
4 5 V
3 6 V
4 6 V
3 6 V
4 6 V
4 6 V
3 6 V
4 6 V
3 7 V
4 6 V
3 7 V
4 7 V
3 7 V
4 7 V
4 7 V
3 7 V
4 7 V
3 8 V
4 8 V
3 7 V
4 8 V
3 8 V
4 8 V
4 8 V
3 9 V
4 8 V
3 8 V
4 9 V
3 9 V
4 9 V
3 9 V
4 9 V
3 9 V
4 9 V
4 10 V
3 10 V
4 9 V
3 10 V
4 10 V
3 10 V
4 11 V
3 10 V
4 11 V
4 10 V
3 11 V
4 11 V
3 11 V
4 11 V
3 12 V
4 11 V
3 12 V
4 11 V
4 12 V
3 12 V
4 13 V
3 12 V
4 13 V
3 12 V
4 13 V
3 13 V
4 13 V
4 14 V
3 13 V
4 14 V
3 14 V
4 14 V
3 14 V
4 14 V
3 15 V
4 14 V
4 15 V
3 15 V
4 15 V
3 16 V
4 15 V
3 16 V
4 16 V
3 16 V
4 17 V
3 16 V
4 17 V
4 17 V
3 17 V
4 17 V
3 18 V
4 18 V
3 18 V
4 18 V
3 18 V
4 19 V
4 19 V
3 19 V
4 19 V
3 19 V
4 20 V
3 19 V
4 20 V
3 21 V
4 20 V
4 21 V
3 20 V
4 21 V
3 21 V
4 21 V
3 22 V
4 21 V
3 22 V
4 21 V
4 22 V
3 22 V
4 22 V
3 21 V
4 22 V
3 22 V
4 21 V
3 22 V
4 21 V
4 20 V
3 21 V
4 20 V
3 19 V
4 19 V
3 18 V
4 16 V
3 16 V
4 15 V
3 13 V
4 11 V
4 9 V
3 7 V
4 4 V
3 1 V
4 -2 V
3 -7 V
4 -12 V
3 -16 V
4 -23 V
4 -30 V
3 -39 V
4 -47 V
3 -57 V
4 -70 V
3 -82 V
4 -96 V
3 -112 V
4 -129 V
stroke
grestore
end
showpage
}
\put(231,2705){\makebox(0,0)[l]{{\fs$\delta$/\%}}}
\put(1708,51){\makebox(0,0){{\fs $\theta$}}}
\put(1617,151){\makebox(0,0){{\fs $170^\circ$}}}
\put(1049,151){\makebox(0,0){{\fs $90^\circ$}}}
\put(480,151){\makebox(0,0){{\fs $10^\circ$}}}
\put(420,2541){\makebox(0,0)[r]{{\fs $10$}}}
\put(420,2159){\makebox(0,0)[r]{{\fs $0$}}}
\put(420,1778){\makebox(0,0)[r]{{\fs $-10$}}}
\put(420,1396){\makebox(0,0)[r]{{\fs $-20$}}}
\put(420,1014){\makebox(0,0)[r]{{\fs $-30$}}}
\put(420,633){\makebox(0,0)[r]{{\fs $-40$}}}
\put(420,251){\makebox(0,0)[r]{{\fs $-50$}}}
\end{picture}
\hspace*{-1.7cm}
\setlength{\unitlength}{0.1bp}
\special{!
/gnudict 40 dict def
gnudict begin
/Color false def
/Solid false def
/gnulinewidth 5.000 def
/vshift -33 def
/dl {10 mul} def
/hpt 31.5 def
/vpt 31.5 def
/M {moveto} bind def
/L {lineto} bind def
/R {rmoveto} bind def
/V {rlineto} bind def
/vpt2 vpt 2 mul def
/hpt2 hpt 2 mul def
/Lshow { currentpoint stroke M
  0 vshift R show } def
/Rshow { currentpoint stroke M
  dup stringwidth pop neg vshift R show } def
/Cshow { currentpoint stroke M
  dup stringwidth pop -2 div vshift R show } def
/DL { Color {setrgbcolor Solid {pop []} if 0 setdash }
 {pop pop pop Solid {pop []} if 0 setdash} ifelse } def
/BL { stroke gnulinewidth 2 mul setlinewidth } def
/AL { stroke gnulinewidth 2 div setlinewidth } def
/PL { stroke gnulinewidth setlinewidth } def
/LTb { BL [] 0 0 0 DL } def
/LTa { AL [1 dl 2 dl] 0 setdash 0 0 0 setrgbcolor } def
/LT0 { PL [] 0 1 0 DL } def
/LT1 { PL [4 dl 2 dl] 0 0 1 DL } def
/LT2 { PL [2 dl 3 dl] 1 0 0 DL } def
/LT3 { PL [1 dl 1.5 dl] 1 0 1 DL } def
/LT4 { PL [5 dl 2 dl 1 dl 2 dl] 0 1 1 DL } def
/LT5 { PL [4 dl 3 dl 1 dl 3 dl] 1 1 0 DL } def
/LT6 { PL [2 dl 2 dl 2 dl 4 dl] 0 0 0 DL } def
/LT7 { PL [2 dl 2 dl 2 dl 2 dl 2 dl 4 dl] 1 0.3 0 DL } def
/LT8 { PL [2 dl 2 dl 2 dl 2 dl 2 dl 2 dl 2 dl 4 dl] 0.5 0.5 0.5 DL } def
/P { stroke [] 0 setdash
  currentlinewidth 2 div sub M
  0 currentlinewidth V stroke } def
/D { stroke [] 0 setdash 2 copy vpt add M
  hpt neg vpt neg V hpt vpt neg V
  hpt vpt V hpt neg vpt V closepath stroke
  P } def
/A { stroke [] 0 setdash vpt sub M 0 vpt2 V
  currentpoint stroke M
  hpt neg vpt neg R hpt2 0 V stroke
  } def
/B { stroke [] 0 setdash 2 copy exch hpt sub exch vpt add M
  0 vpt2 neg V hpt2 0 V 0 vpt2 V
  hpt2 neg 0 V closepath stroke
  P } def
/C { stroke [] 0 setdash exch hpt sub exch vpt add M
  hpt2 vpt2 neg V currentpoint stroke M
  hpt2 neg 0 R hpt2 vpt2 V stroke } def
/T { stroke [] 0 setdash 2 copy vpt 1.12 mul add M
  hpt neg vpt -1.62 mul V
  hpt 2 mul 0 V
  hpt neg vpt 1.62 mul V closepath stroke
  P  } def
/S { 2 copy A C} def
end
}
\begin{picture}(1800,2592)(0,0)
\special{"
gnudict begin
gsave
50 50 translate
0.100 0.100 scale
0 setgray
/Helvetica findfont 100 scalefont setfont
newpath
-500.000000 -500.000000 translate
LTa
LTb
480 251 M
63 0 V
1074 0 R
-63 0 V
480 578 M
63 0 V
1074 0 R
-63 0 V
480 905 M
63 0 V
1074 0 R
-63 0 V
480 1232 M
63 0 V
1074 0 R
-63 0 V
480 1560 M
63 0 V
1074 0 R
-63 0 V
480 1887 M
63 0 V
1074 0 R
-63 0 V
480 2214 M
63 0 V
1074 0 R
-63 0 V
480 2541 M
63 0 V
1074 0 R
-63 0 V
480 251 M
0 63 V
0 2227 R
0 -63 V
1049 251 M
0 63 V
0 2227 R
0 -63 V
1617 251 M
0 63 V
0 2227 R
0 -63 V
480 251 M
1137 0 V
0 2290 V
-1137 0 V
480 251 L
LT0
480 2275 M
4 -3 V
3 -3 V
4 -4 V
3 -3 V
4 -4 V
3 -4 V
4 -4 V
3 -4 V
4 -4 V
4 -4 V
3 -4 V
4 -4 V
3 -4 V
4 -4 V
3 -4 V
4 -4 V
3 -4 V
4 -4 V
4 -5 V
3 -4 V
4 -4 V
3 -4 V
4 -4 V
3 -4 V
4 -4 V
3 -4 V
4 -4 V
3 -4 V
4 -4 V
4 -4 V
3 -3 V
4 -4 V
3 -4 V
4 -4 V
3 -4 V
4 -4 V
3 -4 V
4 -4 V
4 -4 V
3 -4 V
4 -4 V
3 -3 V
4 -4 V
3 -4 V
4 -4 V
3 -4 V
4 -4 V
4 -3 V
3 -4 V
4 -4 V
3 -4 V
4 -3 V
3 -4 V
4 -4 V
3 -4 V
4 -3 V
4 -4 V
3 -4 V
4 -3 V
3 -4 V
4 -4 V
3 -3 V
4 -4 V
3 -3 V
4 -4 V
4 -3 V
3 -4 V
4 -3 V
3 -4 V
4 -3 V
3 -4 V
4 -3 V
3 -3 V
4 -4 V
3 -3 V
4 -3 V
4 -4 V
3 -3 V
4 -3 V
3 -3 V
4 -3 V
3 -3 V
4 -4 V
3 -3 V
4 -3 V
4 -3 V
3 -3 V
4 -3 V
3 -3 V
4 -3 V
3 -2 V
4 -3 V
3 -3 V
4 -3 V
4 -3 V
3 -2 V
4 -3 V
3 -3 V
4 -2 V
3 -3 V
4 -3 V
3 -2 V
4 -3 V
4 -2 V
3 -3 V
4 -2 V
3 -3 V
4 -2 V
3 -2 V
4 -3 V
3 -2 V
4 -2 V
4 -2 V
3 -3 V
4 -2 V
3 -2 V
4 -2 V
3 -2 V
4 -2 V
3 -2 V
4 -2 V
3 -2 V
4 -2 V
4 -1 V
3 -2 V
4 -2 V
3 -2 V
4 -1 V
3 -2 V
4 -2 V
3 -1 V
4 -2 V
4 -1 V
3 -2 V
4 -1 V
3 -1 V
4 -2 V
3 -1 V
4 -1 V
3 -1 V
4 -1 V
4 -1 V
3 -1 V
4 -1 V
3 -1 V
4 -1 V
3 0 V
4 -1 V
3 -1 V
4 0 V
4 -1 V
3 0 V
4 0 V
3 -1 V
4 0 V
3 0 V
4 0 V
3 0 V
4 0 V
4 0 V
3 1 V
4 0 V
3 0 V
4 1 V
3 0 V
4 1 V
3 1 V
4 0 V
3 1 V
4 1 V
4 1 V
3 1 V
4 1 V
3 1 V
4 1 V
3 1 V
4 2 V
3 1 V
4 1 V
4 2 V
3 1 V
4 2 V
3 1 V
4 2 V
3 1 V
4 2 V
3 2 V
4 1 V
4 2 V
3 2 V
4 2 V
3 1 V
4 2 V
3 2 V
4 2 V
3 2 V
4 2 V
4 2 V
3 2 V
4 3 V
3 2 V
4 2 V
3 2 V
4 2 V
3 3 V
4 2 V
3 2 V
4 3 V
4 2 V
3 2 V
4 3 V
3 2 V
4 3 V
3 2 V
4 3 V
3 2 V
4 3 V
4 3 V
3 2 V
4 3 V
3 3 V
4 3 V
3 2 V
4 3 V
3 3 V
4 3 V
4 3 V
3 3 V
4 3 V
3 3 V
4 3 V
3 3 V
4 3 V
3 3 V
4 3 V
4 3 V
3 3 V
4 3 V
3 3 V
4 4 V
3 3 V
4 3 V
3 3 V
4 4 V
4 3 V
3 3 V
4 4 V
3 3 V
4 4 V
3 3 V
4 3 V
3 4 V
4 3 V
3 4 V
4 3 V
4 4 V
3 3 V
4 4 V
3 3 V
4 4 V
3 4 V
4 3 V
3 4 V
4 3 V
4 4 V
3 4 V
4 3 V
3 4 V
4 4 V
3 3 V
4 4 V
3 3 V
4 4 V
4 4 V
3 3 V
4 4 V
3 4 V
4 3 V
3 4 V
4 4 V
3 3 V
4 4 V
4 3 V
3 4 V
4 3 V
3 4 V
4 3 V
3 4 V
4 3 V
3 4 V
4 3 V
4 4 V
3 3 V
4 4 V
3 3 V
4 4 V
3 3 V
4 3 V
3 4 V
4 3 V
3 3 V
4 4 V
4 3 V
3 3 V
4 3 V
3 4 V
4 3 V
3 3 V
4 3 V
3 4 V
4 3 V
4 3 V
3 3 V
4 3 V
3 3 V
4 3 V
3 3 V
4 3 V
3 2 V
4 3 V
LT1
480 2177 M
4 -11 V
3 -12 V
4 -12 V
3 -11 V
4 -12 V
3 -11 V
4 -12 V
3 -11 V
4 -12 V
4 -11 V
3 -12 V
4 -11 V
3 -11 V
4 -12 V
3 -11 V
4 -11 V
3 -11 V
4 -11 V
4 -11 V
3 -11 V
4 -11 V
3 -11 V
4 -11 V
3 -10 V
4 -11 V
3 -10 V
4 -11 V
3 -10 V
4 -10 V
4 -10 V
3 -10 V
4 -10 V
3 -10 V
4 -10 V
3 -9 V
4 -10 V
3 -9 V
4 -10 V
4 -9 V
3 -9 V
4 -9 V
3 -9 V
4 -9 V
3 -9 V
4 -9 V
3 -8 V
4 -9 V
4 -8 V
3 -8 V
4 -9 V
3 -8 V
4 -8 V
3 -8 V
4 -8 V
3 -7 V
4 -8 V
4 -8 V
3 -7 V
4 -8 V
3 -7 V
4 -7 V
3 -7 V
4 -7 V
3 -7 V
4 -7 V
4 -7 V
3 -7 V
4 -7 V
3 -6 V
4 -7 V
3 -6 V
4 -6 V
3 -7 V
4 -6 V
3 -6 V
4 -6 V
4 -6 V
3 -6 V
4 -6 V
3 -5 V
4 -6 V
3 -6 V
4 -5 V
3 -5 V
4 -6 V
4 -5 V
3 -5 V
4 -5 V
3 -5 V
4 -5 V
3 -5 V
4 -5 V
3 -5 V
4 -4 V
4 -5 V
3 -5 V
4 -4 V
3 -4 V
4 -5 V
3 -4 V
4 -4 V
3 -4 V
4 -4 V
4 -4 V
3 -4 V
4 -4 V
3 -4 V
4 -3 V
3 -4 V
4 -3 V
3 -4 V
4 -3 V
4 -4 V
3 -3 V
4 -3 V
3 -3 V
4 -3 V
3 -3 V
4 -3 V
3 -3 V
4 -3 V
3 -2 V
4 -3 V
4 -2 V
3 -3 V
4 -2 V
3 -3 V
4 -2 V
3 -2 V
4 -2 V
3 -2 V
4 -2 V
4 -2 V
3 -2 V
4 -2 V
3 -2 V
4 -1 V
3 -2 V
4 -1 V
3 -2 V
4 -1 V
4 -1 V
3 -2 V
4 -1 V
3 -1 V
4 -1 V
3 -1 V
4 -1 V
3 0 V
4 -1 V
4 -1 V
3 0 V
4 -1 V
3 0 V
4 0 V
3 0 V
4 -1 V
3 0 V
4 0 V
4 0 V
3 1 V
4 0 V
3 0 V
4 1 V
3 0 V
4 1 V
3 0 V
4 1 V
3 1 V
4 0 V
4 1 V
3 1 V
4 1 V
3 1 V
4 2 V
3 1 V
4 1 V
3 2 V
4 1 V
4 2 V
3 1 V
4 2 V
3 2 V
4 2 V
3 2 V
4 1 V
3 3 V
4 2 V
4 2 V
3 2 V
4 2 V
3 3 V
4 2 V
3 3 V
4 2 V
3 3 V
4 3 V
4 2 V
3 3 V
4 3 V
3 3 V
4 3 V
3 3 V
4 3 V
3 4 V
4 3 V
3 3 V
4 4 V
4 3 V
3 4 V
4 4 V
3 3 V
4 4 V
3 4 V
4 4 V
3 4 V
4 4 V
4 4 V
3 4 V
4 5 V
3 4 V
4 5 V
3 4 V
4 5 V
3 4 V
4 5 V
4 5 V
3 5 V
4 5 V
3 5 V
4 5 V
3 5 V
4 5 V
3 6 V
4 5 V
4 5 V
3 6 V
4 6 V
3 5 V
4 6 V
3 6 V
4 6 V
3 6 V
4 6 V
4 6 V
3 6 V
4 7 V
3 6 V
4 7 V
3 6 V
4 7 V
3 6 V
4 7 V
3 7 V
4 7 V
4 7 V
3 7 V
4 7 V
3 8 V
4 7 V
3 8 V
4 7 V
3 8 V
4 8 V
4 7 V
3 8 V
4 8 V
3 8 V
4 9 V
3 8 V
4 8 V
3 9 V
4 8 V
4 9 V
3 8 V
4 9 V
3 9 V
4 9 V
3 9 V
4 9 V
3 10 V
4 9 V
4 9 V
3 10 V
4 10 V
3 9 V
4 10 V
3 10 V
4 10 V
3 10 V
4 10 V
4 10 V
3 10 V
4 11 V
3 10 V
4 11 V
3 10 V
4 11 V
3 11 V
4 10 V
3 11 V
4 11 V
4 11 V
3 10 V
4 11 V
3 11 V
4 11 V
3 11 V
4 11 V
3 10 V
4 11 V
4 11 V
3 11 V
4 10 V
3 11 V
4 10 V
3 11 V
4 10 V
3 10 V
4 10 V
LT2
480 1872 M
4 -28 V
3 -28 V
4 -27 V
3 -26 V
4 -26 V
3 -26 V
4 -25 V
3 -25 V
4 -24 V
4 -24 V
3 -23 V
4 -23 V
3 -22 V
4 -22 V
3 -21 V
4 -21 V
3 -21 V
4 -20 V
4 -20 V
3 -19 V
4 -19 V
3 -19 V
4 -18 V
3 -18 V
4 -18 V
3 -17 V
4 -17 V
3 -17 V
4 -17 V
4 -16 V
3 -16 V
4 -15 V
3 -16 V
4 -15 V
3 -15 V
4 -14 V
3 -15 V
4 -14 V
4 -14 V
3 -14 V
4 -13 V
3 -14 V
4 -13 V
3 -13 V
4 -12 V
3 -13 V
4 -12 V
4 -13 V
3 -12 V
4 -11 V
3 -12 V
4 -12 V
3 -11 V
4 -11 V
3 -11 V
4 -11 V
4 -11 V
3 -10 V
4 -11 V
3 -10 V
4 -10 V
3 -10 V
4 -10 V
3 -9 V
4 -10 V
4 -9 V
3 -10 V
4 -9 V
3 -9 V
4 -9 V
3 -9 V
4 -8 V
3 -9 V
4 -8 V
3 -9 V
4 -8 V
4 -8 V
3 -8 V
4 -8 V
3 -7 V
4 -8 V
3 -7 V
4 -8 V
3 -7 V
4 -7 V
4 -7 V
3 -7 V
4 -7 V
3 -7 V
4 -6 V
3 -7 V
4 -6 V
3 -6 V
4 -6 V
4 -7 V
3 -5 V
4 -6 V
3 -6 V
4 -6 V
3 -5 V
4 -6 V
3 -5 V
4 -5 V
4 -6 V
3 -5 V
4 -5 V
3 -4 V
4 -5 V
3 -5 V
4 -4 V
3 -5 V
4 -4 V
4 -4 V
3 -5 V
4 -4 V
3 -4 V
4 -4 V
3 -3 V
4 -4 V
3 -4 V
4 -3 V
3 -3 V
4 -4 V
4 -3 V
3 -3 V
4 -3 V
3 -3 V
4 -3 V
3 -3 V
4 -2 V
3 -3 V
4 -2 V
4 -3 V
3 -2 V
4 -2 V
3 -2 V
4 -2 V
3 -2 V
4 -2 V
3 -1 V
4 -2 V
4 -2 V
3 -1 V
4 -1 V
3 -2 V
4 -1 V
3 -1 V
4 -1 V
3 -1 V
4 -1 V
4 0 V
3 -1 V
4 -1 V
3 0 V
4 -1 V
3 0 V
4 0 V
3 0 V
4 0 V
4 0 V
3 0 V
4 0 V
3 0 V
4 1 V
3 0 V
4 0 V
3 1 V
4 1 V
3 1 V
4 0 V
4 1 V
3 1 V
4 1 V
3 2 V
4 1 V
3 1 V
4 2 V
3 1 V
4 2 V
4 2 V
3 2 V
4 1 V
3 2 V
4 2 V
3 3 V
4 2 V
3 2 V
4 3 V
4 2 V
3 3 V
4 3 V
3 2 V
4 3 V
3 3 V
4 3 V
3 4 V
4 3 V
4 3 V
3 4 V
4 3 V
3 4 V
4 4 V
3 4 V
4 4 V
3 4 V
4 4 V
3 4 V
4 5 V
4 4 V
3 5 V
4 4 V
3 5 V
4 5 V
3 5 V
4 5 V
3 5 V
4 5 V
4 5 V
3 6 V
4 5 V
3 6 V
4 6 V
3 6 V
4 6 V
3 6 V
4 6 V
4 6 V
3 7 V
4 6 V
3 7 V
4 6 V
3 7 V
4 7 V
3 7 V
4 7 V
4 7 V
3 8 V
4 7 V
3 8 V
4 8 V
3 7 V
4 8 V
3 8 V
4 8 V
4 9 V
3 8 V
4 9 V
3 8 V
4 9 V
3 9 V
4 9 V
3 9 V
4 10 V
3 9 V
4 9 V
4 10 V
3 10 V
4 10 V
3 10 V
4 10 V
3 11 V
4 10 V
3 11 V
4 11 V
4 11 V
3 11 V
4 11 V
3 12 V
4 11 V
3 12 V
4 12 V
3 12 V
4 13 V
4 12 V
3 13 V
4 13 V
3 13 V
4 13 V
3 14 V
4 13 V
3 14 V
4 15 V
4 14 V
3 15 V
4 14 V
3 15 V
4 16 V
3 15 V
4 16 V
3 16 V
4 17 V
4 16 V
3 17 V
4 18 V
3 17 V
4 18 V
3 18 V
4 19 V
3 19 V
4 19 V
3 20 V
4 20 V
4 20 V
3 21 V
4 21 V
3 21 V
4 23 V
3 22 V
4 23 V
3 23 V
4 24 V
4 24 V
3 25 V
4 25 V
3 26 V
4 26 V
3 26 V
4 27 V
3 28 V
4 27 V
stroke
grestore
end
showpage
}
\put(1708,51){\makebox(0,0){{\fs $\theta$}}}
\put(1617,151){\makebox(0,0){{\fs $170^\circ$}}}
\put(1049,151){\makebox(0,0){{\fs $90^\circ$}}}
\put(480,151){\makebox(0,0){{\fs $10^\circ$}}}
\put(420,2541){\makebox(0,0)[r]{{\fs $10$}}}
\put(420,2214){\makebox(0,0)[r]{{\fs $0$}}}
\put(420,1887){\makebox(0,0)[r]{{\fs $-10$}}}
\put(420,1560){\makebox(0,0)[r]{{\fs $-20$}}}
\put(420,1232){\makebox(0,0)[r]{{\fs $-30$}}}
\put(420,905){\makebox(0,0)[r]{{\fs $-40$}}}
\put(420,578){\makebox(0,0)[r]{{\fs $-50$}}}
\put(420,251){\makebox(0,0)[r]{{\fs $-60$}}}
\end{picture}
\hspace*{-1.7cm}
\setlength{\unitlength}{0.1bp}
\special{!
/gnudict 40 dict def
gnudict begin
/Color false def
/Solid false def
/gnulinewidth 5.000 def
/vshift -33 def
/dl {10 mul} def
/hpt 31.5 def
/vpt 31.5 def
/M {moveto} bind def
/L {lineto} bind def
/R {rmoveto} bind def
/V {rlineto} bind def
/vpt2 vpt 2 mul def
/hpt2 hpt 2 mul def
/Lshow { currentpoint stroke M
  0 vshift R show } def
/Rshow { currentpoint stroke M
  dup stringwidth pop neg vshift R show } def
/Cshow { currentpoint stroke M
  dup stringwidth pop -2 div vshift R show } def
/DL { Color {setrgbcolor Solid {pop []} if 0 setdash }
 {pop pop pop Solid {pop []} if 0 setdash} ifelse } def
/BL { stroke gnulinewidth 2 mul setlinewidth } def
/AL { stroke gnulinewidth 2 div setlinewidth } def
/PL { stroke gnulinewidth setlinewidth } def
/LTb { BL [] 0 0 0 DL } def
/LTa { AL [1 dl 2 dl] 0 setdash 0 0 0 setrgbcolor } def
/LT0 { PL [] 0 1 0 DL } def
/LT1 { PL [4 dl 2 dl] 0 0 1 DL } def
/LT2 { PL [2 dl 3 dl] 1 0 0 DL } def
/LT3 { PL [1 dl 1.5 dl] 1 0 1 DL } def
/LT4 { PL [5 dl 2 dl 1 dl 2 dl] 0 1 1 DL } def
/LT5 { PL [4 dl 3 dl 1 dl 3 dl] 1 1 0 DL } def
/LT6 { PL [2 dl 2 dl 2 dl 4 dl] 0 0 0 DL } def
/LT7 { PL [2 dl 2 dl 2 dl 2 dl 2 dl 4 dl] 1 0.3 0 DL } def
/LT8 { PL [2 dl 2 dl 2 dl 2 dl 2 dl 2 dl 2 dl 4 dl] 0.5 0.5 0.5 DL } def
/P { stroke [] 0 setdash
  currentlinewidth 2 div sub M
  0 currentlinewidth V stroke } def
/D { stroke [] 0 setdash 2 copy vpt add M
  hpt neg vpt neg V hpt vpt neg V
  hpt vpt V hpt neg vpt V closepath stroke
  P } def
/A { stroke [] 0 setdash vpt sub M 0 vpt2 V
  currentpoint stroke M
  hpt neg vpt neg R hpt2 0 V stroke
  } def
/B { stroke [] 0 setdash 2 copy exch hpt sub exch vpt add M
  0 vpt2 neg V hpt2 0 V 0 vpt2 V
  hpt2 neg 0 V closepath stroke
  P } def
/C { stroke [] 0 setdash exch hpt sub exch vpt add M
  hpt2 vpt2 neg V currentpoint stroke M
  hpt2 neg 0 R hpt2 vpt2 V stroke } def
/T { stroke [] 0 setdash 2 copy vpt 1.12 mul add M
  hpt neg vpt -1.62 mul V
  hpt 2 mul 0 V
  hpt neg vpt 1.62 mul V closepath stroke
  P  } def
/S { 2 copy A C} def
end
}
\begin{picture}(1800,2592)(0,0)
\special{"
gnudict begin
gsave
50 50 translate
0.100 0.100 scale
0 setgray
/Helvetica findfont 100 scalefont setfont
newpath
-500.000000 -500.000000 translate
LTa
LTb
480 251 M
63 0 V
1074 0 R
-63 0 V
480 537 M
63 0 V
1074 0 R
-63 0 V
480 823 M
63 0 V
1074 0 R
-63 0 V
480 1110 M
63 0 V
1074 0 R
-63 0 V
480 1396 M
63 0 V
1074 0 R
-63 0 V
480 1682 M
63 0 V
1074 0 R
-63 0 V
480 1969 M
63 0 V
1074 0 R
-63 0 V
480 2255 M
63 0 V
1074 0 R
-63 0 V
480 2541 M
63 0 V
1074 0 R
-63 0 V
480 251 M
0 63 V
0 2227 R
0 -63 V
1049 251 M
0 63 V
0 2227 R
0 -63 V
1617 251 M
0 63 V
0 2227 R
0 -63 V
480 251 M
1137 0 V
0 2290 V
-1137 0 V
480 251 L
LT0
480 2122 M
4 7 V
3 6 V
4 5 V
3 3 V
4 2 V
3 2 V
4 1 V
3 0 V
4 0 V
4 -1 V
3 -1 V
4 -2 V
3 -2 V
4 -2 V
3 -2 V
4 -2 V
3 -3 V
4 -2 V
4 -3 V
3 -2 V
4 -3 V
3 -3 V
4 -3 V
3 -3 V
4 -2 V
3 -3 V
4 -3 V
3 -3 V
4 -2 V
4 -3 V
3 -3 V
4 -3 V
3 -3 V
4 -2 V
3 -3 V
4 -3 V
3 -2 V
4 -3 V
4 -3 V
3 -2 V
4 -3 V
3 -3 V
4 -2 V
3 -3 V
4 -3 V
3 -2 V
4 -3 V
4 -3 V
3 -2 V
4 -3 V
3 -2 V
4 -3 V
3 -3 V
4 -2 V
3 -3 V
4 -2 V
4 -3 V
3 -2 V
4 -3 V
3 -2 V
4 -3 V
3 -2 V
4 -3 V
3 -2 V
4 -3 V
4 -2 V
3 -3 V
4 -2 V
3 -3 V
4 -2 V
3 -3 V
4 -2 V
3 -2 V
4 -3 V
3 -2 V
4 -2 V
4 -3 V
3 -2 V
4 -2 V
3 -3 V
4 -2 V
3 -2 V
4 -2 V
3 -3 V
4 -2 V
4 -2 V
3 -2 V
4 -2 V
3 -2 V
4 -2 V
3 -2 V
4 -2 V
3 -2 V
4 -2 V
4 -2 V
3 -2 V
4 -2 V
3 -2 V
4 -2 V
3 -2 V
4 -2 V
3 -2 V
4 -1 V
4 -2 V
3 -2 V
4 -1 V
3 -2 V
4 -2 V
3 -1 V
4 -2 V
3 -2 V
4 -1 V
4 -2 V
3 -1 V
4 -2 V
3 -1 V
4 -1 V
3 -2 V
4 -1 V
3 -1 V
4 -2 V
3 -1 V
4 -1 V
4 -1 V
3 -1 V
4 -2 V
3 -1 V
4 -1 V
3 -1 V
4 -1 V
3 -1 V
4 -1 V
4 -1 V
3 0 V
4 -1 V
3 -1 V
4 -1 V
3 -1 V
4 0 V
3 -1 V
4 -1 V
4 0 V
3 -1 V
4 0 V
3 -1 V
4 0 V
3 -1 V
4 0 V
3 -1 V
4 0 V
4 0 V
3 -1 V
4 0 V
3 0 V
4 0 V
3 0 V
4 0 V
3 0 V
4 -1 V
4 0 V
3 0 V
4 1 V
3 0 V
4 0 V
3 0 V
4 0 V
3 0 V
4 1 V
3 0 V
4 0 V
4 1 V
3 0 V
4 0 V
3 1 V
4 0 V
3 1 V
4 0 V
3 1 V
4 1 V
4 0 V
3 1 V
4 1 V
3 0 V
4 1 V
3 1 V
4 1 V
3 1 V
4 1 V
4 1 V
3 1 V
4 1 V
3 1 V
4 1 V
3 1 V
4 1 V
3 1 V
4 1 V
4 2 V
3 1 V
4 1 V
3 1 V
4 2 V
3 1 V
4 1 V
3 2 V
4 1 V
3 2 V
4 1 V
4 2 V
3 1 V
4 2 V
3 2 V
4 1 V
3 2 V
4 2 V
3 1 V
4 2 V
4 2 V
3 2 V
4 1 V
3 2 V
4 2 V
3 2 V
4 2 V
3 2 V
4 2 V
4 2 V
3 2 V
4 2 V
3 2 V
4 2 V
3 2 V
4 2 V
3 2 V
4 2 V
4 2 V
3 2 V
4 2 V
3 2 V
4 2 V
3 3 V
4 2 V
3 2 V
4 2 V
4 2 V
3 2 V
4 3 V
3 2 V
4 2 V
3 2 V
4 3 V
3 2 V
4 2 V
3 2 V
4 2 V
4 3 V
3 2 V
4 2 V
3 2 V
4 2 V
3 3 V
4 2 V
3 2 V
4 2 V
4 2 V
3 2 V
4 3 V
3 2 V
4 2 V
3 2 V
4 2 V
3 2 V
4 2 V
4 2 V
3 2 V
4 2 V
3 2 V
4 2 V
3 2 V
4 2 V
3 2 V
4 1 V
4 2 V
3 2 V
4 2 V
3 2 V
4 1 V
3 2 V
4 2 V
3 1 V
4 2 V
4 2 V
3 1 V
4 2 V
3 2 V
4 1 V
3 2 V
4 1 V
3 2 V
4 1 V
3 2 V
4 1 V
4 2 V
3 1 V
4 1 V
3 2 V
4 1 V
3 1 V
4 1 V
3 1 V
4 0 V
4 0 V
3 0 V
4 0 V
3 -1 V
4 -1 V
3 -2 V
4 -3 V
3 -3 V
4 -5 V
LT1
480 2176 M
4 -10 V
3 -11 V
4 -10 V
3 -10 V
4 -11 V
3 -10 V
4 -10 V
3 -10 V
4 -10 V
4 -10 V
3 -10 V
4 -10 V
3 -10 V
4 -10 V
3 -10 V
4 -10 V
3 -10 V
4 -10 V
4 -9 V
3 -10 V
4 -10 V
3 -9 V
4 -10 V
3 -10 V
4 -9 V
3 -10 V
4 -9 V
3 -10 V
4 -9 V
4 -9 V
3 -10 V
4 -9 V
3 -9 V
4 -9 V
3 -9 V
4 -9 V
3 -9 V
4 -9 V
4 -9 V
3 -8 V
4 -9 V
3 -9 V
4 -8 V
3 -8 V
4 -9 V
3 -8 V
4 -8 V
4 -8 V
3 -8 V
4 -8 V
3 -8 V
4 -8 V
3 -8 V
4 -7 V
3 -8 V
4 -7 V
4 -8 V
3 -7 V
4 -7 V
3 -8 V
4 -7 V
3 -7 V
4 -7 V
3 -6 V
4 -7 V
4 -7 V
3 -6 V
4 -7 V
3 -6 V
4 -7 V
3 -6 V
4 -6 V
3 -6 V
4 -6 V
3 -6 V
4 -6 V
4 -6 V
3 -6 V
4 -5 V
3 -6 V
4 -5 V
3 -6 V
4 -5 V
3 -5 V
4 -6 V
4 -5 V
3 -5 V
4 -5 V
3 -5 V
4 -4 V
3 -5 V
4 -5 V
3 -4 V
4 -5 V
4 -4 V
3 -5 V
4 -4 V
3 -4 V
4 -4 V
3 -4 V
4 -4 V
3 -4 V
4 -4 V
4 -3 V
3 -4 V
4 -4 V
3 -3 V
4 -4 V
3 -3 V
4 -3 V
3 -3 V
4 -4 V
4 -3 V
3 -3 V
4 -3 V
3 -2 V
4 -3 V
3 -3 V
4 -2 V
3 -3 V
4 -2 V
3 -3 V
4 -2 V
4 -2 V
3 -3 V
4 -2 V
3 -2 V
4 -2 V
3 -1 V
4 -2 V
3 -2 V
4 -2 V
4 -1 V
3 -2 V
4 -1 V
3 -2 V
4 -1 V
3 -1 V
4 -1 V
3 -1 V
4 -1 V
4 -1 V
3 -1 V
4 -1 V
3 0 V
4 -1 V
3 -1 V
4 0 V
3 -1 V
4 0 V
4 0 V
3 0 V
4 0 V
3 -1 V
4 1 V
3 0 V
4 0 V
3 0 V
4 0 V
4 1 V
3 0 V
4 1 V
3 0 V
4 1 V
3 1 V
4 1 V
3 0 V
4 1 V
3 1 V
4 2 V
4 1 V
3 1 V
4 1 V
3 2 V
4 1 V
3 2 V
4 1 V
3 2 V
4 2 V
4 2 V
3 2 V
4 2 V
3 2 V
4 2 V
3 2 V
4 2 V
3 2 V
4 3 V
4 2 V
3 3 V
4 3 V
3 2 V
4 3 V
3 3 V
4 3 V
3 3 V
4 3 V
4 3 V
3 3 V
4 3 V
3 4 V
4 3 V
3 4 V
4 3 V
3 4 V
4 4 V
3 3 V
4 4 V
4 4 V
3 4 V
4 4 V
3 4 V
4 4 V
3 5 V
4 4 V
3 4 V
4 5 V
4 4 V
3 5 V
4 5 V
3 5 V
4 4 V
3 5 V
4 5 V
3 5 V
4 5 V
4 6 V
3 5 V
4 5 V
3 6 V
4 5 V
3 6 V
4 5 V
3 6 V
4 6 V
4 5 V
3 6 V
4 6 V
3 6 V
4 6 V
3 6 V
4 7 V
3 6 V
4 6 V
4 7 V
3 6 V
4 7 V
3 6 V
4 7 V
3 7 V
4 6 V
3 7 V
4 7 V
3 7 V
4 7 V
4 7 V
3 7 V
4 8 V
3 7 V
4 7 V
3 8 V
4 7 V
3 8 V
4 7 V
4 8 V
3 7 V
4 8 V
3 8 V
4 8 V
3 8 V
4 7 V
3 8 V
4 8 V
4 8 V
3 8 V
4 9 V
3 8 V
4 8 V
3 8 V
4 8 V
3 9 V
4 8 V
4 8 V
3 8 V
4 9 V
3 8 V
4 8 V
3 9 V
4 8 V
3 8 V
4 9 V
4 8 V
3 8 V
4 8 V
3 8 V
4 9 V
3 8 V
4 8 V
3 8 V
4 8 V
3 7 V
4 8 V
4 8 V
3 7 V
4 8 V
3 7 V
4 7 V
3 7 V
4 7 V
3 7 V
4 7 V
4 7 V
3 6 V
4 7 V
3 6 V
4 6 V
3 6 V
4 6 V
3 7 V
4 6 V
LT2
480 1938 M
4 -27 V
3 -28 V
4 -26 V
3 -27 V
4 -26 V
3 -26 V
4 -25 V
3 -25 V
4 -25 V
4 -24 V
3 -24 V
4 -23 V
3 -23 V
4 -23 V
3 -22 V
4 -21 V
3 -22 V
4 -21 V
4 -20 V
3 -20 V
4 -20 V
3 -19 V
4 -19 V
3 -19 V
4 -18 V
3 -18 V
4 -18 V
3 -17 V
4 -17 V
4 -17 V
3 -17 V
4 -16 V
3 -16 V
4 -15 V
3 -16 V
4 -15 V
3 -14 V
4 -15 V
4 -14 V
3 -14 V
4 -14 V
3 -14 V
4 -13 V
3 -13 V
4 -13 V
3 -13 V
4 -12 V
4 -13 V
3 -12 V
4 -12 V
3 -11 V
4 -12 V
3 -11 V
4 -12 V
3 -11 V
4 -10 V
4 -11 V
3 -11 V
4 -10 V
3 -10 V
4 -10 V
3 -10 V
4 -9 V
3 -10 V
4 -9 V
4 -10 V
3 -9 V
4 -9 V
3 -8 V
4 -9 V
3 -9 V
4 -8 V
3 -8 V
4 -8 V
3 -8 V
4 -8 V
4 -8 V
3 -7 V
4 -8 V
3 -7 V
4 -7 V
3 -7 V
4 -7 V
3 -7 V
4 -7 V
4 -7 V
3 -6 V
4 -6 V
3 -7 V
4 -6 V
3 -6 V
4 -6 V
3 -5 V
4 -6 V
4 -6 V
3 -5 V
4 -5 V
3 -6 V
4 -5 V
3 -5 V
4 -5 V
3 -5 V
4 -4 V
4 -5 V
3 -4 V
4 -5 V
3 -4 V
4 -4 V
3 -4 V
4 -4 V
3 -4 V
4 -4 V
4 -4 V
3 -3 V
4 -4 V
3 -3 V
4 -3 V
3 -3 V
4 -3 V
3 -3 V
4 -3 V
3 -3 V
4 -3 V
4 -2 V
3 -3 V
4 -2 V
3 -2 V
4 -2 V
3 -3 V
4 -1 V
3 -2 V
4 -2 V
4 -2 V
3 -1 V
4 -2 V
3 -1 V
4 -2 V
3 -1 V
4 -1 V
3 -1 V
4 -1 V
4 -1 V
3 0 V
4 -1 V
3 0 V
4 -1 V
3 0 V
4 0 V
3 -1 V
4 0 V
4 0 V
3 1 V
4 0 V
3 0 V
4 1 V
3 0 V
4 1 V
3 0 V
4 1 V
4 1 V
3 1 V
4 1 V
3 1 V
4 2 V
3 1 V
4 2 V
3 1 V
4 2 V
3 2 V
4 2 V
4 1 V
3 3 V
4 2 V
3 2 V
4 2 V
3 3 V
4 2 V
3 3 V
4 3 V
4 3 V
3 3 V
4 3 V
3 3 V
4 3 V
3 3 V
4 4 V
3 3 V
4 4 V
4 4 V
3 3 V
4 4 V
3 4 V
4 5 V
3 4 V
4 4 V
3 4 V
4 5 V
4 5 V
3 4 V
4 5 V
3 5 V
4 5 V
3 5 V
4 6 V
3 5 V
4 5 V
3 6 V
4 6 V
4 5 V
3 6 V
4 6 V
3 6 V
4 6 V
3 7 V
4 6 V
3 7 V
4 6 V
4 7 V
3 7 V
4 7 V
3 7 V
4 7 V
3 7 V
4 8 V
3 7 V
4 8 V
4 7 V
3 8 V
4 8 V
3 8 V
4 8 V
3 9 V
4 8 V
3 9 V
4 8 V
4 9 V
3 9 V
4 9 V
3 9 V
4 9 V
3 9 V
4 10 V
3 10 V
4 9 V
4 10 V
3 10 V
4 10 V
3 10 V
4 11 V
3 10 V
4 11 V
3 11 V
4 10 V
3 11 V
4 12 V
4 11 V
3 11 V
4 12 V
3 12 V
4 11 V
3 12 V
4 13 V
3 12 V
4 12 V
4 13 V
3 13 V
4 12 V
3 13 V
4 14 V
3 13 V
4 13 V
3 14 V
4 14 V
4 14 V
3 14 V
4 14 V
3 15 V
4 14 V
3 15 V
4 15 V
3 15 V
4 16 V
4 15 V
3 16 V
4 16 V
3 16 V
4 16 V
3 16 V
4 17 V
3 17 V
4 17 V
4 17 V
3 17 V
4 17 V
3 18 V
4 18 V
3 18 V
4 18 V
3 19 V
4 18 V
3 19 V
4 19 V
4 19 V
3 19 V
4 20 V
3 20 V
4 19 V
3 20 V
4 20 V
3 20 V
4 20 V
4 21 V
3 20 V
4 20 V
3 20 V
4 21 V
3 20 V
4 20 V
3 20 V
4 19 V
stroke
grestore
end
showpage
}
\put(1708,51){\makebox(0,0){{\fs $\theta$}}}
\put(1617,151){\makebox(0,0){{\fs $170^\circ$}}}
\put(1049,151){\makebox(0,0){{\fs $90^\circ$}}}
\put(480,151){\makebox(0,0){{\fs $10^\circ$}}}
\put(420,2541){\makebox(0,0)[r]{{\fs $10$}}}
\put(420,1969){\makebox(0,0)[r]{{\fs $0$}}}
\put(420,1396){\makebox(0,0)[r]{{\fs $-10$}}}
\put(420,823){\makebox(0,0)[r]{{\fs $-20$}}}
\put(420,251){\makebox(0,0)[r]{{\fs $-30$}}}
\end{picture}}
\end{picture}
\caption{Differential lowest-order \css\ and relative 
corrections for the \css\ with opposite photon helicities}
\label{fi:difcsmm}
\end{figure}
Whenever the differential \cs\ is sizable, $\delta$ is of the
order of $10\%$. The corrections are in particular small in the forward
and backward direction for all \css\ that involve $t$- and $u$-channel
poles in lowest order. On the other hand, the corrections get very large
when the lowest-order \cs\ is suppressed or tends to zero, in particular
for $\rd\si_{\pm\pm\rL\rL}/\rd\Omega$ 
at high energies and intermediate scattering angles.
The maximal corrections are usually reached for central values of the
scattering angle.
In accordance with the discussion in \refse{se:notcon}, the corrections are
forward-backward symmetric for equal photon helicities.
For opposite photon helicities they
include an asymmetric contribution originating from box diagrams and
$AWW$ vertex corrections involving fermion loops.
The corrections for two negative helicity photons are equal to those for
two positive helicity photons. As a consequence of Bose symmetry, the
corrections to $\rd\si_{+-}/\rd\Omega$ are obtained from those to 
$\rd\si_{-+}/\rd\Omega$ upon 
exchanging $u$ and $t$, \ie $\theta\leftrightarrow 180^\circ-\theta$. 
Thus, the unpolarized \cs\ is forward--backward-symmetric.
 
\begin{table}
\footnotesize
\newdimen\digitwidth
\setbox0=\hbox{0}
\digitwidth=\wd0
\catcode`!=\active 
\def!{\kern\digitwidth}
\newdimen\minuswidth
\setbox0=\hbox{$-$}
\minuswidth=\wd0
\catcode`?=\active 
\def?{\kern\minuswidth}
\begin{center}
\arraycolsep 6pt
$$\begin{array}{|c|c||c|c|c|c|c|c|}
\hline
\sqrt{s}/\mathrm{GeV} & \theta & \sigma^{\mathrm{Born}}/\mathrm{pb} &
\delta_{\Delta E = 0.1E} /\% & 
\delta_{\cut} /\% & 
\delta_{\Delta E = E} /\% & 
\delta_{\bos}/\% &
\delta_{\ferm}/\% \\
\hline\hline
       & !5^\circ &  98.13  & !?0.02 &  -2.79 & ?!2.81 & ?!1.49 &  ?1.32 \\
 \cline{2-8}
       & 20^\circ &  26.04  & !-2.68 &  -2.79 & ?!0.11 & !-0.08 &  ?0.19 \\
\cline{2-8}
  !500 & 90^\circ &  0.724  & -10.79 &  -2.79 & !-8.00 & !-5.62 &  -2.38 \\
\cline{2-8}
       & !0^\circ< \theta <180^\circ &  77.55     & !-3.38 &   -2.79 
& !-0.59 & !-0.65 &   ?0.06  \\
\cline{2-8}
       & 10^\circ< \theta <170^\circ &  60.74     & !-4.27 &   -2.79 &
  !-1.48 & !-1.21 &   -0.27  \\
\cline{2-8}
       & 20^\circ< \theta <160^\circ &  36.67     & !-6.06 &  -2.79 
& !-3.27 & !-2.39 &   -0.89 \\
\hline\hline
       & !5^\circ &  291.9  & !-2.06 &  -4.31 & ?!2.25 & ?!1.04 &  ?1.21 \\
\cline{2-8}
       & 20^\circ &  15.61  & -11.90 &  -4.31 & !-7.59 & !-6.37 &  -1.22 \\
\cline{2-8}
 1000  & 90^\circ &  0.193  & -31.64 &  -4.31 & -27.33 & -21.93 &  -5.40 \\
\cline{2-8}
       & !0^\circ< \theta <180^\circ &  80.05    & !-7.08 &  -4.31 
& !-2.77 & !-2.71 &   -0.06 \\ 
\cline{2-8}
       & 10^\circ< \theta <170^\circ &  37.06    & -12.26 &  -4.31 
& !-7.95 & !-6.65 &   -1.30 \\
\cline{2-8}
       & 20^\circ< \theta <160^\circ & 14.16     & -19.29 &  -4.31 
& -14.98 & -12.20 &  -2.78 \\
\hline\hline
       & !5^\circ &  418.8 & !-7.14 &  -5.80 & !-1.33 & !-1.59 &  ?0.25 \\
\cline{2-8}
       & 20^\circ &  5.163 & -30.31 &  -5.80 & -24.51 & -20.96 &  -3.55 \\
\cline{2-8}
 2000  & 90^\circ &  0.049 & -59.59 &  -5.80 & -53.78 & -45.47 &  -8.32 \\ 
\cline{2-8}
       & !0^\circ< \theta <180^\circ &  80.59     & !-9.85 &  -5.80 
& !-4.04 & !-3.95 &  -0.09 \\
\cline{2-8}
       & 10^\circ< \theta <170^\circ &  14.14     & -27.15 &  -5.80 
& -21.35 & -18.34 &  -3.01 \\
\cline{2-8}
       & 20^\circ< \theta <160^\circ &  4.068     & -41.22 &  -5.80 
& -35.41 & -30.12 &  -5.29 \\
\hline
\end{array}$$
\caption{Lowest-order \css\ and relative corrections for unpolarized
particles}
\label{ta:born}
\end{center}
\end{table}
In \refta{ta:born} we list the unpolarized \cs\ and the corresponding 
corrections for several energies and scattering angles. We include the
corrections for a soft-photon-energy cut-off $\Delta E = 0.1E$, the 
cut-off-dependent corrections $\de_{\cut}$ from \refeq{eq:cut}, and the 
individual (gauge-invariant) fermionic $\delta_{\ferm}$ 
and bosonic corrections  $\delta_{\bos}$.
The fermionic corrections consist of all loop diagrams and \ct\ 
contributions involving fermion loops, all other contributions form the 
bosonic corrections.
The fermionic corrections stay below 5--10\% even for high energies. 
On the other hand, the bosonic contributions are responsible for the
large corrections at high energies, in
particular in the central angular region. 
 
In \citere{Ye91} various observables have been investigated in view of
their sensitivity to anomalous couplings, 
involving the total \cs\ and the following ratios%
\footnote{Note that we do not perform a convolution with a realistic
\label{fo:3}%
photon spectrum but consider the incoming photons as monochromatic.}
\beqar
R_{\mathrm{IO}}&=& \frac{\si(|\!\cos\theta|<0.4)}{\si(|\!\cos\theta|<0.8)}, \\
R_{\mathrm{LT}}&=& \frac{\si_{\rL\rL}}{\si_{\rT\rT}}, \\
R_{\mathrm{02}}&=& \frac{\si_{++}}{\si_{+-}} .
\eeqar
We list the lowest-order predictions together with the $\Oa$-corrected
ones and the relative corrections for these observables in
\refta{ta:obs} using $|\!\cos\theta_{\cut}|=0.8$.
\begin{table}
\newdimen\digitwidth
\setbox0=\hbox{0}
\digitwidth=\wd0
\catcode`!=\active 
\def!{\kern\digitwidth}
\newdimen\minuswidth
\setbox0=\hbox{$-$}
\minuswidth=\wd0
\catcode`?=\active 
\def?{\kern\minuswidth}
\begin{center}
\arraycolsep 6pt
$$\begin{array}{|c|c||c|c|c|c|}
\hline
\sqrt{s}/\mathrm{GeV} & & \sigma/\mathrm{pb} &
R_{\mathrm{IO}} & R_{\mathrm{LT}} & R_{02} \\
\hline\hline
       & \mathrm{Born~level}  & ?15.74  & ?0.265  &  0.0308  & ?1.934 \\
\cline{2-6}
 !500  & \mathrm{corrected}   & ?14.82  & ?0.259  &  0.0325  & ?1.950 \\
\cline{2-6}
       & \mathrm{corrections/\%} & !-5.83  & !-2.02  &  !5.43!  & ?0.78! \\
\hline\hline
       & \mathrm{Born~level}  & ?4.659  & ?0.241  &  0.0235  & ?2.229 \\
\cline{2-6}
 1000  & \mathrm{corrected}   & ?3.617  & ?0.227  &  0.0276  & ?2.184 \\
\cline{2-6}
       & \mathrm{corrections/\%} & -22.36  & !-5.64  &  17.08!  & -2.05! \\
\hline\hline
       & \mathrm{Born~level}  & ?1.218  & ?0.234  &  0.0220  & ?2.307 \\
\cline{2-6}
 2000  & \mathrm{corrected}   & ?0.647  & ?0.207  &  0.0321  & ?2.168 \\
\cline{2-6}
       & \mathrm{corrections/\%} & -46.86  &  -11.53  &  46.11!  & -6.02!\\
\hline
\end{array}$$
\caption{Lowest-order predictions and corresponding  corrections for various
observables and $|\!\cos\theta_{\cut}|=0.8$}
\label{ta:obs}
\end{center}
\end{table}

In \reftas{ta:MTvar.mw} -- \ref{ta:MHvar.gf} we show the variation of the 
SM corrections with the top-quark and Higgs-boson masses at
$\sqrt{s}=500\GeV$ 
in per cent of the \cs\ for our standard set of parameters \refeq{eq:par}.
We have determined this variation by searching the maximum and minimum
\css\ in the range $130\GeV < \Mt < 210\GeV$ for the variation with
$\Mt$ and in the ranges $60\GeV < \MH < 400\GeV$ and $600\GeV < \MH <
1000\GeV$ for the variation with $\MH$. The range $400\GeV < \MH <
600\GeV$ has been left out as there the Higgs-mass dependence is
dominated by the Higgs resonance. Because the resonance dominates
$\si_{\pm\pm\rL\rL}$ in an even wider range, we have omitted this
\cs\ in the tables for the Higgs dependence.

\btab
$$ \begin{array}{|c||*6{c|}}
\hline
$\MW$ \mathrm{~fixed}
&\rU\rU\rU\rU & +{+}\rT\rT & +{+}\rL\rL & +{-}\rT\rT & 
+{-}\rL\rL & +{-}(\rL\rT+\rT\rL) \\
\hline
\theta=20^\circ & 0.15\% & 0.18\%  & 0.46\%  & 0.14\% &  0.53\% &  0.06\% \\
\theta=90^\circ & 0.62\% & 0.58\%  & 3.52\% & 0.44\% & 1.50\% &  0.29\% \\
\mbox{integrated over} &  & & &  &  & \\
10^\circ<\theta<170^\circ
&  0.22\% & 0.26\% & 1.06\% & 0.16\% & 1.23\% &  0.25\% \\
20^\circ<\theta<160^\circ 
&  0.29\% & 0.33\% & 1.48\% & 0.21\% & 1.23\% &  0.26\% \\
\hline
\earr $$
\caption{Variation of various polarized \css\ at $\ECMS = 500\GeV$
for fixed $\MW$
with the top-quark mass in the range $130\GeV < \Mt < 210\GeV$ in per cent 
of the \cs\ for $\Mt = 174\GeV$}
\label{ta:MTvar.mw}
\etab
\btab
$$ \begin{array}{|c||*5{c|}}
\hline
$\MW$ \mathrm{~fixed}
&\rU\rU\rU\rU & +{+}\rT\rT & +{-}\rT\rT & 
+{-}\rL\rL & +{-}(\rL\rT+\rT\rL) \\
\hline
\theta=20^\circ & 0.16\% & 0.18\%  & 0.13\%  & 0.37\% &  0.81\%  \\
\theta=90^\circ & 0.44\% & 0.21\%  & 0.34\% & 2.62\% & 1.51\%  \\
\mbox{integrated over} &  & & &  &  \\
10^\circ<\theta<170^\circ
&  0.09\% & 0.12\% & 0.10\% & 2.04\% & 0.37\% \\
20^\circ<\theta<160^\circ 
&  0.07\% & 0.09\% & 0.06\% & 2.06\% & 0.52\% \\
\hline
\earr $$
\caption{Variation of various polarized \css\ at $\ECMS = 500\GeV$
for fixed $\MW$
with the Higgs-boson mass
in the ranges $60\GeV < \MH < 400\GeV$ and $600\GeV < \MH <
1000\GeV$ in per cent
of the \cs\ for $\MH = 250\GeV$}
\label{ta:MHvar.mw}
\etab
In \reftas{ta:MTvar.mw} and \ref{ta:MHvar.mw} the $\PW$-boson mass is
kept fixed at $\MW=80.22\GeV$. Then,
as argued in the previous section, the variation is 
small owing to the absence of large top- and Higgs-mass-dependent
corrections. The larger variation of the \css\ 
involving longitudinal \PW~bosons
is due to terms proportional to $\Mt^2/\MW^2$ or $\MH^2/\MW^2$ 
arising as a remnant of the unitarity cancellations for
$\sqrt{s}\gg\Mt,\MH$. 
These terms induce a sizable variation of these \css\
with $\Mt$ and $\MW$ at higher energies.

\btab
$$ \begin{array}{|c||*6{c|}}
\hline
$\GF$ \mathrm{~fixed}
&\rU\rU\rU\rU & +{+}\rT\rT & +{+}\rL\rL & +{-}\rT\rT & 
+{-}\rL\rL & +{-}(\rL\rT+\rT\rL) \\
\hline
\theta=20^\circ & 1.20\% & 1.24\%  & 1.39\%  & 1.17\% &  0.45\% &  0.05\% \\
\theta=90^\circ & 0.18\% & 0.06\%  & 3.23\%  & 0.20\% & 1.56\% &  1.22\% \\
\mbox{integrated over} &  & & &  &  & \\
10^\circ<\theta<170^\circ
                & 0.98\% & 0.94\% & 1.62\% & 1.12\% & 1.17\% &  0.84\% \\
20^\circ<\theta<160^\circ 
                & 0.52\% & 0.51\% & 2.03\% & 0.67\% & 1.19\% &  0.95\% \\
\hline
\earr $$
\caption{Same as in \protect\refta{ta:MTvar.mw} but now for fixed $\GF$}
\label{ta:MTvar.gf}
\etab
\btab
$$ \begin{array}{|c||*5{c|}}
\hline
$\GF$ \mathrm{~fixed}
&\rU\rU\rU\rU & +{+}\rT\rT & +{-}\rT\rT & 
+{-}\rL\rL & +{-}(\rL\rT+\rT\rL) \\
\hline
\theta=20^\circ & 0.33\% & 0.33\%  & 0.35\% &  0.73\% &  0.81\% \\
\theta=90^\circ & 0.56\% & 0.36\%  & 0.55\% & 2.63\% &  1.26\% \\
\mbox{integrated over} &  & &  &  & \\
10^\circ<\theta<170^\circ
&  0.36\% & 0.34\% & 0.38\% & 2.09\% & 0.25\% \\
20^\circ<\theta<160^\circ 
&  0.31\% & 0.27\% & 0.32\% & 2.10\% & 0.37\% \\
\hline
\earr $$
\caption{Same as in \protect\refta{ta:MHvar.mw} but now for fixed $\GF$}
\label{ta:MHvar.gf}
\etab
The variations of the corrections for fixed $\GF$ are shown in
\reftas{ta:MTvar.gf} and \ref{ta:MHvar.gf}. 
It is larger in particular for the \css\ for purely transverse
\PW~bosons. This fact results
from the dependence of $\MW$ on $\Mt$ and $\MH$ that involves
logarithmic top- and Higgs-mass-dependent terms and terms  proportional
to $\Mt^2/\MW^2$. 

To visualize the Higgs resonance, we plot in \reffi{fi:intcshiggs} the \cs\ 
including \Oa\ corrections
integrated over $20^\circ <\theta <160^\circ$ for various values of the
Higgs-boson mass. 
\begin{figure}
\setlength{\unitlength}{1mm}
\begin{picture}(160,110)(0,0)
\put(0,10){\input{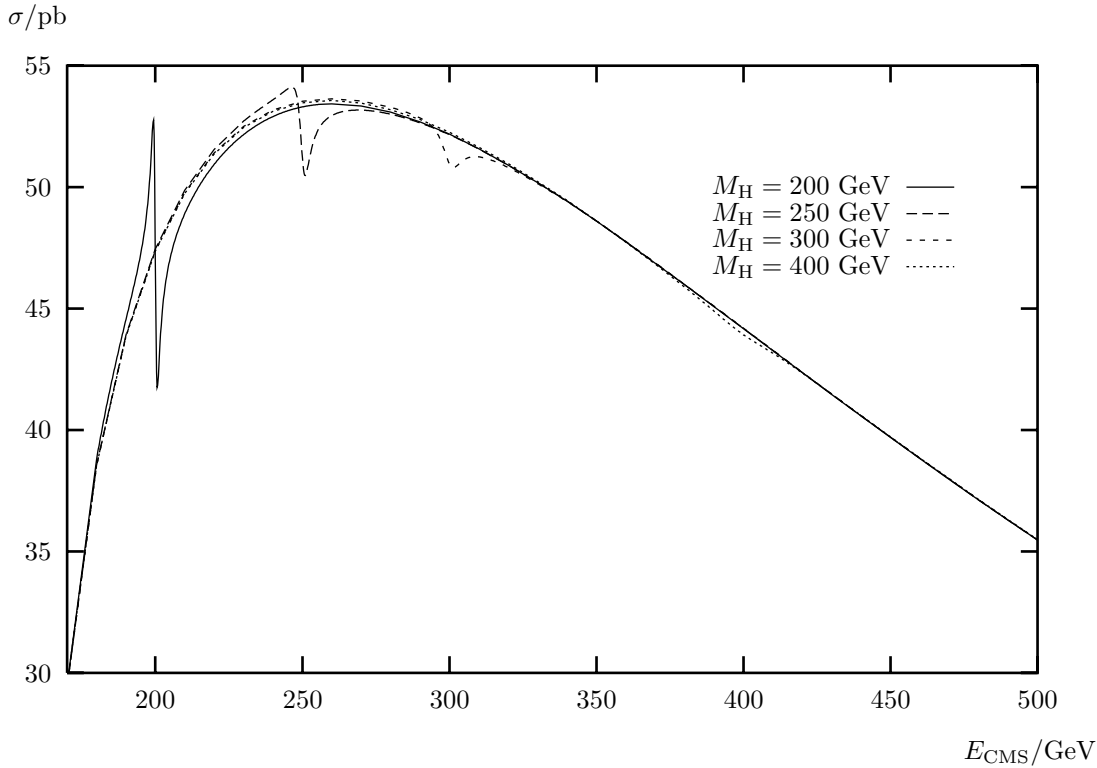}}
\end{picture}
\caption{Integrated unpolarized \cs\ including $\Oa$ corrections
for various Higgs-boson masses ($20^\circ<\theta<160^\circ$)}
\label{fi:intcshiggs}
\end{figure}
Our results agree qualitatively%
\footnote{See footnote \ref{fo:3}.}
well with those of \citere{Mo94}.
While the resonance
is comparably sharp at small energies, it is washed out by the large width 
of the Higgs boson at high energies. Already for $\MH=400\GeV$ 
the Higgs resonance is hardly visible in $\AAWW$.

\section{Summary}
 
The process $\AAWW$ will be one of the most interesting reactions at future
$\ga\ga$ colliders.
In particular, it is very useful to study triple and quartic 
non-Abelian gauge couplings.
 
We have calculated the one-loop radiative corrections to $\AAWW$ within the
electroweak Standard Model in the soft-photon approximation for arbitrary
polarizations of the photons and \PW~bosons. 
By using a non-linear gauge-fixing term the number of contributing
diagrams can be reduced by roughly a factor of two.
An interesting peculiarity of $\AAWW$
is the absence of most (universal) leading corrections,
such as leading logarithms of light quark masses associated with the
running of $\al$ and leading logarithms associated with collinear
bremsstrahlung. 
Therefore, the theoretical predictions are very clean.
 
In the heavy mass limit no leading $\Mt^2$- and 
$\log\Mt$-terms and $\log\MH$-terms exist.
Consequently, 
the variation of the \css\ with the top-quark and Higgs-boson masses
is small if $\MW$ is kept fixed with the exception of the \css\
involving longitudinal \PW~bosons at high energies. For fixed $\GF$
the variation arises mainly from the variation of $\MW$ with 
$\Mt$ and $\MH$ and is thus of similar origin like 
the one of $\eeff$ or $\eeWW$.

We have presented a detailed numerical discussion of the lowest-order
\css\ and the virtual and soft-photonic corrections to $\AAWW$.
The soft-photon-cut-off-independent radiative corrections to the
total \cs\ are of the order of 10\%. They are 
increased at high energies if the forward and backward regions are 
excluded by an angular cut.
This is due to the fact that at high energies the radiative corrections
reach several
10\% for intermediate scattering angles whereas they
are at the level of several per cent in the forward and backward direction 
which dominate the total \cs. 
The large corrections are caused by bosonic loop diagrams whereas the
effects of the fermionic diagrams are of the order of 5--10\%.

\section*{Acknowledgement}
We are grateful to M.~B\"ohm for useful discussions.

\end{document}